\documentclass[journal]{IEEEtran}

\usepackage{epsfig,amsmath,amssymb,epsf,amsthm,scalefnt,multirow,subfig}
\usepackage{xcolor}
\usepackage{float}
\usepackage{cite}
\usepackage{psfrag}
\usepackage{graphics,amsmath,amssymb,graphicx,amsthm,scalefnt,multirow,dsfont,subfig}
\usepackage{algorithm}
\usepackage{algpseudocode}
\usepackage{latexsym,amsmath}
\usepackage{stfloats}

\newcommand{\argmax}{\mathop{\mathrm{argmax}}}
\newcommand{\argmin}{\mathop{\mathrm{argmin}}}

\newtheorem{lemma}{Lemma}
\newtheorem*{lemma*}{Lemma}

\def\argmin{\mathop{\mathrm{argmin}}}
\def\argmax{\mathop{\mathrm{argmax}}}

\def\b0{{\pmb{0}}} 

\def\ba{{\mathbf{a}}}  \def\bc{{\mathbf{c}}} \def\bd{{\mathbf{d}}}
\def\bee{{\mathbf{e}}} \def\bff{{\mathbf{f}}} \def\bg{{\mathbf{g}}} 
   
 \def\bn{{\mathbf{n}}}  
\def\bq{{\mathbf{q}}} \def\br{{\mathbf{r}}}  
\def\bu{{\mathbf{u}}} \def\bv{{\mathbf{v}}} \def\bw{{\mathbf{w}}} \def\bx{{\mathbf{x}}}
\def\by{{\mathbf{y}}} \def\bz{{\mathbf{z}}}  

\def\bA{{\mathbf{A}}} \def\bB{{\mathbf{B}}}  \def\bD{{\mathbf{D}}}
 \def\bF{{\mathbf{F}}}  \def\bH{{\mathbf{H}}}
\def\bI{{\mathbf{I}}}  \def\bK{{\mathbf{K}}} 
   
 \def\bR{{\mathbf{R}}}  
 \def\bV{{\mathbf{V}}} \def\bW{{\mathbf{W}}} \def\bX{{\mathbf{X}}}
\def\bY{{\mathbf{Y}}} \def\bZ{{\mathbf{Z}}}

\def\red{{}}

\DeclareMathOperator{\E}{\mathbb{E}}

\begin{document}
	
	%\documentclass{singlecolumn}
	%\onecolumn
	% paper title
	\title{Beam Designs for Millimeter-Wave Backhaul with Dual-Polarized Uniform Planar Arrays
	\thanks{This work was supported by the National Research Foundation (NRF) grant funded by the MSIT of the Korea government (2019R1C1C1003638).}	
	}
	
	\author{Sucheol Kim,~\IEEEmembership{Student Member,~IEEE}, Junil Choi,~\IEEEmembership{Senior Member,~IEEE} and Jiho Song,~\IEEEmembership{Member,~IEEE}
		%	\IEEEauthorblockA{vxvxvxx (Wireless Networking and Communications Group)\\
		%		vxvxvxvxvxx (Pohang University of Science and Technology, *University of Ulsan)\\
		%		Emails: \{recusmik,junil\}@postech.ac.kr, *jihosong@ulsan.ac.kr}
		%	\IEEEauthorblockA{$^{\dag}$Department of Electrical and Computer Engineering, The University of Texas at Austin, Austin, TX}
		% <-this % stops a space
		\thanks{S. Kim and J. Choi are with the Department of Electrical Engineering, Korea Advanced Institute of Science and Technology (e-mail: \{loehcusmik, junil\}@kaist.ac.kr).}
		\thanks{J. Choi is the corresponding author.}
		\thanks{J. Song is with the Department of Electrical Engineering, University of Ulsan (e-mail: jihosong@ulsan.ac.kr).}}	
	%author names and IEEE memberships
	\maketitle
	
	%\squeezeup\squeezeup\squeezeup\squeezeup
	\begin{abstract}
		This paper proposes hybrid beamforming designs for millimeter-wave (mmWave) multiple-input multiple-output (MIMO) backhaul systems equipped with uniform planar arrays (UPAs) of dual-polarization antennas at both the transmit and receive base stations. 
		The proposed beamforming designs are to near-optimally solve optimization problems taking the dual-polarization UPA structure into account. 
		Based on the solutions of optimization problems, this paper shows it is possible to generate the optimal dual-polarization beamformer from the optimal single-polarization beamformer sharing the same optimality.
		As specific examples, squared error and magnitude of inner product are considered respectively for optimization criteria. 
		To optimize proposed beamformers, partial channel information is needed, and the use of low overhead pilot sequences is also proposed to figure out the required information.
%		For each optimization criterion, the generation of the dual-polarization beamformers form single-polarization beamformers are derived.
		Simulation results verify that the resulting beamformers have the most uniform gain (with the squared error criterion) or the highest average gain (with the magnitude of inner product criterion) in the covering region with the UPA of dual-polarization antennas.
	\end{abstract}
	
	
	\section{Introduction}\label{sec1}
	
	%Backhaul is a communication link between base stations (BSs). The radius of cell is usually several kilometers, and the backhaul communication should support the distance twice the cell radius. Small cell networks (literally with small cell radius), however, are considered as a promising strategy to reply the rising demand for high data rates \cite{S.Hur:2013, S.Hur:2011, X.Ge:2014}. A macro-cell is divided into multiple cells, and each cell has its own BS. In other words, small cell networks adopt multiple BSs in a macro-cell. This reduces the number of users in a cell and the distance between a user and a BS, and these reduction contribute to attaining high data rate. Although the communication distance shortens, the number of backhauls increase. Most of the current backhauls are mechanical links supporting high data rates, i.e. conventional optical fiber \cite{Z.Gao:2015}, but mechanical links for numerous backhauls cost impractically \cite{Z.Gao:2015, S.Chia:2009}. Besides, the installation, movement, and demolition of BS in small cell networks would be more frequent than in the ordinary large cell networks. In terms of the cost efficiency, a solution is wireless links. Millimeter-wave (mmWave) wireless communication, especially, can be considered \cite{S.Hur:2013, R.Taori:2015, X.Ge:2014} for both high data rate and bearable cost. 
	Cell densification has been a most direct and practical way of supporting the exponential growth of mobile devices and data rates \cite{J.G.Andrews:2014,S.Hur:2013,X.Ge:2014}. With more cells, both macro and small cells, it is important to deploy cost-efficient backhauls among cells since conventional backhauls using optical fibers are expensive. In terms of cost efficiency, a simple solution is using wireless links for backhauls. Especially, millimeter-wave (mmWave) wireless communications can be considered for backhauls satisfying both high data rates and bearable cost \cite{S.Hur:2013, R.Taori:2015, X.Ge:2014}.
	
	MmWave communications use a carrier frequency of 30 to 300 GHz and the corresponding wavelength of 1 to 10 mm. With its huge bandwidth, mmWave communications can support enormous data rates \cite{T.S.Rapport:2011, Z.Pi:2011, H.Huang:2015}. An important characteristic of mmWave systems is its high attenuation \cite{H.Zhang:2010,H.Huang:2015}, which requires sharp beam patterns to concentrate signal power and compensate for the attenuation. To enjoy the benefit of the large bandwidth, hence, beamforming techniques are necessary to overcome harsh mmWave environments. 
	
	Many beamforming designs have been proposed under various environments. Beamformers were designed to maximize capacity using limited information of channels at the transmitter in \cite{S.A.Jafar:2001,J.Choi:2005}. 
%	In \cite{J.Choi:2005}, the transmitter constructs a beamformer from the limited information about the optimal beamformer, which is obtained at the receiver using channel information and fed back to the transmitter. 
	These beamformers were based on an assumption of rich scattering environments, e.g., sub-6 GHz spectrum, and the resulting beamformers rarely have \textit{physical} beam-like patterns. In the mmWave systems, on the contrary, channels can be represented with their dominant line-of-sight (LOS) component \cite{J.Song:2017} or sum of a few dominant components \cite{T.Kim:2013,A.Alkhateeb:2014}. This leads the beamformer design problem to consider graphical or geometrical shape. The codebook-based beamformings in \cite{S.Hur:2013,L.Chen:2011}, for example, find the best beamformer by gradually narrowing the beamwidth of possible beamformers. In \cite{Z.Xiao:2016,S.Noh:2017,J.Zhang:2017,Z.Xiao:2017}, based on the small numbers of channel components, codebook designs for the beam alignment and channel estimation are proposed. 
	In this paper, we also approximate the channel as its LOS component and design beamformers based on physical beam patterns. 

	Since the beamforming depends on the combination of weights for each antenna, use of multiple antennas is essential.
	In mmWave system, by virtue of its extremely small wavelength, it is possible to deploy large numbers of antennas of uniform linear arrays (ULAs) or uniform planar arrays (UPAs) within a small form factor. 
	%When the digital beamforming is adopted, however, each antenna requires a dedicated RF chain \cite{O.E.Ayach:2014}. In this case, exploiting multiple antennas involves use of the same number of RF chains, and rise of power consumption and cost follows. Considering the use of numerous antennas in mmWave systems, the digital beamforming is practically infeasible \cite{C.H.Doan:2004,J.Nsenga:2010}. 
	Considering the use of numerous antennas in mmWave systems, the digital beamforming is practically infeasible due to the cost and power consumption of RF chains \cite{O.E.Ayach:2014,C.H.Doan:2004,J.Nsenga:2010}.
	%A feasible solution is the analog beamforming, which requires only one RF chain in return for abandonment of amplitude control. The following constant modulus property of analog beamforming supports single data stream only \cite{S.Hur:2013} and largely reduces the variety of beam pattern shape. 
	One feasible solution is the analog beamforming, which requires only one RF chain in return for abandonment of amplitude control and variety of beam pattern shapes \cite{S.Hur:2013,H.Huang:2015}.
	%To accomplish both feasibility and diversity of beam pattern shape, hybrid beamforming can be suggested. The hybrid beamforming reduces the use of RF chains by conjoining the digital beamforming and the analog beamforming. This conjunction enables control of both amplitude and phase, which allows precoding multiple data streams \cite{A.Alkhateeb:2013}.
	The hybrid beamforming, which reduces the use of RF chains by conjoining the digital and analog beamformings, is another practical solution to accomplish both the feasibility and diversity of beam pattern shapes \cite{A.Alkhateeb:2014,X.Yu:2016,A.Alkhateeb:2015,S.He:2017,Y.Lee:2015,C.Rusu:2016,S.Noh:2017,Z.Xiao:2016,Z.Xiao:2017,J.Zhang:2017}.
	For example, the algorithm in \cite{J.Zhang:2017} iteratively updates a beamformer to reduce the ripple and raise average gain of beam pattern in the covering region, and the beamformer in \cite{Z.Xiao:2017} is designed to minimize the mean squared error (MSE) of its beam pattern in the covering region. The two designs, however, are focused on the single-polarization ULA structure.
	
	Further increase of the number of the antennas in a limited form factor is possible by exploiting dual-polarization antennas \cite{C.Oestges:2008}. The channel of dual-polarization antennas, though, is fundamentally different from that of single-polarization antennas \cite{B.Clerckx:2008}. For example, the imbalance of channel gains and the orientation difference between the transmit (Tx) and receive (Rx) antennas need to be considered, each of which changes the structure of channel and causes additional randomness. \red{The complicated channel structure hinders the decomposition of Tx and Rx beamforming gain, making it difficult to design Tx and Rx beamformers.}
%	Therefore, the beamforming for dual-polarization antennas requires joint design of Tx and Rx beamforming or specific approach to decompose Tx and Rx beamforming gain while considering additional characteristics of channel.
	%The beamforming for dual-polarization antennas needs to consider additional characteristics of channel, which come from the dual-polarization. 
	In \cite{T.Kim:2010}, cross-polarization discrimination (XPD), which is a parameter about the dual-polarization, is considered in the beamformer design. In \cite{B.Clerckx:2008}, the beamformer of dual-polarization is generated by modifying the beamformer of single-polarization based on the structure of the dual-polarization channel. Most of previous works on the dual-polarization beamforming, however, are based on the digital beamforming with ULAs.
	
	In this paper, we propose the hybrid beamforming designs for mmWave multiple-input multiple-output (MIMO) backhaul systems with dual-polarization UPAs. To the best of authors' knowledge, the hybrid beamforming design for mmWave MIMO system with dual-polarization UPAs has not been considered before. 
	We propose flexible beamforming design methods that can deal with variable backhaul links. Although the backhaul links are usually fixed, and using a prefixed beamformer may be sufficient for communications, the new installation, movement, or demolition of small cell base stations (BSs) would be frequent, which requires a new beamformer for each event. 
	
	To design a beamformer, we first define an ideal beam pattern, and then beamformers are optimized to mimic the ideal beam pattern. As specific examples of optimization criterion, squared error (SE) and magnitude of inner product (MIP) are considered respectively. Based on the common structure of optimization solutions, we propose a unified method to generate the optimal dual-polarization beamformer from the optimal single-polarization beamformer sharing the same optimality. The proposed design methods depend on partial channel information, and we also propose the use of pilot sequences to measure the required channel information.
%	For each criterion, the generation of the dual-polarization beamformers from single-polarization beamformers are derived in this paper. 
	Numerical results show that the SE beamformer has the most uniform beam patterns and the MIP beamformers has the highest average and peak beamforming gain for the dual-polarization UPA structure.
%	The proposed beamformers, i.e., not only the SE and MIP beamformers but also the dual-polarization beamformers generated from single-polarization beamformers, are designed to possess flexibility in order that variable backhaul links can be handled.
%	Although the backhaul links are usually fixed, and using a prefixed beamformer may be sufficient for communications, the new installation, movement, or demolition of small cell base stations (BSs) would be frequent, which requires a new beamformer for each event. 

	In the rest of the paper, system and channel models are described in Section \ref{sec2}. The details of proposed beamforming designs are explained in Section \ref{sec3}. The numerical results of the proposed beamformers are shown in Section \ref{sec4}, and the conclusion follows in Section \ref{sec5}.
	
	$\textbf{Notations}$: $\mathbb{N}$, $\mathbb{R}$, and $\mathbb{C}$ represent set of natural numbers, real numbers, and complex numbers. Matrices and vectors are written in bold face capital letters $\bA$ and bold face small letters $\ba$. $(\cdot)^*$, $(\cdot)^\mathrm{T}$, and $(\cdot)^\mathrm{H}$ mean element-wise conjugate, transpose, and Hermitian of the corresponding matrix or vector. $\otimes$ and $\odot$ represent the Kronecker product and the Hadamard product. 
	% The $a$-th column of the matrix $\bA$ is remarked as $(\bA)_{(:,a)}$, and 
	The $b$-th component of the vector $\ba$ is remarked as $(\ba)_{(b)}$. $\bI_a$ is the $a\times a$ identity matrix, $\bee_{a,b}$ is the $b$-th column of the identity matrix $\bI_a$, and $\boldsymbol{1}_a$ represents the $a\times1$ all one vector. The concatenation of matrices is denoted as $[\bA, \bB]$ where $\bA$ and $\bB$ have the same number of rows. $\Lambda_\mathrm{max} (\cdot)$ is the maximum eigenvalue of the corresponding matrix, and $\boldsymbol{\mathfrak{v}}_\mathrm{max} (\cdot)$ is the maximum eigenvector of the corresponding matrix.

	\section{System and channel models}\label{sec2}
	\subsection{System model}\label{sec2-1}
	
	\begin{figure}[t]
		\includegraphics[width=1\linewidth]{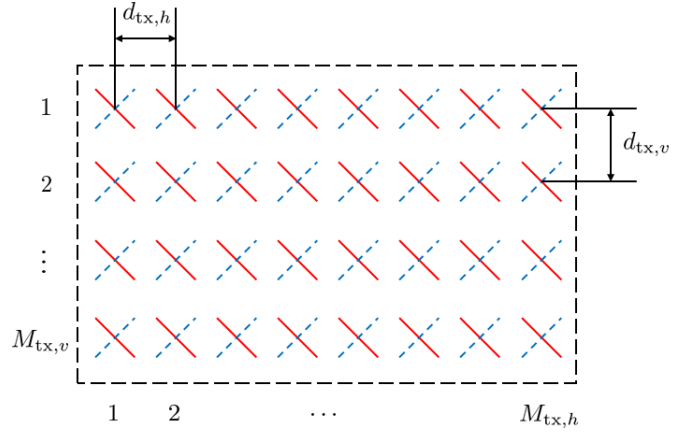}
		\caption{Dual-polarization antennas deployed in UPA.}
		\label{array}
	\end{figure}
	
	We consider MIMO backhaul links from the Tx BS equipped with dual-polarization UPA with dimension $M_{\mathrm{tx},h}\times M_{\mathrm{tx},v}$ to the Rx BS equipped with dual-polarization UPA with dimension $M_{\mathrm{rx},h}\times M_{\mathrm{rx},v}$. The total number of antennas in the Tx BS is $M_{\mathrm{tx}}=2M_{\mathrm{tx},h}M_{\mathrm{tx},v}$, with the spacing of antennas $d_{\mathrm{tx},h}$ and $d_{\mathrm{tx},v}$ as shown in Fig. \ref{array}. The number of antennas in the Rx BS is $M_{\mathrm{rx}}=2M_{\mathrm{rx},h}M_{\mathrm{rx},v}$, which have similar deployment as in the Tx BS. The Tx BS has $N_\mathrm{tx}$ RF chains, where each RF chain is fully connected to all the Tx antennas, and the Rx BS has the same structure with $N_\mathrm{rx}$ RF chains. 
%	The $N_\mathrm{rx}$ RF chains in the Rx BS have the same structure. 
	
	Assuming block fading, a received signal $y\in\mathbb{C}$ is
	\begin{align}\label{Rx signal}
		y =\sqrt{P} \bc_{\mathrm{rx}}^\mathrm{H}\bH \bc_{\mathrm{tx}} s+\bc_{\mathrm{rx}}^\mathrm{H}\bn,
	\end{align}
	where $P\in\mathbb{R}$ is the transmit power, $\bc_{\mathrm{rx}}\in\mathbb{C}^{M_\mathrm{rx}\times1}$ is the unit-norm receive beamformer, $\bH\in\mathbb{C}^{M_\mathrm{rx}\times M_\mathrm{tx}}$ is the channel matrix, $\bc_{\mathrm{tx}}\in\mathbb{C}^{M_\mathrm{tx}\times1}$ is the unit-norm transmit beamformer, $s\in\mathbb{C}$ is the transmit data symbol with a constraint $\E[|s|^2]\le 1$, and $\bn\in\mathbb{C}^{M_\mathrm{rx}\times1}$ is the additive white Gaussian noise (AWGN) vector where the mean and variance of each element are zero and $\sigma^2$. The signal-to-noise ratio (SNR) is ${P}/{\sigma^2}$. The beamformers $\bc_{\mathrm{tx}}$ and $\bc_{\mathrm{rx}}$ are selected from codebooks $\mathcal{C}_\mathrm{tx}=\{\bc^{(1,1)}_\mathrm{tx},\cdots,\bc^{(Q_{\mathrm{tx},h},Q_{\mathrm{tx},v})}_\mathrm{tx} \}$ and $\mathcal{C}_\mathrm{rx}=\{\bc^{(1,1)}_\mathrm{rx},\cdots,\bc^{(Q_{\mathrm{rx},h},Q_{\mathrm{rx},v})}_\mathrm{rx} \}$, where $Q_{\mathrm{tx}}=Q_{\mathrm{tx},h}Q_{\mathrm{tx},v}$ and $Q_{\mathrm{rx}}=Q_{\mathrm{rx},h}Q_{\mathrm{rx},v}$ are the number of codewords in the Tx and Rx codebooks. 
	In the rest of the paper, we will use the terms codeword and beamformer interchangeably. 
	
	Based on the beam alignment as in \cite{S.Noh:2017,S.Hur:2013,J.Song:2015}, the Rx BS selects the codeword pair with the highest received power as 
	\begin{align}\label{align}
		&(\check{p}_\mathrm{tx},\check{q}_\mathrm{tx},\check{p}_\mathrm{rx},\check{q}_\mathrm{rx})
		\notag\\
		&=\argmax_{ 
			p_\mathrm{tx},q_\mathrm{tx},p_\mathrm{rx},q_\mathrm{rx}
			} 
		\bigg\lvert 
		\sqrt{P} \left(\bc_{\mathrm{rx}}^{(p_{\mathrm{rx}},q_{\mathrm{rx}})}\right)^\mathrm{H} \red{\bH} \bc_{\mathrm{tx}}^{(p_{\mathrm{tx}},q_{\mathrm{tx}})} 
		\notag \\
		&\qquad\qquad\qquad\qquad~~
		+\left(\bc_{\mathrm{rx}}^{(p_{\mathrm{rx}},q_{\mathrm{rx}})}\right)^\mathrm{H}   \bn^{(p_{\mathrm{tx}},q_{\mathrm{tx}},p_{\mathrm{rx}},q_{\mathrm{rx}})} 
		\bigg\rvert^2,
	\end{align}
	where $p_\mathrm{ax}\in\{1,\cdots,Q_{\mathrm{ax},h}\}$, $q_\mathrm{ax}\in\{1,\cdots,Q_{\mathrm{ax},v}\}$, $\mathrm{ax}\in\{\mathrm{tx},\mathrm{rx}\}$, and $\bn^{(p_{\mathrm{tx}},q_{\mathrm{tx}},p_{\mathrm{rx}},q_{\mathrm{rx}})}$ is the AWGN vector with zero mean and variance $\sigma^2$ for each element. The index of the transmit codeword is then fed back to the Tx BS, and the beamformers become
	\begin{align}
		\bc_{\mathrm{tx}}=\bc_{\mathrm{tx}}^{(\check{p}_\mathrm{tx},\check{q}_\mathrm{tx})},\qquad \bc_{\mathrm{rx}}=\bc_{\mathrm{rx}}^{(\check{p}_\mathrm{rx},\check{q}_\mathrm{rx})}.
	\end{align}

	For the hybrid beamforming, each codeword consists of a digital beamformer and an analog beamformer as
	\begin{align}
		\bc_\mathrm{ax}=\bF_\mathrm{ax}\bv_\mathrm{ax},
	\end{align}
	where $\mathrm{ax}\in\{{\mathrm{tx},\mathrm{rx}}\}$, $\bF_\mathrm{ax}=[\bff_{\mathrm{ax},1},\cdots,\bff_{\mathrm{ax},N_\mathrm{ax}} ]  \in\mathbb{C}^{M_{\mathrm{ax}}\times N_{\mathrm{ax}}}$ is the analog beamformer, and $\bv_\mathrm{ax}\in\mathbb{C}^{N_{\mathrm{ax}}\times1}$ is the digital beamformer. Each column of the analog beamformer corresponds to phase shifters that are connected to one of the RF chains. The analog beamformer is, hence, constant modulus, i.e., each element in the analog beamformer can be represented as $e^{j\tau}$ for some $\tau\in[0,2\pi)$. In this paper, we assume the full resolution phase shifters, which can be approximated with more than four bits quantization per phase \cite{O.E.Ayach:2014}, for beam pattern designs, but the simulation results in Section \ref{sec4} are based on $4$-bit phase quantization. Under the fully connected hybrid beamforming architecture, the digital beamformer combines columns of the analog beamformer ensuring the unit-norm constraint as $\lVert \bF_\mathrm{ax}\bv_\mathrm{ax}\rVert_2=1$.

	\begin{figure}[t]
		\centering
		\subfloat[Orientation difference $0$]{
			\includegraphics[width=.45\linewidth]{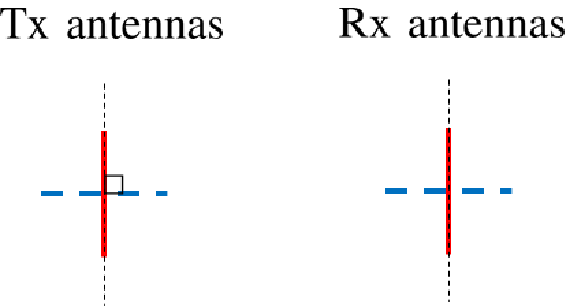}
			\label{orientation_a}
		}~
		\subfloat[Orientation difference $\phi$]{
			\includegraphics[width=.45\linewidth]{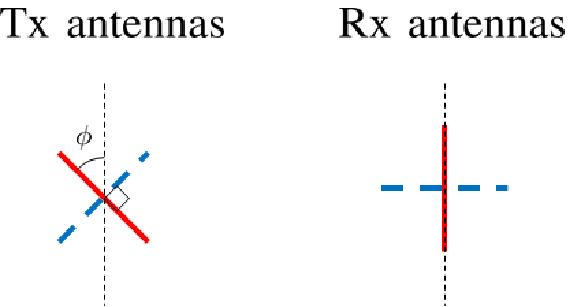}
			\label{orientation_b}
		}\qquad
		\caption{The orientation difference between transmit and receive antennas.}
		\label{orientation}
	\end{figure}
	
	\subsection{Channel model}\label{sec2-2}
	The dual-polarization MIMO channel can be modeled as \cite{L.Jiang:2007}
	\begin{align}\label{channel model}
		&\bH
%		\notag\\
%		&
		=\red{\sqrt{\frac{K}{1+K}\cdot\frac{M_\mathrm{tx}M_\mathrm{rx}}{2^2}}} \bH_{0}
		\notag\\
		&\quad\qquad 
		+\red{\sqrt{\frac{1}{1+K}\cdot\frac{M_\mathrm{tx}M_\mathrm{rx}}{2^2}}} \Bigg( \frac{1}{\sqrt{N_\mathrm{NLOS}}} \sum_{i=1}^{N_\mathrm{NLOS}}\bH_{i} \Bigg),
	\end{align}
	where $K$ is the Rician $K$-factor, $\bH_{0}\in\mathbb{C}^{M_\mathrm{rx}\times M_\mathrm{tx}}$ is the LOS component, $\bH_{i}\in\mathbb{C}^{M_\mathrm{rx}\times M_\mathrm{tx}}$ with $i\in\{1,\cdots N_\mathrm{NLOS}\}$ are the non-line-of-sight (NLOS) components, and $N_\mathrm{NLOS}$ is the number of NLOS components. 
	%	\red{The factor $\sqrt{\frac{M_\text{tx}M_\text{rx}}{2^2}}$ of each term is about the array gains of UPAs at Tx and Rx BSs, where each is represented as $\sqrt{\frac{M_\mathrm{ax}}{2}}=\sqrt{M_{\text{ax},h}M_{\text{ax},v}}$.}
%	\red{The factor $\frac{M_\text{tx}M_\text{rx}}{2^2}$ of each term is about the array gains of UPAs at Tx and Rx BSs, which is determined by the dimension of UPAs $\frac{M_\mathrm{ax}}{2}=M_{\mathrm{ax},h}M_{\mathrm{ax},v}$ with $\mathrm{ax}\in\{\mathrm{tx},\mathrm{rx}\}$ not by the number of antennas $M_\mathrm{ax}$.}
	The denominator $2^2$ in each squared root is for the fact \red{$M_\mathrm{ax}=2M_{\mathrm{ax},h}M_{\mathrm{ax},v}$} with $\mathrm{ax}\in\{\mathrm{tx},\mathrm{rx}\}$, which is about the array gain. 
	\red{This normalization will become clear after explaining $\bH_i$ in detail.}
%	\red{Although the number of antennas is doubled by the dual-polarization structure $2M_{\mathrm{ax},h}M_{\mathrm{ax},v}$, what affects the array gain is the dimension of UPA $M_{\mathrm{ax},h}M_{\mathrm{ax},v}$.}

	The structure of each component can be written as \cite{J.Song:2013,L.Jiang:2007,G.Calcev:2007}	
	\begin{align}
		\bH_{i}
		&=h_{i}  
		\Bigg
		\{	\Bigg(	\begin{bmatrix} 	\sqrt{\frac{1}{1+\chi}} & \sqrt{\frac{\chi}{1+\chi}}\\ \sqrt{\frac{\chi}{1+\chi}} & \sqrt{\frac{1}{1+\chi}} \end{bmatrix}
		\odot \begin{bmatrix} e^{j\angle\zeta_{i}^{vv}} & e^{j\angle\zeta_{i}^{vh}} \\ e^{j\angle\zeta_{i}^{hv}} & e^{j\angle\zeta_{i}^{hh}} \end{bmatrix}\Bigg)
		\notag\\
		&\quad~
		\otimes \ba_{\mathrm{rx}}(\theta_{\mathrm{rx},i}^{\mathrm{az}},\theta_{\mathrm{rx},i}^{\mathrm{el}}) \ba_{\mathrm{tx}} (\theta_{\mathrm{tx},i}^{\mathrm{az}},\theta_{\mathrm{tx},i}^{\mathrm{el}})^\mathrm{H} 
		\Bigg\} 	
		\bR(\phi),\label{channel component}
	\end{align}
	where $i\in\{0,\cdots,N_\mathrm{NLOS}\}$, $h_{i}$ is the complex gain of $i$-th path, $\chi$ is the XPD value that represents the distinction ability of different antenna polarization, the superscripts $v$ and $h$ of $\angle\zeta_{i}$ represent vertical and horizontal polarization, and $\angle\zeta_{i}^{ab}$ is the random phase of $i$-th path from $b$ Tx antenna to the $a$ Rx antenna with $a, b \in \{v,h\}$.
	$\ba_{\mathrm{tx}} (\theta_{\mathrm{tx},i}^{\mathrm{az}},\theta_{\mathrm{tx},i}^{\mathrm{el}})\in\mathbb{C}^{\frac{M_\mathrm{tx}}{2}\times1}$ is the single path array response vector of UPA in the Tx BS with angle-of-departure (AoD) $(\theta_{\mathrm{tx},i}^{\mathrm{az}},\theta_{\mathrm{tx},i}^{\mathrm{el}})$, $\ba_{\mathrm{rx}}(\theta_{\mathrm{rx},i}^{\mathrm{az}},\theta_{\mathrm{rx},i}^{\mathrm{el}})\in\mathbb{C}^{\frac{M_\mathrm{rx}}{2}\times1}$ is the single path array response vector of UPA in the Rx BS with angle-of-arrival (AoA) $(\theta_{\mathrm{rx},i}^{\mathrm{az}},\theta_{\mathrm{rx},i}^{\mathrm{el}})$, and $\bR(\phi)=\begin{bmatrix} \cos\phi &-\sin\phi \\ \sin\phi&\cos\phi\end{bmatrix}\otimes\bI_{M_\mathrm{tx}/2}$ is the Givens rotation matrix with the orientation difference $\phi$ between the Tx and the Rx antennas \cite{L.Jiang:2007,B.Clerckx:2008}. Fig. \ref{orientation_a} shows the zero orientation difference, and Fig. \ref{orientation_b} shows the $\phi$ orientation difference. In the case of backhaul communications, antenna arrays are almost fixed with little movement by wind. Therefore, to design beams, we assume the antenna arrays are stationary with the difference $\phi$. Because the XPD value also does not change much over time, we assume the Tx BS knows $\chi$ and $\phi$ in this and next sections. 
	
	The $i$-th path array response vector of UPA in the Tx BS is
	\begin{align}\label{array response vector Kronecker}
		\ba_{\mathrm{tx}}(\theta_{\mathrm{tx},i}^{\mathrm{az}},\theta_{\mathrm{tx},i}^{\mathrm{el}})=\ba_{\mathrm{tx},h}(\theta_{\mathrm{tx},i}^{\mathrm{az}},\theta_{\mathrm{tx},i}^{\mathrm{el}})\otimes\ba_{\mathrm{tx},v}(\theta_{\mathrm{tx},i}^{\mathrm{el}}),
	\end{align}
	where $\ba_{\mathrm{tx},h} (\theta_{\mathrm{tx},i}^\mathrm{az},\theta_{\mathrm{tx},i}^\mathrm{el} )\in\mathbb{C}^{M_{\mathrm{tx},h}\times1}$ is the array response vector of horizontally arranged ULA, $\ba_{\mathrm{tx},v} (\theta_{\mathrm{tx},i}^\mathrm{el} )\in\mathbb{C}^{M_{\mathrm{tx},v}\times1}$ is the array response vector of vertically arranged ULA, and $\theta_{\mathrm{tx},i}^{\mathrm{az}}$ and $\theta_{\mathrm{tx},i}^{\mathrm{el}}$ are the azimuth and elevation AoDs of $i$-th path. Specifically, $\ba_{\mathrm{tx},h} (\theta_{\mathrm{tx},i}^\mathrm{az},\theta_{\mathrm{tx},i}^\mathrm{el})$ and $\ba_{\mathrm{tx},v} (\theta_{\mathrm{tx},i}^\mathrm{el})$ can be written as
	\begin{align}
		\ba_{\mathrm{tx},h} (\theta_{\mathrm{tx},i}^{\mathrm{az}},\theta_{\mathrm{tx},i}^{\mathrm{el}})
		&
		= \frac{1}{\sqrt{M_{\mathrm{tx},h}}} [1,e^{j \frac{2\pi d_{\mathrm{tx},h}}{\lambda}  \sin \theta_{\mathrm{tx},i}^{\mathrm{az}} \cos\theta_{\mathrm{tx},i}^{\mathrm{el}}  },
		\notag\\
		&\qquad
		\cdots,e^{j \frac{2\pi d_{\mathrm{tx},h}}{\lambda} (M_{\mathrm{tx},h}-1) \sin \theta_{\mathrm{tx},i}^{\mathrm{az}} \cos  \theta_{\mathrm{tx},i}^{\mathrm{el}}  } ]^\mathrm{T}, \label{array response vector h}
		\\
		\ba_{\mathrm{tx},v} (\theta_{\mathrm{tx},i}^{\mathrm{el}})
		&
		= \frac{1}{\sqrt{M_{\mathrm{tx},v}}} [1,e^{j \frac{2\pi d_{\mathrm{tx},v}}{\lambda}  \sin \theta_{\mathrm{tx},i}^{\mathrm{el}} },
		\notag\\
		&\qquad
		\cdots,e^{j \frac{2\pi d_{\mathrm{tx},v}}{\lambda} (M_{\mathrm{tx},v}-1) \sin \theta_{\mathrm{tx},i}^{\mathrm{el}}  } ]^\mathrm{T},\label{array response vector v}
	\end{align}
	where $d_{\mathrm{tx},h}$ is the interval of the horizontal ULA, $d_{\mathrm{tx},v}$ is the interval of the vertical ULA, and $\lambda$ is the wavelength of the carrier frequency. For simplicity, we set $d_{\mathrm{tx},h}=d_{\mathrm{tx},v}=\frac{\lambda}{2}$ in this paper. The array response vector of UPA in the Rx BS has a similar structure with the number of antennas $M_\mathrm{rx}$ and AoA $(\theta_{\mathrm{rx},i}^{\mathrm{az}},\theta_{\mathrm{rx},i}^{\mathrm{el}})$.
	
	\red{Note that the denominator $2^2$ in \eqref{channel model} is to reflect the dual-polarization structure in $\bH_i$. The array response vectors in \eqref{array response vector h} and \eqref{array response vector v} are normalized with $\sqrt{M_{\text{tx},h}}$ and $\sqrt{M_{\text{tx},v}}$, respectively. The dimension of Kronecker product of these two vectors, which represents the single-polarization UPA structure in \eqref{array response vector Kronecker}, is doubled by the Kronecker product in \eqref{channel component} to model the dual-polarization UPA structure. To normalize this doubling effect for both Tx and Rx parts in the channel gain, we need to have the normalization constant $2^2$ in \eqref{channel model}.}
	
	Within the entire angle range $((-\pi,\pi),(-\frac{\pi}{2},\frac{\pi}{2}))$, we focus on a smaller range $((-\frac{\pi}{2},\frac{\pi}{2}),(-\frac{\pi}{4},\frac{\pi}{4}))$ for both AoD and AoA considering practical cell sectorization. Then, the corresponding horizontal and vertical spatial frequencies, i.e., $\psi_{\mathrm{ax},i}^\mathrm{az}=\pi  \sin \theta_{\mathrm{ax},i}^\mathrm{az} \cos\theta_{\mathrm{ax},i}^\mathrm{el}$ and $\psi_{\mathrm{ax},i}^\mathrm{el}=\pi\sin\theta_{\mathrm{ax},i}^\mathrm{el}$, are bounded as
	\begin{align}
		-\pi < \psi_{\mathrm{ax},i}^\mathrm{az} < \pi \label{spatial frequency bound},\quad 
		-\frac{\pi}{\sqrt{2}} < \psi_{\mathrm{ax},i}^\mathrm{el} < \frac{\pi}{\sqrt{2}},
	\end{align}
	where $\mathrm{ax}\in\{\mathrm{tx},\mathrm{rx}\}$. 
	Hence the coverage angle range $((-\frac{\pi}{2},\frac{\pi}{2}),(-\frac{\pi}{4},\frac{\pi}{4}))$ can be handled with the corresponding spatial frequency range $((-\pi,\pi),(-\frac{\pi}{\sqrt{2}},\frac{\pi}{\sqrt{2}}))$. In this paper, we consider unpaired spatial frequencies 
	\begin{align}
		\psi_{\mathrm{ax},i}^\mathrm{az}=\pi \sin \theta_{\mathrm{ax},i}^\mathrm{az},
		\quad 
		\psi_{\mathrm{ax},i}^\mathrm{el}=\pi  \sin\theta_{\mathrm{ax},i}^\mathrm{el}
	\end{align}
	and array response vectors
	\begin{align}
		&\bd_{\mathrm{ax},h} (\psi_{\mathrm{ax},i}^\mathrm{az} )= \frac{1}{\sqrt{M_{\mathrm{ax},h}}}\left[1,e^{j\psi_{\mathrm{ax},i}^\mathrm{az}},\cdots,e^{j\psi_{\mathrm{ax},i}^\mathrm{az} (M_{\mathrm{ax},h}-1) } \right]^\mathrm{T},	\\
		&\bd_{\mathrm{ax},v} (\psi_{\mathrm{ax},i}^\mathrm{el} )= \frac{1}{\sqrt{M_{\mathrm{ax},v}}}\left[1,e^{j\psi_{\mathrm{ax},i}^\mathrm{el}},\cdots,e^{j\psi_{\mathrm{ax},i}^\mathrm{el} (M_{\mathrm{ax},v}-1) } \right]^\mathrm{T},
	\end{align}
	since the difference between the region of unpaired spatial frequencies and that of paired spatial frequencies is negligible \cite{J.Song:2017}. 
%	For the receiver, we also consider unpaired spatial frequencies $\psi_{\mathrm{rx},i}^\mathrm{az},~ \psi_{\mathrm{rx},i}^\mathrm{el}$ and array response vectors $\bd_{\mathrm{rx},h}(\psi_{\mathrm{tx},i}^\mathrm{az}),~ \bd_{\mathrm{rx},v}(\psi_{\mathrm{rx},i}^\mathrm{el})$. 
	Although we consider the unpaired spatial frequencies for beam pattern designs, the numerical results in Section \ref{sec4} are based on the paired spatial frequencies.
	% The bounds of the unpaired spatial frequencies are the same as those of the paired spatial frequencies in \eqref{spatial frequency bound 1}, \eqref{spatial frequency bound 2}.
	
	With the large Rician $K$-factor, mmWave channel can be approximated as its dominant LOS channel component 
%	\begin{align}
		$\bH\approx\bH_0$.
%	\end{align} 
	We design beams using the approximated channel in the following section, but the numerical results in Section \ref{sec4} are examined with the original channel model in \eqref{channel model}.
	
	\section{Beamforming designs}\label{sec3}
	
	\begin{figure*}[!b]
		\hrule
		\begin{align}\label{beamforming gain decomposition}
		%		&
%		gain 
		%		\notag\\
		&
		\left| \bar{\bc}_\mathrm{rx}^\mathrm{H}\bar{\bH}_0\bc_{\mathrm{tx}} \right|^2
		\notag\\
		&
		=\Big| \bar{\bc}_\mathrm{rx}^\mathrm{H} 
		h_0  \Big\{	
		\begin{bmatrix} 	\sqrt{\frac{1}{1+\chi}}e^{j\angle\zeta_{0}^{vv}} & \sqrt{\frac{\chi}{1+\chi}}e^{j\angle\zeta_{0}^{vh}} \end{bmatrix}
		%			\notag\\
		%			&\quad~
		\otimes \left( 
		\bd_{\mathrm{rx}}(\psi_{\mathrm{rx},0}^{\mathrm{az}},\psi_{\mathrm{rx},0}^{\mathrm{el}}) \bd_{\mathrm{tx}} (\psi_{\mathrm{tx},0}^{\mathrm{az}},\psi_{\mathrm{tx},0}^{\mathrm{el}})^\mathrm{H} 		
		%		\ba_{\mathrm{rx}}(\theta_{\mathrm{rx},0}^{\mathrm{az}},\theta_{\mathrm{rx},0}^{\mathrm{el}}) \ba_{\mathrm{tx}} (\theta_{\mathrm{tx},0}^{\mathrm{az}},\theta_{\mathrm{tx},0}^{\mathrm{el}})^\mathrm{H} 
		\right)
		\Big\} 	
		\bR(\phi)\bc_{\mathrm{tx}} \Big|^2
		\notag\\
		&
		=\Big| 
		h_0  
		\begin{bmatrix} 	\sqrt{\frac{1}{1+\chi}}e^{j\angle\zeta_{0}^{vv}}\bar{\bc}_\mathrm{rx}^\mathrm{H}\bd_{\mathrm{rx}}(\psi_{\mathrm{rx},0}^{\mathrm{az}},\psi_{\mathrm{rx},0}^{\mathrm{el}}) \bd_{\mathrm{tx}} (\psi_{\mathrm{tx},0}^{\mathrm{az}},\psi_{\mathrm{tx},0}^{\mathrm{el}})^\mathrm{H}  & \sqrt{\frac{\chi}{1+\chi}}e^{j\angle\zeta_{0}^{vh}}\bar{\bc}_\mathrm{rx}^\mathrm{H}\bd_{\mathrm{rx}}(\psi_{\mathrm{rx},0}^{\mathrm{az}},\psi_{\mathrm{rx},0}^{\mathrm{el}}) \bd_{\mathrm{tx}} (\psi_{\mathrm{tx},0}^{\mathrm{az}},\psi_{\mathrm{tx},0}^{\mathrm{el}})^\mathrm{H}  \end{bmatrix}
		\bR(\phi)\bc_{\mathrm{tx}} \Big|^2
		\notag\\
		&
		=\Big| 
		h_0  \bar{\bc}_\mathrm{rx}^\mathrm{H}\bd_{\mathrm{rx}}(\psi_{\mathrm{rx},0}^{\mathrm{az}},\psi_{\mathrm{rx},0}^{\mathrm{el}})
		\begin{bmatrix} 	\sqrt{\frac{1}{1+\chi}}e^{j\angle\zeta_{0}^{vv}} \bd_{\mathrm{tx}} (\psi_{\mathrm{tx},0}^{\mathrm{az}},\psi_{\mathrm{tx},0}^{\mathrm{el}})^\mathrm{H}  & \sqrt{\frac{\chi}{1+\chi}}e^{j\angle\zeta_{0}^{vh}} \bd_{\mathrm{tx}} (\psi_{\mathrm{tx},0}^{\mathrm{az}},\psi_{\mathrm{tx},0}^{\mathrm{el}})^\mathrm{H}  \end{bmatrix}
		\bR(\phi)\bc_{\mathrm{tx}} \Big|^2
		\notag\\
		&
		=\Big| h_0 \bar{\bc}_\mathrm{rx}^\mathrm{H}
		\bd_{\mathrm{rx}}(\psi_{\mathrm{rx},0}^{\mathrm{az}},\psi_{\mathrm{rx},0}^{\mathrm{el}})
		%		\ba_{\mathrm{rx}}(\theta_{\mathrm{rx},0}^{\mathrm{az}},\theta_{\mathrm{rx},0}^{\mathrm{el}}) 
		%			\notag\\
		%			&\quad 
		%			\cdot 
		\Big(
		\begin{bmatrix} 	\sqrt{\frac{1}{1+\chi}}e^{j\angle\zeta_{0}^{vv}} & \sqrt{\frac{\chi}{1+\chi}}e^{j\angle\zeta_{0}^{vh}} \end{bmatrix}
		%			\notag\\
		%			&\quad~
		\otimes 
		\bd_{\mathrm{tx}} (\psi_{\mathrm{tx},0}^{\mathrm{az}},\psi_{\mathrm{tx},0}^{\mathrm{el}})^\mathrm{H}
		%		\ba_{\mathrm{tx}} (\theta_{\mathrm{tx},0}^{\mathrm{az}},\theta_{\mathrm{tx},0}^{\mathrm{el}})^\mathrm{H}
		\Big)
		\bR(\phi)\bc_{\mathrm{tx}} \Big|^2
		\notag\\
		&
		=|h_0 |^2\underbrace{\left| \bar{\bc}_\mathrm{rx}^\mathrm{H} 
			\bd_{\mathrm{rx}}(\psi_{\mathrm{rx},0}^{\mathrm{az}},\psi_{\mathrm{rx},0}^{\mathrm{el}}) 
			%		\ba_{\mathrm{rx}}(\theta_{\mathrm{rx},0}^{\mathrm{az}},\theta_{\mathrm{rx},0}^{\mathrm{el}}) 
			\right|^2}_\text{Rx side beamforming gain}
		%			\notag\\
		%			&\quad 
		%			\cdot
		\underbrace{ 
			\big|
			\underbrace{ \big(\begin{bmatrix} \rho_{vv} ~ \rho_{vh} \end{bmatrix}	
				\otimes 
				\bd_{\mathrm{tx}} (\psi_{\mathrm{tx},0}^{\mathrm{az}},\psi_{\mathrm{tx},0}^{\mathrm{el}})^\mathrm{H}
				%				\ba_{\mathrm{tx}} (\theta_{\mathrm{tx},0}^{\mathrm{az}},\theta_{\mathrm{tx},0}^{\mathrm{el}})^\mathrm{H}
				\big)
				\bR(\phi)
			}_\text{Tx side channel}
			\bc_{\mathrm{tx}} \big|^2 
		}_\text{Tx side beamforming gain}	\tag{16}
		\end{align}
	\end{figure*}

%	In mmWave system, the large bandwidth ensures the high data rates without using spatial multiplexing, and we propose beamformers for single stream. 
	Since dealing with the dual-polarization beamforming for both the Tx and Rx BSs is highly complicated, we first design Tx beamformers assuming $M_\mathrm{tx}$ dual-polarization Tx antennas and $\frac{M_\mathrm{rx}}{2}$ single-polarization Rx antennas in $v$ polarization. For the given Tx beamformers, then, we design Rx beamformers for the original $M_\mathrm{tx}$ dual-polarization Tx antennas and $M_\mathrm{rx}$ dual-polarization Rx antennas. The channel matrix $\bar{\bH}\in\mathbb{C}^{\frac{M_\mathrm{rx}}{2}\times M_\mathrm{tx}}$ that is considered for the Tx beamforming is
	\begin{align}\label{Tx channel}
		&\bar{\bH}
		\notag\\
		&
		=\sqrt{\frac{KM_\mathrm{tx}M_\mathrm{rx}}{2^2(1+K)}} \bar{\bH}_{0}+\sqrt{\frac{M_\mathrm{tx}M_\mathrm{rx}}{2^2(1+K)}} \Bigg(\frac{1}{\sqrt{N_\mathrm{NLOS}}}\sum_{i=1}^{N_\mathrm{NLOS}}\bar{\bH}_{i}\Bigg),
	\end{align}
	where $\bar{\bH}_0\in\mathbb{C}^{\frac{M_\mathrm{rx}}{2}\times M_\mathrm{tx}}$ and $\bar{\bH}_i\in\mathbb{C}^{\frac{M_\mathrm{rx}}{2}\times M_\mathrm{tx}}$ are the LOS and a NLOS component of the channel. The scaling factors in \eqref{Tx channel} are the same as in \eqref{channel model} since they are related to the Rician $K$-factor and the array gain of UPA structure.
	
	To design a Tx beamformer, we first decompose the overall beamforming gain into Tx and Rx parts and define the ideal Tx beam pattern in Section \ref{sec3-a}. Then, we perform the Tx beamformer optimizations in Section \ref{SE} and Section \ref{MIP} with the SE and MIP criteria based on the ideal beam pattern. In Section \ref{discussion}, we discuss the main contribution of this paper, i.e., how to generate dual-polarization beamformers from single-polarization beamformers. For the given Tx beamformers, the designs of Rx beamformers follow in Section \ref{Rx beamforming}.
	Since the proposed beamforming designs require partial channel information at the Tx BS, we propose the use of pilots to estimate the partial information in Section \ref{sec3-d}.

	\subsection{Preliminary for transmit beamforming designs}\label{sec3-a}
		
	The LOS channel component in \eqref{Tx channel} is given as
	\begin{align}\label{Tx side channel}
		\bar{\bH}_0
		&=h_0  
		\Big\{	
		\left(	\begin{bmatrix} 	\sqrt{\frac{1}{1+\chi}} & \sqrt{\frac{\chi}{1+\chi}} \end{bmatrix}
		\odot \begin{bmatrix} e^{j\angle\zeta_{0}^{vv}} & e^{j\angle\zeta_{0}^{vh}} \end{bmatrix}	\right)
		\notag\\
		&\quad~
		\otimes \bd_{\mathrm{rx}}(\psi_{\mathrm{rx},0}^{\mathrm{az}},\psi_{\mathrm{rx},0}^{\mathrm{el}}) \bd_{\mathrm{tx}} (\psi_{\mathrm{tx},0}^{\mathrm{az}},\psi_{\mathrm{tx},0}^{\mathrm{el}})^\mathrm{H} 
		\Big\} 	
		\bR(\phi).
%		&=h_0  
%		\Big\{	
%		\left(	\begin{bmatrix} 	\sqrt{\frac{1}{1+\chi}} & \sqrt{\frac{\chi}{1+\chi}} \end{bmatrix}
%		\odot \begin{bmatrix} e^{j\angle\zeta_{0}^{vv}} & e^{j\angle\zeta_{0}^{vh}} \end{bmatrix}	\right)
%		\notag\\
%		&\quad~
%		\otimes \ba_{\mathrm{rx}}(\theta_{\mathrm{rx},0}^{\mathrm{az}},\theta_{\mathrm{rx},0}^{\mathrm{el}}) \ba_{\mathrm{tx}} (\theta_{\mathrm{tx},0}^{\mathrm{az}},\theta_{\mathrm{tx},0}^{\mathrm{el}})^\mathrm{H} 
%		\Big\} 	
%		\bR(\phi).
	\end{align}
	Considering the structure of the LOS component in \eqref{Tx side channel}, the beamforming gain of arbitrary Tx and Rx beamformers $\bc_{\mathrm{tx}}\in\mathbb{C}^{M_\mathrm{tx}\times1}$ and $\bar{\bc}_{\mathrm{rx}}\in\mathbb{C}^{\frac{M_\mathrm{rx}}{2}\times1}$ can be decomposed as \eqref{beamforming gain decomposition} on the bottom of this page.
	In \eqref{beamforming gain decomposition}, $\bd_{\mathrm{ax}}=\bd_{\mathrm{ax},h}\otimes\bd_{\mathrm{ax},v}$ with $\mathrm{ax}\in\{\mathrm{tx},\mathrm{rx}\}$, $\rho_{vv}=\sqrt{\frac{1}{1+\chi}}e^{j\angle\zeta_{0}^{vv}}$, and $\rho_{vh}=\sqrt{\frac{\chi}{1+\chi}}e^{j\angle\zeta_{0}^{vh}}$. 
%	The equality $(a)$ is by the property of Kronecker product $(\bW\otimes \bX)(\bY\otimes \bZ)=(\bW\bY)\otimes(\bX\bZ)$. 
	In this and following two subsections, to design Tx beamformers, we consider the Tx side beamforming gain and Tx side channel with proper normalization. From Section \ref{sec3-a} to \ref{discussion}, we omit the subscript $\mathrm{tx}$ for briefness, while we keep the subscript $\mathrm{rx}$. 
	\setcounter{equation}{16}
	
	The Tx side beamforming gain in \eqref{beamforming gain decomposition} implies that the unit-norm Tx beamformer giving the highest gain is the normalized Hermitian of the Tx side channel
	\begin{align}
	\bc_{\mathrm{dual}}
	=b\bR(\phi)^\mathrm{H} \left( \begin{bmatrix} \rho_{vv}^* \\ \rho_{vh}^* \end{bmatrix} 
	\otimes 
	\bd (\psi_{0}^{\mathrm{az}},\psi_{0}^{\mathrm{el}}) 
%	\ba (\theta_{0}^{\mathrm{az}},\theta_{0}^{\mathrm{el}}) 
	\right)
	,
	\end{align}
	where $b=\left\{ |\rho_{vv} |^2+|\rho_{vh}|^2 \right\}^{-\frac{1}{2}}$ is the normalization factor. Similarly, for the same AoD and dimension of UPA, but with single-polarization antennas, the Tx beamformer giving the highest gain is the normalized array response vector $\bc_{\mathrm{single}}=\bd (\psi_{0}^{\mathrm{az}},\psi_{0}^{\mathrm{el}})$. It can be noticed that the dual-polarization beamformer is a function of the single-polarization beamformer as
	\begin{align}
	\bc_{\mathrm{dual}} 	
	&
	=b\bR(\phi)^\mathrm{H} \left( \begin{bmatrix} \rho_{vv}^* \\ \rho_{vh}^* \end{bmatrix} 
	\otimes 
	\bd (\psi_{0}^{\mathrm{az}},\psi_{0}^{\mathrm{el}})
%	\ba (\theta_{0}^{\mathrm{az}},\theta_{0}^{\mathrm{el}})
	 \right) 
	\notag\\
	&
	=b\bR(\phi)^\mathrm{H} \left( \begin{bmatrix} \rho_{vv}^* \\ \rho_{vh}^* \end{bmatrix} 
	\otimes \bc_{\mathrm{single}} \right).
	\end{align}
	This means that for a given optimization criterion, e.g., maximizing the beamforming gain as in this example, the optimal dual-polarization beamformer $\bc_\mathrm{dual}$ can be attained from the optimal single-polarization beamformer $\bc_\mathrm{single}$ sharing the same optimality. 
	
	In this paper, two kinds of optimalities are considered for the Tx beamformer design. One is the squared error (SE) and the other is the magnitude of inner product (MIP) with respect to the ideal Tx beam pattern that will be defined shortly. Based on these criteria, we first derive the optimal digital beamformer for the dual-polarization UPA and apply the orthogonal matched pursuit (OMP) based algorithm to generate the corresponding hybrid beamformer.

	\begin{figure}[t]
		\includegraphics[width=1\linewidth]{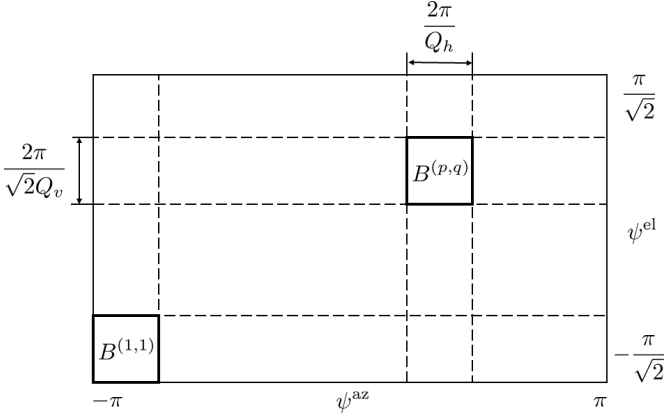}
		\caption{Quantized spatial frequency regions of entire covering region.}
		\label{region}
	\end{figure}
	
	To derive the optimal digital beamformer, we first divide the spatial frequency range into $Q_h\times Q_v$ regions as in Fig. \ref{region}. Each quantized region is defined as
	\begin{align}
	B^{(p,q)}
	&
	=\bigg\{(\psi^\mathrm{az},\psi^\mathrm{el} ): -\pi+\frac{2\pi(p-1)}{Q_h} \le \psi^\mathrm{az} < -\pi+\frac{2\pi p}{Q_h} ,\notag\\
	&\qquad -\frac{\pi}{\sqrt{2}}+\frac{2\pi(q-1)}{\sqrt{2} Q_v} \le \psi^\mathrm{el} < -\frac{\pi}{\sqrt{2}}+\frac{2\pi q}{\sqrt{2} Q_v} \bigg\},
	\end{align}
	where $p\in\{1,\cdots,Q_h \}$ and $q\in\{1,\cdots,Q_v \}$. To minimize the number of required codewords to cover the entire region, we let each quantized region be covered by one of $Q=Q_h Q_v$ codewords. For each quantized region, as in \cite{J.Song:2017,S.Noh:2017}, the ideal beam pattern is defined to have an equal gain within the quantized region of interest and zero gain outside the region. The digital beamformer design goal is that each of codewords has the most similar beam pattern to the ideal beam pattern at each quantized region.

%	The shape of ideal beam pattern is distinguished, and it holds for the dual-polarization antenna structure. The equal gain value of the ideal beam pattern, however, is not defined for the dual-polarization antenna structure. To derive a proper equal gain of the ideal beam pattern at each quantized region,
	Since a proper equal gain of the ideal beam pattern is essential to design beamformers, we derive the gain following similar steps in \cite{J.Song:2017}. 
%	As a preliminary, 
	We first consider the expected data rate conditioned on $\lvert h_0 \rvert^2$ and an arbitrary Rx beamformer $\bar{\bc}_{\mathrm{rx}}\in\mathbb{C}^{\frac{M_\mathrm{rx}}{2}\times1}$
	\begin{align}\label{data rate}
	R_\mathrm{data}
	&
%	={\E}_{\psi^\mathrm{el},\psi^\mathrm{el}} 
	=\E
	\left[ \log_2 \left(1+\frac{P}{\sigma^2} |\bar{\bc}_{\mathrm{rx}}^\mathrm{H}\bar{\bH} \bc|^2 \right)\bigg|\bar{\bc}_{\mathrm{rx}},|h_0| \right] 
	\notag\\	
	&
%	=\underset{\psi^\mathrm{el},\psi^\mathrm{el}}{\E}
	=\E
	\bigg[ \log_2\bigg( 1+\frac{P}{b^2\sigma^2}  |h_0|^2\left|\bar{\bc}_{\mathrm{rx}}^\mathrm{H} 
	\bd_{\mathrm{rx}}(\psi_{\mathrm{rx},0}^\mathrm{az},\psi_{\mathrm{rx},0}^\mathrm{el})
%	\ba_{\mathrm{rx}}(\theta_{\mathrm{rx},0}^\mathrm{az},\theta_{\mathrm{rx},0}^\mathrm{el})
	\right|^2  
	\notag\\
	&\qquad\qquad\quad
	\cdot g_\mathrm{ref} (\psi_{0}^\mathrm{az},\psi_0^\mathrm{el},\bc) \bigg) \bigg| \bar{\bc}_{\mathrm{rx}},|h_0| \bigg],
	\end{align}
	where the expectation is taken over $\psi_0^\mathrm{el}$ and $\psi_0^\mathrm{az}$. In \eqref{data rate}, $g_\mathrm{ref} (\psi_0^\mathrm{az},\psi_0^\mathrm{el},\bc)$ is the reference gain of a Tx beamformer $\bc$, which is defined as
	\begin{align}\label{reference gain}
	&
	g_\mathrm{ref} (\psi^\mathrm{az},\psi^\mathrm{el},\bc)				
	\notag\\
	&\quad
	=\left| b\left\{ \begin{bmatrix} \rho_{vv} & \rho_{vh} \end{bmatrix} \otimes \big(\bd_h (\psi^\mathrm{az} )\otimes\bd_v (\psi^\mathrm{el} ) \big)^\mathrm{H} \right\} \bR(\phi) \bc \right|^2.
	\end{align}
	Dealing with the reference gain, we derive the following two lemmas that will be used to drive the equal gain value of the ideal Tx beam pattern.

	\begin{lemma}\label{lemma 1}
		For the dual-polarization channels, the integral of the reference gain $g_\mathrm{ref} (\psi^\mathrm{az},\psi^\mathrm{el},\bc)$ over the entire direction $[-\pi,\pi)\times[-\pi,\pi)$ is bounded as
		\begin{align} 
		\int_{-\pi}^\pi \int_{-\pi}^\pi g_\mathrm{ref} (\psi^\mathrm{az},\psi^\mathrm{el},\bc)d\psi^\mathrm{az}  d\psi^\mathrm{el} \le \frac{(2\pi)^2}{M_h M_v}
		\end{align} 
		for any unit-norm Tx beamformer $\bc\in\mathbb{C}^{M\times1}$, and the equality holds when $\bc$ is the linear combination of the vectors $\left[\rho_{vv}^* \bee_{\frac{M}{2},\ell}^\mathrm{T},\rho_{vh}^* \bee_{\frac{M}{2},\ell}^\mathrm{T} \right]^\mathrm{T}$, $\ell\in\left\{1,\cdots,\frac{M}{2}\right\}$.
	\end{lemma}
	\begin{IEEEproof}
		The integral of the reference gain is
		\begingroup
		\allowdisplaybreaks
		\begin{align}
		&
		\int_{-\pi}^\pi \int_{-\pi}^\pi g_\mathrm{ref} (\psi^\mathrm{az},\psi^\mathrm{el},\bc) d\psi^\mathrm{az} d\psi^\mathrm{el} 		
		\notag \\
		&
		\stackrel{(a)}{=} \int_{-\pi}^\pi \int_{-\pi}^\pi \bigg| \frac{b\rho_{vv}}{\sqrt{M_h}} \sum_{\ell=1}^{M_h} e^{j\psi^\mathrm{az} (\ell-1) } \bd_v^\mathrm{H} (\psi^\mathrm{el} ) {{\bc'}}_{vv,\ell} 
		\notag\\
		&\qquad\qquad +\frac{b\rho_{vh}}{\sqrt{M_h}} \sum_{\ell=1}^{M_h}e^{j\psi^\mathrm{az} (\ell-1) } \bd_v^\mathrm{H} (\psi^\mathrm{el} ) {{\bc'}}_{vh,\ell} \bigg|^2 d\psi^\mathrm{az} d\psi^\mathrm{el} 
		\notag \\
%		&
%		=\int_{-\pi}^\pi \int_{-\pi}^\pi \frac{b^2}{M_h}  \Bigg| \sum_{\ell=1}^{M_h}e^{-j\psi^\mathrm{az} (\ell-1) } \big(\rho_{vv} \bd_v^\mathrm{H} (\psi^\mathrm{el} ) {{\bc'}}_{vv,\ell} 
%		\notag\\
%		&\qquad\qquad +\rho_{vh} \bd_v^\mathrm{H} (\psi^\mathrm{el} ) {{\bc'}}_{vh,\ell} \big) \Bigg|^2 d\psi^\mathrm{az} d\psi^\mathrm{el} 		
%		\notag \\
		&
		\stackrel{(b)}{=}  \frac{2\pi b^2}{M_h}  \sum_{\ell=1}^{M_h}\int_{-\pi}^\pi  \Bigg|  \frac{1}{\sqrt{M_v}}  \sum_{m=1}^{M_v}e^{j\psi^\mathrm{el} (m-1) } (\rho_{vv} {{\bc'}}_{vv,\ell}
		\notag\\
		&\qquad\qquad +\rho_{vh} {{\bc'}}_{vh,\ell} )_{(m)} \Bigg|^2 d\psi^\mathrm{el} 		
		\notag\\
		&
		\stackrel{(c)}{=}   \frac{(2\pi)^2 b^2}{M_hM_v} \sum_{\ell=1}^{M_h} \sum_{m=1}^{M_v} | (\rho_{vv} {{\bc'}}_{vv,\ell}+\rho_{vh} {{\bc'}}_{vh,\ell} )_{(m)} |^2  
		\notag \\
		&
		=\frac{(2\pi)^2 b^2}{M_hM_v} \left\lVert \left[ \rho_{vv} \bI_{\frac{M}{2}}, \rho_{vh} \bI_{\frac{M}{2}} \right] [{{\bc'}}_{vv}^\mathrm{T}   {{\bc'}}_{vh}^\mathrm{T} ]^\mathrm{T} \right\rVert_2^2                  
		\notag\\
		&
		\le \frac{(2\pi)^2 b^2}{M_hM_v} \Lambda_\mathrm{max} \left( \left[ \rho_{vv} \bI_{\frac{M}{2}}, \rho_{vh} \bI_{\frac{M}{2}} \right]^\mathrm{H} \left[ \rho_{vv} \bI_{\frac{M}{2}}, \rho_{vh} \bI_{\frac{M}{2}} \right] \right)  
		\notag\\
		&
		\stackrel{(d)}{=}  \frac{(2\pi)^2 b^2}{M_hM_v} (|\rho_{vv} |^2+|\rho_{vh} |^2 )                            
		\notag\\    
		&
		=\frac{(2\pi)^2}{M_h M_v}.					 
		\end{align}
		\endgroup
		The equality $(a)$ is derived by the replacement ${\bc'}=\bR(\phi)\bc = [{{\bc'}}_{vv}^\mathrm{T},{{\bc'}}_{vh}^\mathrm{T} ]^\mathrm{T}$ where ${{\bc'}}_{vv}\in\mathbb{C}^{\frac{M}{2}\times1}$ is the first half elements of ${\bc'}$ and ${{\bc'}}_{vh}\in\mathbb{C}^{\frac{M}{2}\times1}$ is the last half elements of ${\bc'}$, and each vector is further divided equally as ${{\bc'}}_{vx}=[{{\bc'}}_{vx,1}^\mathrm{T},\ {{\bc'}}_{vx,2}^\mathrm{T},\cdots,\ {{\bc'}}_{vx,M_h}^\mathrm{T} ]^\mathrm{T},\ {{\bc'}}_{vx,\ell}\in\mathbb{C}^{M_v\times1},\ vx\in\{vv,vh\}$. Equalities $(b)$ and $(c)$ are derived based on the Parseval's theorem \cite{H.Baher:2001}. The equality $(d)$ is obtained by the fact that the eigenvectors corresponding to the zero eigenvalue are the linear combination of the vectors $[-\rho_{vh} \bee_{\frac{M}{2},k}^\mathrm{T},\rho_{vv} \bee_{\frac{M}{2},k}^\mathrm{T} ]^\mathrm{T}$, $k\in\{1,\cdots,\frac{M}{2}\}$, and the eigenvectors corresponding to the eigenvalue $|\rho_{vv}|^2+|\rho_{vh}|^2$ are the linear combination of the vectors $[\rho_{vv}^* \bee_{\frac{M}{2},k}^\mathrm{T},\rho_{vh}^* \bee_{\frac{M}{2},k}^\mathrm{T} ]^\mathrm{T}$, $k\in\{1,\cdots,\frac{M}{2}\}$. 
	\end{IEEEproof}
	
	In the following lemma, we use Lemma \ref{lemma 1} to give an upper bound of the date rate \eqref{data rate} and define the equal gain of the ideal beam pattern to achieve this upper bound. 
	
	\begin{lemma}\label{lemma 2}
		In the region $B^{(p,q)}$, the ideal beam pattern
		\begin{align}\label{ideal beam pattern}
		g^{(p,q)}_\mathrm{ideal} (\psi^\mathrm{az},\psi^\mathrm{el} )=\begin{cases}
		\frac{Q\sqrt{2}}{M_h M_v },   &(\psi^\mathrm{az},\psi^\mathrm{el} )\in B^{(p,q)}    \\0,      &(\psi^\mathrm{az},\psi^\mathrm{el} )\notin B^{(p,q)} 	\end{cases}
		\end{align}
		achieves the upper bound of the expected data rate \eqref{data rate}
		\begin{align}\label{upper bound}
		R_\mathrm{data}^\mathrm{upper}=\log_2\left(1+\frac{P}{b^2\sigma^2}  |h_0|^2G_\mathrm{rx} \frac{Q\sqrt{2}}{M_h M_v}\right),
		\end{align}
		where $G_\mathrm{rx}=\left| \bar{\bc}_\mathrm{rx}^\mathrm{H}\bd_{\mathrm{rx}}(\psi_{\mathrm{rx},0}^{\mathrm{az}},\psi_{\mathrm{rx},0}^{\mathrm{el}}) \right|^2$.
	\end{lemma}
	\begin{IEEEproof}
		We first consider an arbitrary reference gain $t(\psi^\mathrm{az},\psi^\mathrm{el} )$ with the upper bound of its integral as given in Lemma \ref{lemma 1}. Assuming uniform distribution of $\psi^\mathrm{az}$ and $\psi^\mathrm{el}$ in the region $B^{(p,q)}$, the conditioned data rate of the reference gain is 
		\begin{align}
		&R_\mathrm{data}
		\notag\\
		&~
		=
%		\underset{\psi^\mathrm{el},\psi^\mathrm{el}}{\E}
		{\E}
		\bigg[ \log_2\bigg( 1+\frac{P}{b^2\sigma^2}  |h_0|^2G_\mathrm{rx} 
		t(\psi^\mathrm{az},\psi^\mathrm{el}) \bigg) \bigg| \bar{\bc}_{\mathrm{rx}},|h_0| \bigg]
		\notag\\
		&~
		=\frac{\iint_{B^{(p,q)}}\log_2\bigg( 1+\frac{P}{b^2\sigma^2}  |h_0|^2G_\mathrm{rx} 
			t(\psi^\mathrm{az},\psi^\mathrm{el}) \bigg) d\psi^\mathrm{az} d\psi^\mathrm{el}}{\iint_{B^{(p,q)}} d\psi^\mathrm{az} d\psi^\mathrm{el}}\notag\\
		&~
		\stackrel{(a)}{\le}  \log_2\left(1+  \frac{P}{b^2\sigma^2}  |h_0|^2G_\mathrm{rx} \frac{ \iint_{B^{(p,q)}}
			t(\psi^\mathrm{az},\psi^\mathrm{el})  d\psi^\mathrm{az} d\psi^\mathrm{el} }{ \iint_{B^{(p,q)}} d\psi^\mathrm{az} d\psi^\mathrm{el} }\right) \notag\\
%		&~
%		=  \log_2\left(1+\frac{P}{b^2\sigma^2}  |h_0|^2G_\mathrm{rx} \frac{  \iint_{B^{(p,q)}} t(\psi^\mathrm{az},\psi^\mathrm{el} )  d\psi^\mathrm{az} d\psi^\mathrm{el} }{ \frac{2\pi}{Q_h}\frac{2\pi}{\sqrt{2}Q_v}}\right) \notag\\
		&~
		\stackrel{(b)}{\le}  \log_2\left(1+\frac{P}{b^2\sigma^2}  |h_0|^2G_\mathrm{rx} \frac{Q\sqrt{2}}{M_h M_v}\right),	      
		\end{align}
		where $(a)$ is derived by Jensen's inequality, and $(b)$ is by Lemma \ref{lemma 1} and the fact that $\iint_{B^{(p,q)}}$ $ d\psi^\mathrm{az} d\psi^\mathrm{el}=\frac{2\pi}{Q_h}\frac{2\pi}{\sqrt{2}Q_v}$. The equalities of $(a)$ and $(b)$ hold when
		\begin{align} 
		t(\psi^\mathrm{az},\psi^\mathrm{el} )=
		\begin{cases}
		\frac{Q\sqrt{2}}{M_h M_v},   &(\psi^\mathrm{az},\psi^\mathrm{el} )\in B^{(p,q)}\\
		0,      &(\psi^\mathrm{az},\psi^\mathrm{el} )\notin B^{(p,q)} 	      
		\end{cases},
		\end{align} 
		which finishes the proof.
	\end{IEEEproof}
	
	To complete the definition of the ideal beam pattern, we apply an additional constraint on the reference gain
	\begin{align}
	&g_\mathrm{ref} (\psi^\mathrm{az},\psi^\mathrm{el},\bc)\notag \\
	&~
	\le 
	\left\lVert b\left\{ \begin{bmatrix} \rho_{vv} & \rho_{vh} \end{bmatrix} \otimes \big(\bd_h (\psi^\mathrm{az} )\otimes\bd_v (\psi^\mathrm{el} ) \big)^\mathrm{H} \right\} \bR(\phi)\right\rVert_2^2 \lVert\bc \rVert_2^2
	\notag\\
	&~
	=1.
	\end{align}
	This constraint, however, is trivially satisfied when ${M_h M_v} > {Q\sqrt{2}}$, which usually holds for mmWave systems using a large number of antennas, and we set the ideal beam pattern as \eqref{ideal beam pattern}.
	
	\subsection{Optimization criterion 1: squared error (SE)}\label{SE}

%	In \cite{J.Zhang:2017}, the ripple and average gain of beamforming gain is considered to optimize a beamformer, where additional parameters for the optimization and iterative updates are need to be selected. The resulting performance of the beamformer in \cite{J.Zhang:2017} varies over the parameters. The SE criterion in this section, on the contrary, requires one parameters to be defined, and the performance is similar over the different parameter values.
	With the defined ideal Tx beam pattern in \eqref{ideal beam pattern}, we can find a codeword that has the most similar beam pattern to the ideal beam pattern at each quantized region. We assess the similarity by the SE with the ideal beam pattern and find an optimal codeword of which the beam pattern has the minimum SE with the ideal beam pattern.  
	
	To calculate the SE between two beam patterns, we consider vector forms of beam patterns. By dividing each quantized region into $L=L_h L_v$ sections with $L_h$ columns and $L_v$ rows, we can define a vector with  each component being the gain of each section
	\begin{align}
	\label{beam pattern vector ideal}
	\bg^{(p,q)}_\mathrm{ideal}
	&=G \bee_{Q_h,p} \otimes \bee_{Q_v,q} \otimes \boldsymbol{1}_L,                \\
	\label{beam pattern vector codeword}
	\bg(\bc)
	&=\left\lvert b\left( \begin{bmatrix} \rho_{vv}& \rho_{vh} \end{bmatrix} \otimes \bD^\mathrm{H}\right)\bR(\phi) \bc \right\rvert^2   ,
	\end{align}
	where $\bg^{(p,q)}_\mathrm{ideal}$ is the ideal beam pattern vector at region $B^{(p,q)}$, $\bg(\bc)$ is the beam pattern vector of the codeword $\bc$, $G=\frac{Q\sqrt{2}}{M_h M_v}$ is the equal gain of the ideal beam pattern, $p\in\{1,\cdots,Q_h \}$, $q\in\{1,\cdots,Q_v \}$, and $\bD=\bD_h\otimes\bD_v$ with
	\begin{align}
	&
	\bD_h=\bigg[\bd_h \left(-\pi+\frac{\pi}{Q_h L_h}\right),\bd_h \left(-\pi+\frac{\pi}{Q_h L_h}+\frac{2\pi}{Q_h L_h}\right),\notag\\
	&\qquad\quad\ \ \cdots,\bd_h \left(-\pi+\frac{\pi}{Q_h L_h} +\frac{2\pi (Q_h L_h-1)}{Q_h L_h}\right)\bigg], \\
	&
	\bD_v=\bigg[\bd_v \left(-\frac{\pi}{\sqrt{2}}+\frac{\pi}{\sqrt{2}Q_v L_v}\right),\notag\\
	&\qquad\quad\ \ \bd_v \left(-\frac{\pi}{\sqrt{2}}+\frac{\pi}{\sqrt{2}Q_v L_v}+\frac{2\pi}{\sqrt{2}Q_v L_v} \right),\notag\\
	&\qquad\quad\ \ \cdots,\bd_v \left(-\frac{\pi}{\sqrt{2}}+\frac{\pi}{\sqrt{2}Q_v L_v}+\frac{2\pi (Q_v L_v-1)}{\sqrt{2}Q_v L_v} \right)\bigg]   .
	\end{align} 
	The columns of $\bD$ are the array response vectors of which angles are directing to the center of $QL$ sections of entire quantized regions. With the vector forms of the ideal beam pattern and the beam pattern of a given codeword, the SE optimal codeword is
	\begin{align}\label{dual opt}
	\bc^{(p,q)}_{SE,\mathrm{dual}}=\argmin_{\bc\in\mathbb{C}^{M\times1}} \lVert\bg^{(p,q)}_\mathrm{ideal}-\bg(\bc)\rVert_2^2.
	\end{align}
	We focus on the region $B^{(1,1)}$ in this paper and use $\bc_{SE,\mathrm{dual}}$ to represent the optimal beamformer $\bc^{(1,1)}_{SE,\mathrm{dual}}$ as a brief notation. The optimal beamformer for other regions can be obtained using the information of $\bc_{SE,\mathrm{dual}}$, which will be described later in this subsection.

	To have a closed form solution, which does not exist in \eqref{dual opt}, we reformulate the vector forms of beam patterns in \eqref{beam pattern vector ideal} and \eqref{beam pattern vector codeword} as
	\begin{align}
	\bg^{(1,1)}_\mathrm{ideal}  
	&=G \bee_{Q_h,1}\otimes\bee_{Q_v,1}\otimes\boldsymbol{1}_L	\notag\\
	&=G \bee_{Q_h,1}\otimes \boldsymbol{1}_{L_h}\otimes\bee_{Q_v,1}\otimes\boldsymbol{1}_{L_v}\notag\\		
	&=\left\{\sqrt{G}  \bee_{Q_h,1}\otimes \bq_{L_h}\otimes \bee_{Q_v,1}\otimes\bq_{L_v} \right\} \notag\\
	&\quad \odot\left\{\sqrt{G}  \bee_{Q_h,1}\otimes\bq_{L_h}\otimes\bee_{Q_v,1}\otimes\bq_{L_v} \right\}^*    \label{ideal vector},
	\\
	\bg(\bc)&=\left\{b\left( \begin{bmatrix} \rho_{vv}& \rho_{vh} \end{bmatrix} \otimes \bD ^\mathrm{H}\right) \bR(\phi) \bc\right\}\notag\\
	&\quad \odot\left\{b \left( \begin{bmatrix} \rho_{vv}& \rho_{vh} \end{bmatrix} \otimes\bD^\mathrm{H}\right) \bR(\phi) \bc\right\}^*, \label{beamformer vector}
	\end{align}		
	where $\bq_{L_a}\in\mathbb{C}^{L_a\times1}$, $a\in\{h,v\}$ is an arbitrary vector satisfying $\bq_{L_a}\otimes\bq_{L_a}^*=\boldsymbol{1}_{L_a}$. With the reformulated forms, we replace the original objective function in \eqref{dual opt} with another function that usually gives a suboptimal solution
	\begin{align} \label{SE opt}
	\bc_{SE,\mathrm{dual}}
	&=\argmin_{\bc\in\mathbb{C}^{M\times1}} \Bigg\lVert \gamma \left\{b \left( \begin{bmatrix} \rho_{vv} & \rho_{vh} \end{bmatrix} \otimes \bD^\mathrm{H}\right) \bR(\phi) \bc\right\} \notag\\
	&\quad -\left\{\sqrt{G} (\bee_{Q_h,1}\otimes\bq_{L_h}\otimes\bee_{Q_v,1}\otimes\bq_{L_v} ) \right\}\Bigg\rVert_2^2 ,
	\end{align} 
	where $\gamma\in\mathbb{C}$ is a complex normalization constant. To find the constant $\gamma$, we use the Wirtinger derivative of the objective function \cite{R.Hunger:2007}, where the derivative becomes zero with 
%	\begin{align} \label{Wirtinger derivative}
%		&\frac{\partial}{\partial\gamma^*} \Bigg\lVert \gamma \left\{b \left( \begin{bmatrix} \rho_{vv} & \rho_{vh} \end{bmatrix} \otimes\bD^\mathrm{H}\right) \bR(\phi) \bc\right\} \notag\\
%		&\qquad -\left\{\sqrt{G} (\bee_{Q_h,1}\otimes\bq_{L_h}\otimes\bee_{Q_v,1}\otimes\bq_{L_v} )\right\} \Bigg\rVert_2^2 
%		\notag
%%		\\
%	\end{align}
%	\begin{align}
%		&
%		=\gamma \left\lVert b\left( \begin{bmatrix} \rho_{vv} & \rho_{vh} \end{bmatrix} \otimes \bD^\mathrm{H} \right) \bR(\phi) \bc \right\rVert_2^2 
%		\notag\\
%		&\qquad
%		-\Bigg\{b \left( \begin{bmatrix} \rho_{vv} & \rho_{vh} \end{bmatrix} \otimes \bD^\mathrm{H}\right) \bR(\phi) \bc\Bigg\}^\mathrm{H} 
%		\notag\\ 
%		&\qquad  
%		\cdot \left\{ \sqrt{G} (\bee_{Q_h,1} \otimes \bq_{L_h} \otimes \bee_{Q_v,1} \otimes \bq_{L_v} ) \right\}.
%	\end{align}
%	The constant, then, is found as a complex number that makes the derivative \eqref{Wirtinger derivative} equal to zero
	\begin{align} \label{gamma value}
	&
	\gamma
	=  \frac{ \left\{b \left( \begin{bmatrix} \rho_{vv} & \rho_{vh} \end{bmatrix} \otimes \bD^\mathrm{H}\right) \bR(\phi) \bc\right\}^\mathrm{H}}
	{\left\lVert b \left( \begin{bmatrix} \rho_{vv} & \rho_{vh} \end{bmatrix} \otimes \bD^\mathrm{H}\right) \bR(\phi) \bc\right\rVert_2^2}  \notag\\
	&
	\qquad\ \ \cdot \sqrt{G}(\bee_{Q_h,1} \otimes \bq_{L_h} \otimes \bee_{Q_v,1} \otimes \bq_{L_v} ).
	\end{align}

	\begin{figure*}[!b]
		\hrule
		\begin{align} \label{SE last}
		&
		\max_{\tilde{\boldsymbol{\mu}}\in\mathbb{C}^{\frac{M}{2}\times1}} \Bigg| \sqrt{G} (\bee_{Q_h,1} \otimes \bq_{L_h} \otimes \bee_{Q_v,1} \otimes \bq_{L_v} )^\mathrm{H}
		\cdot \frac{ \left( \begin{bmatrix} \rho_{vv}& \rho_{vh} \end{bmatrix} \otimes \bD ^\mathrm{H}\right) [\rho_{vv}^*\tilde{\boldsymbol{\mu}}^\mathrm{T},\rho_{vh}^*\tilde{\boldsymbol{\mu}}^\mathrm{T} ]^\mathrm{T} }
		{\left\lVert \left( \begin{bmatrix} \rho_{vv}& \rho_{vh} \end{bmatrix} \otimes \bD ^\mathrm{H}\right) [\rho_{vv}^*\tilde{\boldsymbol{\mu}}^\mathrm{T},\rho_{vh}^*\tilde{\boldsymbol{\mu}}^\mathrm{T} ]^\mathrm{T} \right\rVert_2 } \Bigg|^2 		
		\notag\\
		%		&
		%		=\max_{\tilde{\boldsymbol{\mu}}\in\mathbb{C}^{\frac{M}{2}\times1}} \Bigg| \sqrt{G} (\bee_{Q_h,1} \otimes \bq_{L_h} \otimes \bee_{Q_v,1} \otimes \bq_{L_v} )^\mathrm{H}
		%		\cdot \frac{ \left( \begin{bmatrix} \rho_{vv}& \rho_{vh} \end{bmatrix} \otimes \bD ^\mathrm{H}\right) \left( \begin{bmatrix} \rho_{vv}^*\\ \rho_{vh}^* \end{bmatrix} \otimes \tilde{\boldsymbol{\mu}} \right) }
		%		{\left\lVert \left( \begin{bmatrix} \rho_{vv}& \rho_{vh} \end{bmatrix} \otimes \bD ^\mathrm{H}\right) \left( \begin{bmatrix} \rho_{vv}^*\\ \rho_{vh}^* \end{bmatrix} \otimes \tilde{\boldsymbol{\mu}} \right) \right\rVert_2 } \Bigg|^2 		
		%		\notag\\
		&
		\stackrel{(a)}{=}\max_{\tilde{\boldsymbol{\mu}}\in\mathbb{C}^{\frac{M}{2}\times1}} \Bigg| \sqrt{G} (\bee_{Q_h,1} \otimes \bq_{L_h} \otimes \bee_{Q_v,1} \otimes \bq_{L_v} )^\mathrm{H} 
		\cdot \frac{(|\rho_{vv}|^2+|\rho_{vh}|^2)\bD^\mathrm{H}\tilde{\boldsymbol{\mu}}} {\lVert (|\rho_{vv}|^2+|\rho_{vh}|^2)\bD^\mathrm{H} \tilde{\boldsymbol{\mu}} \rVert_2} \Bigg|^2 
		\notag\\
		%		&
		%		=\max_{\tilde{\boldsymbol{\mu}}\in\mathbb{C}^{\frac{M}{2}\times1}} \Bigg| \sqrt{G} (\bee_{Q_h,1} \otimes \bq_{L_h} \otimes \bee_{Q_v,1} \otimes \bq_{L_v} )^\mathrm{H} 
		%		\cdot \frac{(\bD_h\otimes\bD_v )^\mathrm{H}\tilde{\boldsymbol{\mu}}}{\lVert \bD^\mathrm{H} \tilde{\boldsymbol{\mu}} \rVert_2} \Bigg|^2 
		%		\notag\\
		&
		\stackrel{(b)}{=}\max_{\tilde{\boldsymbol{\mu}}\in\mathbb{C}^{\frac{M}{2}\times1}} \left| \frac{\sqrt{G}(\bD_{h,1}\bq_{L_h}\otimes\bD_{v,1}\bq_{L_v})^\mathrm{H}\tilde{\boldsymbol{\mu}} }{\lVert\bD^\mathrm{H} \tilde{\boldsymbol{\mu}}\rVert_2 }\right|^2\tag{42}
		\end{align}
	\end{figure*}

	To describe the problem with a simple form, we consider the effective codeword ${\bc'}=\bR(\phi) \bc$. The codeword ${\bc'}\in\mathbb{C}^{M\times1}$ also has the unit-norm because $\bR(\phi)$ is a unitary matrix, and the original codeword can be recovered by $\bc={\left(\bR(\phi)\right)} ^{-1} {\bc'}=\bR(\phi)^\mathrm{H}{\bc'}$. Substituting the effective codeword and $\gamma$, the reformulated objective function of \eqref{SE opt} is written as
	\begin{align}\label{SE max}
	&\argmax_{{\bc'}\in\mathbb{C}^{M\times1}} \Bigg|\sqrt{G} (\bee_{Q_h,1} \otimes \bq_{L_h} \otimes \bee_{Q_v,1} \otimes \bq_{L_v} )^\mathrm{H} \notag \\
	&\qquad \qquad \cdot\frac{\left\{ \left( \begin{bmatrix} \rho_{vv} & \rho_{vh} \end{bmatrix}\otimes \bD ^\mathrm{H}\right){\bc'} \right\} }
	{\left\lVert  \left( \begin{bmatrix} \rho_{vv} & \rho_{vh} \end{bmatrix}\otimes \bD ^\mathrm{H}\right){\bc'}  \right\rVert_2}\Bigg|^2.
	\end{align} 
	In the objective function \eqref{SE max}, we first consider the denominator
	\begin{align}\label{SE denominator}
	&\left\lVert  \left( \begin{bmatrix} \rho_{vv} & \rho_{vh} \end{bmatrix}\otimes \bD ^\mathrm{H}\right){\bc'}  \right\rVert_2^2 \notag\\
	&\quad 
	={\bc'}^\mathrm{H} \left\{\left( \begin{bmatrix} \rho_{vv} & \rho_{vh} \end{bmatrix}\otimes \bD^\mathrm{H} \right)^\mathrm{H}\left( \begin{bmatrix} \rho_{vv} & \rho_{vh} \end{bmatrix}\otimes \bD ^\mathrm{H}\right) \right\} {\bc'}\notag\\
	&\quad
	={\bc'}^\mathrm{H} \bK{\bc'},
	\end{align} 
	where $\bK=\left( \begin{bmatrix} \rho_{vv} & \rho_{vh} \end{bmatrix}\otimes \bD^\mathrm{H} \right)^\mathrm{H}\left( \begin{bmatrix} \rho_{vv} & \rho_{vh} \end{bmatrix}\otimes \bD ^\mathrm{H}\right)$. The term in \eqref{SE denominator} depends on the eigenvalues of $\bK$. We can consider a set $\Omega=\left\{ \left[-\rho_{vh} \tilde{\boldsymbol{\nu}}^\mathrm{T},\rho_{vv} \tilde{\boldsymbol{\nu}}^\mathrm{T} \right]^\mathrm{T}: \tilde{\boldsymbol{\nu}}\in\mathbb{C}^{\frac{M}{2}\times1} \right\}$ of which the elements are the eigenvectors of $\bK$ with corresponding zero eigenvalue. The vectors in $\Omega$ span half of the vector space of dimension $\mathbb{C}^{M\times1}$. Another set that we can consider is $\Gamma=\left\{\left[ \rho_{vv}^*\tilde{\boldsymbol{\mu}}^\mathrm{T},\rho_{vh}^*\tilde{\boldsymbol{\mu}}^\mathrm{T} \right]^\mathrm{T}: \tilde{\boldsymbol{\mu}}\in\mathbb{C}^{\frac{M}{2}\times1} \right\}$ where the vectors in the set span the rest half of the vector space. All the vectors in $\Gamma$ are orthogonal to all the vectors in $\Omega$, and both sets are closed under vector additions. 
	Hence, the sum of two vectors, one from $\Gamma$ and the other from $\Omega$, can represent all the codewords
	\begin{align} 
	{\bc'}=x\boldsymbol{\mu}+z\boldsymbol{\nu}, ~~ \boldsymbol{\mu}\in\Gamma,\ \boldsymbol{\nu}\in\Omega, 	  
	\end{align} 
	where $x\in\mathbb{C}$ and $z\in\mathbb{C}$ are weights satisfying $\lVert x\boldsymbol{\mu}+z\boldsymbol{\nu} \rVert_2^2=1$. By substituting the sum of two vectors for the effective codeword, we can rewrite the objective function \eqref{SE max} as
	\begin{align} \label{SE replace}
	&\max 
	\Bigg| 
	\sqrt{G} (\bee_{Q_h,1} \otimes \bq_{L_h} \otimes \bee_{Q_v,1} \otimes \bq_{L_v} )^\mathrm{H} \notag\\
	&\qquad \cdot \frac{\left( \begin{bmatrix} \rho_{vv} & \rho_{vh} \end{bmatrix} \otimes \bD ^\mathrm{H}\right) (x\boldsymbol{\mu}+z\boldsymbol{\nu})  }{\left\lVert \left( \begin{bmatrix} \rho_{vv} & \rho_{vh} \end{bmatrix} \otimes \bD ^\mathrm{H}\right) (x\boldsymbol{\mu}+z\boldsymbol{\nu}) \right\rVert_2}
	\Bigg|^2 	\notag\\
	&
	=\max
	\Bigg|
	\sqrt{G} (\bee_{Q_h,1} \otimes \bq_{L_h} \otimes \bee_{Q_v,1} \otimes \bq_{L_v} )^\mathrm{H}\notag\\
	&\qquad \cdot \frac{ \left( \begin{bmatrix} \rho_{vv}& \rho_{vh} \end{bmatrix} \otimes \bD^\mathrm{H}\right) \boldsymbol{\mu}} { 
		\left\lVert \left( \begin{bmatrix} \rho_{vv} &\rho_{vh} \end{bmatrix}\otimes \bD ^\mathrm{H}\right) \boldsymbol{\mu}\right\rVert_2 } 
	\Bigg|^2.
	\end{align}

	\setcounter{equation}{42}

	We can safely assume that the beamformer ${\bc'}$ is equal to $x\boldsymbol{\mu}$ because the objective function is independent from $z$ and $\boldsymbol{\nu}$. Based on the structure of $\boldsymbol{\mu} = [\rho_{vv}^*\tilde{\boldsymbol{\mu}}^\mathrm{T}, \rho_{vh}^*\tilde{\boldsymbol{\mu}}^\mathrm{T} ]^\mathrm{T},\ \tilde{\boldsymbol{\mu}}\in\mathbb{C}^{\frac{M}{2}\times1}$, we can consider sufficient conditions $\lVert\tilde{\boldsymbol{\mu}}\rVert_2^2=1$ and $x=b$ instead of the unit-norm constraint $\lVert{\bc'}\rVert_2^2=1$. With the sufficient conditions, the following simplified objective function is as \eqref{SE last} on the bottom of the previous page.
	In the equation, $\bD_{h,1}  = \bD_h\cdot (\bee_{Q_h,1}\otimes \bI_{L_h})$, $\bD_{v,1}=\bD_v\cdot (\bee_{Q_v,1}\otimes\bI_{L_v})$, and $(a)$ and $(b)$ are by the property of the Kronecker product $(\bW\otimes \bX)(\bY\otimes \bZ)=(\bW\bY)\otimes(\bX\bZ)$ and the structure $\boldsymbol{\mu}=\begin{bmatrix} \rho_{vv}^*\\ \rho_{vh}^* \end{bmatrix} \otimes \tilde{\boldsymbol{\mu}}$. The objective function \eqref{SE last} is the same as the reformulated objective function of single-polarization beamformer in \cite{J.Song:2017}, and the solutions of two objective functions are obviously the same. With the solution $\tilde{\boldsymbol{\mu}}_{opt}$ of the objective function in \eqref{SE last}, the optimal dual-polarization beamformer is 
	\begin{align} \label{SE final}
		\bc_{SE,\mathrm{dual}}
		&=\bR(\phi)^\mathrm{H} \bc_\mathrm{opt}'	
		\notag\\
		&
		=b\bR(\phi)^\mathrm{H}[\rho_{vv}^*\tilde{\boldsymbol{\mu}}_\mathrm{opt}^\mathrm{T},\rho_{vh}^*\tilde{\boldsymbol{\mu}}_\mathrm{opt}^\mathrm{T} ]^\mathrm{T}  
		\notag \\
		&
		=b\bR(\phi)^\mathrm{H}\left( \begin{bmatrix} \rho_{vv}^* \\ \rho_{vh}^*\end{bmatrix} \otimes\bc_{SE,\mathrm{single}} \right),
	\end{align}
	where $\tilde{\boldsymbol{\mu}}_\mathrm{opt}=\bc_{SE,\mathrm{single}}=\frac{\bD_{h,1} \bq_{L_h}\otimes\bD_{v,1} \bq_{L_v}}{\lVert\bD_{h,1} \bq_{L_h}\otimes\bD_{v,1} \bq_{L_v} \rVert_2}$ is the solution of the objective function of single-polarization beamformer \cite{J.Song:2017}. Notice that the beamformer depends on the values $(\bq_{L_h},\bq_{L_v})$. We use the notation $\bc_{SE,\mathrm{dual}} (\bq_{L_h},\bq_{L_v} )$ to represent this dependency and call the beamformer as the SE beamformer candidate. The final hybrid SE beamformer will be chosen among the candidates after applying the OMP-based algorithm.

	%\alglanguage{pseudocode}
	\begingroup
	\begin{algorithm}[!t]
		\caption{Hybrid SE beamformer candidate design based on OMP}
		\label{SE algorithm}
		\textbf{Initialize}: for a given $\left(\bq_{L_h},\bq_{L_v}\right)\in \mathcal{G}_h\times\mathcal{G}_v$
		\begin{algorithmic}[1]
			\State Optimal beamformer candidate: 
			\begin{align*}
			\bc_{SE,\mathrm{dual}}(\bq_{L_h},\bq_{L_v})
			=b\bR&(\phi)^\mathrm{H}\bigg( \begin{bmatrix} \rho_{vv}^*\\ \rho_{vh}^* \end{bmatrix} \\
			&\otimes \frac{\bD_{h,1}\bq_{L_h}\otimes\bD_{v,1}\bq_{L_v}}{\left\lVert \bD_{h,1}\bq_{L_h}\otimes\bD_{v,1}\bq_{L_v} \right\rVert_2}\bigg)
			\end{align*}
			\State Define $\bF_0$ as an empty matrix and initialize residual: 
			\begin{align*} 
			\br_0=\bc_{SE,\mathrm{dual}}(\bq_{L_h},\bq_{L_v})
			\end{align*}
		\end{algorithmic}
		\textbf{Hybrid beamformer update}
		\begin{algorithmic}[1]
			\addtocounter{ALG@line}{+2}		% 알고리즘번호 2개 올리기
			\State \textbf{For} $1\le n \le N$
			\State Update analog beamformer: 
			\begin{align*}
			\bF_n=[\bF_{n-1},\ \bff_n]\in \mathbb{C}^{M\times N},
			\end{align*}
			where $\bff_n=\frac{1}{\sqrt{M}}e^{j\angle\br_{n-1}}\in\mathbb{C}^{M\times1}$.
			\State Update digital beamformer: 
			\begin{align*} 
			%		\bv_n=\frac{\bu_n}{\left\lVert\bF_n\bu_n\right\rVert_2}\in\mathbb{C}^{N\times 1},
			\bv_n=\frac{\bu_n}{\left\lVert\bF_n\bu_n\right\rVert_2}\in\mathbb{C}^{N\times 1},
			\end{align*}
			where $\bu_n = (\bF_n^\mathrm{H}\bF_n)^{-1}\bF_n^\mathrm{H}\bc_{SE,\mathrm{dual}}(\bq_{L_h},\bq_{L_v})$
			%		$\bu_n=\boldsymbol{\mathfrak{v}}_\mathrm{max}\left( \left( \bF_n^\mathrm{H}\bX\bX^\mathrm{H}\bF_n \right)^{-1} \bF_n^\mathrm{H} \left( \bX\boldsymbol{\eta}\boldsymbol{\eta}^\mathrm{H}\bX^\mathrm{H} \right)\bF_n \right)\in\mathbb{C}^{N\times1}$, $\boldsymbol{\eta}=\bee_{Q_h,1}\otimes\bq_{L_h}\otimes\bee_{Q_v,1}\otimes\bq_{L_v}$, and $\bX=\bR(\phi)\left( \begin{bmatrix} \rho_{vv}& \rho_{vh} \end{bmatrix} \otimes \bD \right)$.
			\State Update residual:
			\begin{align*}
			\br_n=\bc_{SE,\mathrm{dual}}(\bq_{L_h},\bq_{L_v})-\bF_n\bv_n
			\end{align*}
			\State \textbf{End for}
			\State Update analog and digital beamformers:
			\begin{align*}
			\bF(\bq_{L_h},\bq_{L_v})=\bF_N, \quad \bv(\bq_{L_h},\bq_{L_v})=\bv_N
			\end{align*}
		\end{algorithmic}
		\textbf{Compute hybrid SE beamformer candidate}
		\begin{algorithmic}[1]
			\addtocounter{ALG@line}{+8}
			\State Hybrid SE beamformer candidate:
			\begin{align*}
			\bF(\bq_{L_h},\bq_{L_v})\bv(\bq_{L_h},\bq_{L_v})
			\end{align*}
		\end{algorithmic}
	\end{algorithm} 
	\endgroup
	
	The hybrid SE beamformer candidate is obtained by operating Algorithm \ref{SE algorithm}, which is based on the OMP algorithm \cite{O.E.Ayach:2014,Y.C.Pati:1993}, on the SE beamformer candidate. For each candidate $\bc_{SE,\mathrm{dual}} (\bq_{L_h},\bq_{L_v} )$ with $(\bq_{L_h},\bq_{L_v} )\in \mathcal{G}_h\times \mathcal{G}_v$, where $\mathcal{G}_h$ and $\mathcal{G}_v$ are the sets of $\bq_{L_h}$ and $\bq_{L_v}$, the algorithm iteratively updates the digital beamformer $\bv\in\mathbb{C}^{N\times1}$ and each column of the analog beamformer $\bF=[\bff_1,\bff_2,\cdots,\bff_N ]\in\mathbb{C}^{M\times N}$ one by one. In the algorithm, we first compute the SE beamformer candidate for a given $(\bq_{L_h},\bq_{L_v} )$. We, then, initialize the analog beamformer as an empty matrix and residual $\br_n$ as the SE beamformer candidate. In the hybrid SE beamformer update, we iteratively update digital and analog beamformers and residual. The updated $n$-th column of the analog beamformer is $\bff_n=\frac{1}{\sqrt{M}}  e^{j\angle \br_{n-1}}$, where $(\angle\br_{n-1})_{(i)}$ is the phase of $(\br_{n-1})_{(i)}$ and we take $\angle \br_{n-1}\in [0,2\pi)^M$ for the concrete expression. For the update of the digital beamformer, we consider a vector $\bu_n\in\mathbb{C}^{N\times1}$, which constitutes the digital beamformer $\bv_n=\frac{\bu_n}{\lVert \bF_n \bu_n\rVert_2}$, to satisfy the unit-norm constraint $\lVert \bF_n \bv_n \rVert_2=1$. The vector $\bu_n$ is designed to project the SE beamformer candidate on the analog beamformer space as
	%By substituting ${\bc'}$ in \eqref{SE max} with $\bR(\phi) \bF_n \bv_n$, $\bu_n$ is obtained as the solution of the objective function \eqref{SE max}
	%\begin{align} \label{SE digital beamformer}
	%	\bu_n 
	%	&
	%	= \argmax_{\bu\in\mathbb{C}^{N\times1}} \sqrt{G}\Bigg|  (\bee_{Q_h,1} \otimes \bq_{L_h} \otimes \bee_{Q_v,1} \otimes \bq_{L_v} )^\mathrm{H} \notag\\
	%	&\quad \cdot \frac{ \left\{ \left( \begin{bmatrix} \rho_{vv}\\ \rho_{vh}\end{bmatrix} \otimes \bD^\mathrm{H}\right) \bR(\phi) \bF_n \bu\right\} } 
	%	{ \left\lVert \left( \begin{bmatrix} \rho_{vv}\\ \rho_{vh}\end{bmatrix} \otimes \bD^\mathrm{H}\right) \bR(\phi) \bF_n \bu \right\rVert_2}  \Bigg|^2 \notag \\
	%	&
	%	=\argmax_{\bu\in\mathbb{C}^{N\times1}} \frac{ \bu^\mathrm{H} \bF_n^\mathrm{H} \bX\boldsymbol{\eta}\boldsymbol{\eta}^\mathrm{H} \bX^\mathrm{H} \bF_n \bu}{\bu^\mathrm{H} \bF_n^\mathrm{H} \bX\bX^\mathrm{H} \bF_n \bu} \notag\\ 		
	%	&
	%	=\boldsymbol{\mathfrak{v}}_\mathrm{max} \left( {(\bF_n^\mathrm{H} \bX\bX^\mathrm{H} \bF_n )}^{-1} \bF_n^\mathrm{H} (\bX\boldsymbol{\eta}\boldsymbol{\eta}^\mathrm{H} \bX^\mathrm{H} ) \bF_n \right),
	%\end{align}
	%where $\boldsymbol{\eta}=\bee_{Q_h,1}\otimes\bq_{L_h}\otimes\bee_{Q_v,1}\otimes\bq_{L_v}$, and $\bX=\bR(\phi)\left(\begin{bmatrix} \rho_{vv}& \rho_{vh} \end{bmatrix} \otimes \bD\right)$. 
	\begin{align} \label{SE digital beamformer}
	\bu_n 
	&
	= (\bF_n^\mathrm{H}\bF_n)^{-1}\bF_n^\mathrm{H}\bc_{SE,\mathrm{dual}}.
	\end{align}
	Lastly, the residual is updated as the difference between the SE beamformer candidate and the $n$-th updated hybrid beamformer. After the iterative update, the hybrid SE beamformer is obtained as the $N$-th updated hybrid beamformer.

	Among the hybrid SE beamformer candidates, we select the final hybrid SE beamformer for region $B^{(1,1)}$ as 
	\begin{align} 
	\bc_{SE,\mathrm{dual}}^{(1,1)}&=\bF^{(1,1)} \bv^{(1,1)} ,\\
	\quad\bF^{(1,1)}&=\bF(\bq_{L_h,\mathrm{opt}},\bq_{L_v,\mathrm{opt}} ) , \\ \bv^{(1,1)}&=\bv(\bq_{L_h,\mathrm{opt}},\bq_{L_v,\mathrm{opt}} ), 
	\end{align}
	where $(\bq_{L_h,\mathrm{opt}},\bq_{L_v,\mathrm{opt}} )$ is chosen to minimize the SE with the ideal beam pattern vector
	\begin{align}
	&(\bq_{L_h,\mathrm{opt}},\bq_{L_v,\mathrm{opt}} )  					\notag\\
	&\quad =\argmin_{\bq_{L_h}\in \mathcal{G}_h,\bq_{L_v}\in \mathcal{G}_v} \lVert \bg^{(1,1)}_\mathrm{ideal}-\bg(\bF(\bq_{L_h},\bq_{L_v} )\bv(\bq_{L_h},\bq_{L_v} ))\rVert_2^2 .
	\end{align}

	The SE beamformers in other regions $B^{(p,q)}\neq B^{(1,1)}$ can be derived by using the transform function proposed in \cite{J.Song:2017} as
	\begin{align} 
	\bc_{SE,\mathrm{dual}}^{(p,q)}
	&
	=\bR(\phi)^\mathrm{H}\left( \begin{bmatrix} \rho_{vv}^*\\ \rho_{vh}^*\end{bmatrix}\otimes \bc_{SE,\mathrm{single}}^{(p,q)} \right),\notag	\\
	\bc_{SE,\mathrm{single}}^{(p,q)} 
	&
	=T(\bc_{SE,\mathrm{single}}^{(1,1)},\Delta_h^p,\Delta_v^q )\notag\\
	&
	=\bc_{SE,\mathrm{single}}^{(1,1)}\odot \tilde{\bd}(\Delta_h^p,\Delta_v^q ),         
	\end{align}
	where $\Delta_h^p=\frac{2\pi(p-1)}{Q_h}$ , $\Delta_v^q=\frac{2\pi(q-1)}{\sqrt{2} Q_v}$, $\tilde{\bd}(\Delta_h^p,\Delta_v^q )=\frac{\bd(\Delta_h^p,\Delta_v^q )}{\lVert\bd(\Delta_h^p,\Delta_v^q )\odot\bd(\Delta_h^p,\Delta_v^q )\rVert_2}$, and $\bd(\Delta_h^p,\Delta_v^q )=\bd_h (\Delta_h^p )\otimes\bd_v (\Delta_v^q )$. The optimality of $\bc_{SE,\mathrm{dual}}^{(p,q)}$ in other regions can be proved similarly as in $B^{(1,1)}$ with the optimality of the transformed single-polarization SE optimal beamformer $\bc_{SE,\mathrm{single}}^{(p,q)}$.

	\begin{figure*}[!b]
		\hrule
		\begin{align} \label{MIP codeword 3}
		\bg(\bc)^\mathrm{H} \bg^{(1,1)}_\mathrm{ideal}
		%	 \notag\\
		&
		\stackrel{(a)}{=} \Bigg[ \left\{b \left( \begin{bmatrix} \rho_{vv}& \rho_{vh} \end{bmatrix} \otimes\bD^\mathrm{H}\right) \bR(\phi) \bc\right\} 
		%	\notag\\
		%	&\qquad 
		\odot \left\{ b\left( \begin{bmatrix} \rho_{vv}& \rho_{vh} \end{bmatrix} \otimes\bD^\mathrm{H}\right) \bR(\phi) \bc\right\}^* \Bigg]^\mathrm{H} 
		\notag\\
		&\qquad \qquad
		\Big[\Big\{\sqrt{G}(\bee_{Q_h,1}\otimes\bq_{L_h}\otimes\bee_{Q_v,1}\otimes\bq_{L_v})\Big\} 
		%	\notag\\
		%	&\qquad 
		\odot  \Big\{\sqrt{G}
		(\bee_{Q_h,1}\otimes\bq_{L_h}\otimes\bee_{Q_v,1}\otimes\bq_{L_v})\Big\}^* \Big] 	\notag\\
		&
		\stackrel{(b)}{=}\Bigg[ \left\{b \left( \begin{bmatrix} \rho_{vv}& \rho_{vh} \end{bmatrix} \otimes\bD^\mathrm{H}\right) \bR(\phi) \bc\right\} 
		%	\notag\\
		%	&\qquad 
		\odot \Big\{ \sqrt{G}(\bee_{Q_h,1}\otimes\bq_{L_h}\otimes\bee_{Q_v,1}\otimes\bq_{L_v}) \Big\} \Bigg]^\mathrm{H} 
		\notag\\
		&\qquad \qquad
		\Bigg[ \left\{b \left( \begin{bmatrix} \rho_{vv}& \rho_{vh} \end{bmatrix} \otimes\bD^\mathrm{H}\right) \bR(\phi) \bc\right\}  
		%	\notag\\
		%	&\qquad 
		\odot  \Big\{\sqrt{G}(\bee_{Q_h,1}\otimes\bq_{L_h}\otimes\bee_{Q_v,1}\otimes\bq_{L_v})\Big\} \Bigg] 	\notag\\
		&
		= \Bigg\lVert \left\{b \left( \begin{bmatrix} \rho_{vv}& \rho_{vh} \end{bmatrix} \otimes\bD^\mathrm{H}\right) \bR(\phi) \bc\right\} 
		%	\notag\\
		%	&\qquad 
		\odot \Big\{\sqrt{G}(\bee_{Q_h,1}\otimes\bq_{L_h}\otimes\bee_{Q_v,1}\otimes\bq_{L_v})\Big\} \Bigg\rVert^2_2 \tag{51}
		\end{align} 
	\end{figure*}

%%%%%%%%%%%%%%%%%%%%%%%%%%%%%%%%%%%%%%%%%%%%%%%%%%%%%%%%%%%%%
	
	\subsection{Optimization criterion 2: magnitude of inner product (MIP)}\label{MIP}

	The MIP is another criterion that yields a codeword generating a beam pattern similar to the ideal beam pattern. The MIP criterion requires the maximum magnitude of the inner product, and the large magnitude of the inner product means geometrical similarity between the two vectors. In the region $B^{(p,q)}$, the MIP optimal dual-polarization codeword is
	\begin{align} \label{MIP codeword eq}
	\bc^{(p,q)}_{MIP,\mathrm{dual}}=\argmax_{\bc\in\mathbb{C}^{M\times1}} |\bg(\bc)^\mathrm{H} \bg^{(p,q)}_\mathrm{ideal} |,
	\end{align}
	where $\bg_\mathrm{ideal}^{(p,q)}$ and $\bg(\bc)$ are defined in \eqref{beam pattern vector ideal} and \eqref{beam pattern vector codeword}. Considering the region $B^{(1,1)}$, we use $\bc_{MIP,\mathrm{dual}}$ to briefly represent the optimal dual-polarization beamformer $\bc^{(1,1)}_{MIP,\mathrm{dual}}$. 
	
	The inner product can be rewritten by representing each vector with the Hadamard product as \eqref{MIP codeword 3} on the bottom of this page, where the equality $(a)$ is derived by \eqref{ideal vector} and \eqref{beamformer vector}, and $(b)$ is by the fact that $(\bx\odot\bx^*)^\mathrm{H}(\bz\odot\bz^*)=(\bx\odot\bz)^\mathrm{H}(\bx\odot\bz)$. The optimal codeword \eqref{MIP codeword eq} is, then, represented by
	\setcounter{equation}{51}
	\begin{align} \label{MIP codeword 2}
	\bc_{MIP,\mathrm{dual}}&=\argmax_{\bc\in\mathbb{C}^{M\times1}} |\bg(\bc)^\mathrm{H} \bg^{(1,1)}_\mathrm{ideal} | \notag\\
	&=\argmax_{\bc\in\mathbb{C}^{M\times1}} \Bigg\lVert \left\{b \left( \begin{bmatrix} \rho_{vv}& \rho_{vh} \end{bmatrix} \otimes\bD ^\mathrm{H}\right) \bR(\phi) \bc\right\} \notag\\
	&\qquad\quad \odot \left\{\sqrt{G}(\bee_{Q_h,1}\otimes\bq_{L_h}\otimes\bee_{Q_v,1}\otimes\bq_{L_v})\right\} \Bigg\rVert^2_2.
	\end{align}
	To further simplify the objective function in \eqref{MIP codeword 2}, we first derive the following lemma.

	\begin{lemma}\label{lemma 3}
		If a vector $\bx\in\mathbb{C}^{k\times1}$, $k\in\mathbb{N}$ has components with magnitude $1$ or $0$ and a vector $\bz\in\mathbb{C}^{k\times1}$ is $\ell_2$-norm bounded, the $\ell_2$-norm of the Hadamard product of two vectors is
		\begin{align} 
		\lVert\bx\odot\bz\rVert_2^2=\sum_{i\in \mathcal{I}_x} |(\bz)_{(i)} |^2 ,
		\end{align} 
		where $\mathcal{I}_x=\{i\in\{1,\cdots,k\}:|(x)_{(i)} |=1\}$.
	\end{lemma} 
	\begin{IEEEproof} 
		The $\ell_2$-norm of the Hadamard product is 
		\begin{align} 
		\lVert\bx\odot\bz\rVert^2_2
		&=\sum_{i=1}^k |(x)_{(i)} (z)_{(i)} |^2 \notag\\
		& =\sum_{i=1}^k |(x)_{(i)} |^2 |(z)_{(i)} |^2 \notag\\
		&=\sum_{i\in \mathcal{I}_x} |(z)_{(i)} |^2 .
		\end{align}
	\end{IEEEproof}

	\begin{figure*}[!b]
		\hrule
		\begin{align}
		\label{MIP single}
		&
		\bc_{MIP,\mathrm{single}}
		\notag\\
		&
		=\argmax_{\bw\in\mathbb{C}^{\frac{M}{2}\times1}} |\bg_\mathrm{single} (\bw)^\mathrm{H} \bg^{(1,1)}_\mathrm{ideal} |                                      
		\notag\\ 
		&
		\stackrel{(a)}{=}\argmax_{\bw\in\mathbb{C}^{\frac{M}{2}\times1}} \Big[ \left( \bD^\mathrm{H} \bw\right) \odot \left(\bD^\mathrm{H} \bw \right)^* \Big]^\mathrm{H} 
		\Big[\left\{ \sqrt{G} (\bee_{Q_h,1}\otimes\bq_{L_h}\otimes\bee_{Q_v,1}\otimes\bq_{L_v} )\right\} 	
		\odot \Big\{ \sqrt{G}(\bee_{Q_h,1}\otimes\bq_{L_h} \otimes\bee_{Q_v,1}\otimes\bq_{L_v})\Big\}^* \Big] 	\notag\\
%		&
%		=\argmax_{\bw\in\mathbb{C}^{\frac{M}{2}\times1}} \big[\left(\bD^\mathrm{H} \bw\right)	
%		\odot \left\{\sqrt{G} (\bee_{Q_h,1}\otimes \bq_{L_h}\otimes\bee_{Q_v,1}\otimes\bq_{L_v} ) \right\}\Big]^\mathrm{H} 	
%		\Big[\left(\bD^\mathrm{H} \bw\right)		
%		\odot \left\{\sqrt{G} (\bee_{Q_h,1}\otimes \bq_{L_h}\otimes\bee_{Q_v,1}\otimes\bq_{L_v} )\right\}\Big]	\notag
%		\\
		&
		=\argmax_{\bw\in\mathbb{C}^{\frac{M}{2}\times1}} \Big\lVert \left(\bD^\mathrm{H} \bw\right)	
		\odot \left\{\sqrt{G} (\bee_{Q_h,1}\otimes \bq_{L_h}\otimes\bee_{Q_v,1}\otimes\bq_{L_v} ) \right\}\Big\rVert_2^2 	\notag\\
		&
		\stackrel{(b)}{=}  \argmax_{\bw\in\mathbb{C}^{\frac{M}{2}\times1}}G \sum_{i\in \mathcal{I}_{(1,1)}} \left|\left(\bD^\mathrm{H} \bw\right)_{(i)} \right|^2   	
		\quad	=\argmax_{\bw\in\mathbb{C}^{\frac{M}{2}\times1}} G\lVert \bD_1^\mathrm{H}\bw\rVert^2_2		%\notag
		%	\\
		%	&
		\quad =\argmax_{\bw\in\mathbb{C}^{\frac{M}{2}\times1}}G \bw^\mathrm{H} \bD_1 \bD_1^\mathrm{H}\bw		\notag\\  	
		&
		=\boldsymbol{\mathfrak{v}}_\mathrm{max} \left(\bD_1\bD_1^\mathrm{H} \right) \tag{59}
		\end{align} 
	\end{figure*}

	Using Lemma \ref{lemma 3} and the representation of the effective codeword ${\bc'}=\bR(\phi)\bc$, the objective function can be reformulated
%	\begingroup
%	\allowdisplaybreaks
	\begin{align}\label{reformulated objective function}
		&\max_{{\bc'}\in\mathbb{C}^{M\times1}}\ \Bigg\lVert \left\{ b\left( \begin{bmatrix} \rho_{vv}& \rho_{vh} \end{bmatrix} \otimes \bD ^\mathrm{H}\right) {\bc'} \right\} 
		\notag\\
		&\qquad\quad 
		\odot \left\{ \sqrt{G} (\bee_{Q_h,1}\otimes\bq_{L_h}\otimes\bee_{Q_v,1}\otimes\bq_{L_v} ) \right\} \Bigg\rVert_2^2	
		\notag\\
		&\quad
		\stackrel{(a)}{=}   \max_{{\bc'}\in\mathbb{C}^{M\times1}} \ b^2G \sum_{i\in \mathcal{I}_{(1,1)}} \left| \left( \left( \begin{bmatrix} \rho_{vv}& \rho_{vh} \end{bmatrix} \otimes \bD^\mathrm{H}\right) {\bc'} \right)_{(i)} \right|^2  
		\notag\\
		&\quad
		\stackrel{(b)}{=}\max_{{\bc'}\in\mathbb{C}^{M\times1}} b^2G\left\lVert \left( \begin{bmatrix} \rho_{vv}& \rho_{vh}\end{bmatrix} \otimes\bD_1 ^\mathrm{H}\right) {\bc'} \right \rVert_2^2     
		\notag\\
		&\quad
		\stackrel{(c)}{=}\max_{{\bc'}\in\mathbb{C}^{M\times1}} b^2G {\bc'}^\mathrm{H} \bK_1{\bc'},
	\end{align}
%	\endgroup
	where $(a)$ is derived by Lemma \ref{lemma 3} with $\mathcal{I}_{(1,1)}=\Big\{ i\in\{1,\cdots,QL\}: \big| (\bee_{Q_h,1}\otimes\bq_{L_h}\otimes\bee_{Q_v,1}\otimes\bq_{L_v})_{(i)} \big|^2=1 \Big\}$. The equalities $(b)$ and $(c)$ are by representing $\bD_1=\bD_{h,1}\otimes\bD_{v,1}$ and $\bK_1=\left( \begin{bmatrix} \rho_{vv}& \rho_{vh}\end{bmatrix} \otimes\bD_1^\mathrm{H} \right)^\mathrm{H}\left( \begin{bmatrix} \rho_{vv}&\rho_{vh}\end{bmatrix} \otimes\bD_1^\mathrm{H} \right)$. As in Section \ref{SE}, we again consider the two vector sets $\Omega$ and $\Gamma$. Each of two sets spans half of the vector space of dimension $\mathbb{C}^{M\times1}$ exclusively, and the elements in $\Omega$ are eigenvectors of $\bK_1$ with corresponding zero eigenvalue. We represent the codeword as the sum of two vectors each from one of vector sets as in the SE criterion case. The objective function \eqref{reformulated objective function} can be rewritten by using the representation of the codeword ${\bc'}=x\boldsymbol{\mu}+z\boldsymbol{\nu}, ~~ \boldsymbol{\mu}\in\Gamma,\ \boldsymbol{\nu}\in\Omega$
	\begin{align}\label{MIP objective}
%		&\max~
		b^2G(x\boldsymbol{\mu}+z\boldsymbol{\nu})^\mathrm{H} \bK_1
%		\left( \begin{bmatrix} \rho_{vv}& \rho_{vh} \end{bmatrix} \otimes\bD_1^\mathrm{H} \right)^\mathrm{H} 			
%		\notag\\
%		&\qquad\quad 
%		\cdot\left( \begin{bmatrix} \rho_{vv}& \rho_{vh} \end{bmatrix} \otimes\bD_1^\mathrm{H} \right) 
		(x\boldsymbol{\mu}+z\boldsymbol{\nu}) 	
%		\notag\\
%		&\quad
		=
%		\max~
		b^2|x|^2G \boldsymbol{\mu}^\mathrm{H} \bK_1
%		\left( \begin{bmatrix} \rho_{vv}& \rho_{vh} \end{bmatrix} \otimes\bD_1 \right) 
%		\notag\\
%		&\quad\qquad\qquad 
%		\cdot \left( \begin{bmatrix} \rho_{vv}& \rho_{vh} \end{bmatrix} \otimes\bD_1 \right)^\mathrm{H}
		 \boldsymbol{\mu},
	\end{align}
	where $x\in\mathbb{C}$ and $z\in\mathbb{C}$ are weights satisfying $\lVert x\boldsymbol{\mu}+z\boldsymbol{\nu}\rVert_2^2=1$. Because the objective function \eqref{MIP objective} is independent of $z$ and $\boldsymbol{\nu}$, the objective function can be further simplified by writing $\boldsymbol{\mu}=[\rho_{vv}^*\tilde{\boldsymbol{\mu}}^\mathrm{T},\rho_{vh}^*\tilde{\boldsymbol{\mu}}^\mathrm{T} ]^\mathrm{T},~ \tilde{\boldsymbol{\mu}}\in\mathbb{C}^{\frac{M}{2}\times1}$
	\begin{align}\label{simplified objective function}
		\max_{\tilde{\boldsymbol{\mu}}\in\mathbb{C}^{\frac{M}{2}\times1}} 
		G\tilde{\boldsymbol{\mu}}^\mathrm{H} \bD_1  \bD_1^\mathrm{H} \tilde{\boldsymbol{\mu}} ,
	\end{align}
	where $\tilde{\boldsymbol{\mu}}$ is under the constraint $\lVert \tilde{\boldsymbol{\mu}} \rVert_2^2=1$ with $x=b$ satisfying the unit-norm condition of the Tx codeword. The vector that maximizes the above function is $\tilde{\boldsymbol{\mu}}_\mathrm{opt}=\boldsymbol{\mathfrak{v}}_\mathrm{max} \left( \bD_1\bD_1^\mathrm{H} \right)$. The resulting MIP optimal dual-polarization codeword is 
	\begin{align} 
		\bc_{MIP,\mathrm{dual}}
		&
		=\bR(\phi)^\mathrm{H} \bc_\mathrm{opt}'
		\notag\\
		&
		=b\bR(\phi)^\mathrm{H}[\rho_{vv}^* \tilde{\boldsymbol{\mu}}_\mathrm{opt}^\mathrm{T},\rho_{vh}^* \tilde{\boldsymbol{\mu}}_\mathrm{opt}^\mathrm{T} ]^\mathrm{T}		
		\notag\\
		&
		=b\bR(\phi)^\mathrm{H}\left( \begin{bmatrix} \rho_{vv}^*\\ \rho_{vh}^*\end{bmatrix}\otimes\tilde{\boldsymbol{\mu}}_\mathrm{opt} \right). 
	\end{align}

	To represent the dual-polarization beamformer as a function of single-polarization beamformer, we examine the MIP optimal single-polarization beamformer, which has not been done before up to the authors' knowledge. In the region $B^{(1,1)}$, using a unit-norm vector variable $\bw\in\mathbb{C}^{\frac{M}{2}\times1}$, the MIP optimal single-polarization codeword is calculated as \eqref{MIP single} on the bottom of this page. In the equation, $\bg_\mathrm{single} (\bw)=(\bD^\mathrm{H} \bw)\odot(\bD^\mathrm{H} \bw)^*$, and $(a)$ is by \eqref{ideal vector} and the fact that $(\bx\odot\bx^*)^\mathrm{H}(\bz\odot\bz^*)=(\bx\odot\bz)^\mathrm{H}(\bx\odot\bz)$. The equality $(b)$ is derived by Lemma \ref{lemma 3}. The other steps can be acquired in similar ways as in \eqref{MIP codeword 3}. The objective function in \eqref{MIP single} is equivalent to that of \eqref{simplified objective function}, thus, the solutions of two objective functions $\bc_{MIP,\mathrm{single}}$ and $\tilde{\boldsymbol{\mu}}_\mathrm{opt}$ are the same. The equality of the two solutions gives the following expression of the MIP optimal dual-polarization beamformer	
	\setcounter{equation}{59}
	\begin{align} \label{MIP final}
		\bc_{MIP,\mathrm{dual}}=b\bR(\phi)^\mathrm{H} \left( \begin{bmatrix} \rho_{vv}^*\\ \rho_{vh}^* \end{bmatrix} \otimes\bc_{MIP,\mathrm{single}} \right).
	\end{align}

	For other regions $B^{(p,q)}\neq B^{(1,1)}$, the MIP beamformer is computed as
	\begin{align}\label{MIP solution}
	\bc_{MIP,\mathrm{dual}}^{(p,q)}&=b\bR(\phi)^\mathrm{H} \left( \begin{bmatrix} \rho_{vv}^*\\ \rho_{vh}^* \end{bmatrix} \otimes \bc_{MIP,\mathrm{single}}^{(p,q)} \right) ,	\notag\\
	\bc_{MIP,\mathrm{single}}^{(p,q)}&=\boldsymbol{\mathfrak{v}}_\mathrm{max} \left( (\bD_{h,p}\otimes\bD_{v,q} ) (\bD_{h,p}\otimes\bD_{v,q} )^\mathrm{H} \right),
	\end{align}
	where $\bD_{h,p}=\bD_h\cdot(\bee_{Q_h,p}\otimes\bI_{L_h})$, and $\bD_{v,q}=\bD_v\cdot(\bee_{Q_v,q}\otimes\bI_{L_v})$.
	The final hybrid MIP beamformer in each region is, then, obtained by applying an OMP-based algorithm on the MIP beamformer. The algorithm is the same as Algorithm \ref{SE algorithm} for the hybrid SE beamformer case except a few steps. The key difference is the number of candidates. With the sole candidate for the MIP beamformer, the direct computation of the MIP beamformer and application of the OMP-based algorithm at each region is possible under reasonable computation complexity. The hybrid MIP beamformer in other regions, of course, can also be obtained from the hybrid MIP beamformer for $B^{(1,1)}$ as in SE criterion case. This can be derived by representing the single-polarization MIP beamformer for $B^{(p,q)}$ with the transform function $T(\cdot,\cdot,\cdot)$ and $\bc^{(1,1)}_{MIP,\mathrm{single}}$. 
	
	Note that the codebook design complexity is not a serious problem since the codebooks can be attained offline. The offline designs allow to have sufficiently large numbers of sections $L=L_hL_v$ to optimize the beam patterns.

%%%%%%%%%%%%%%%%%%%%%%%%%%%%%%%%%%%%%%%

\subsection{Discussion}\label{discussion}
So far, we have shown that the optimal dual-polarization Tx beamformers can be obtained from the optimal single-polarization Tx beamformers under the same criterion with the transformation
\begin{align}\label{dualize_tx}
	\bc_\mathrm{dual}=\bV_\mathrm{dual} (\bc_\mathrm{single} )=b\bR(\phi)^\mathrm{H}\left( \begin{bmatrix} \rho_{vv}^*\\ \rho_{vh}^* \end{bmatrix} \otimes\bc_\mathrm{single} \right).
\end{align}
This structure makes sense because the output of the function satisfies the necessary condition in Lemma \ref{lemma 1} to achieve the upper bound of data rate in \eqref{upper bound}. Therefore, instead of designing dual-polarization beamformers from scratch by solving the optimization problems in \eqref{dual opt} or \eqref{MIP codeword eq}, it is possible to directly generate dual-polarization beamformers from single-polarization beamformers.

%In the SE and MIP criteria, an option to consider the dual-polarization antenna structure is constraints on optimization problem of dual-polarization beamformer as \eqref{dual opt}, \eqref{MIP codeword eq}. Another option is applying the function  on the single-polarization beamformer $\bc_\mathrm{opt,single}$, which is optimal under the same criterion. The function form of $V_\mathrm{dual}(\cdot)$ is common in both SE and MIP codewords \eqref{SE final}, \eqref{MIP final}. This function structure makes sense because the output of the function satisfies the necessary condition in Lemma \ref{lemma 1} to achieve the upper bound of data rate \eqref{upper bound}. 

A common point of a single-polarization beamformer and a dual-polarization beamformer in \eqref{dualize_tx} can be found in their beam patterns. With proper normalization, the beam patterns of two beamformers are the same. This equality holds for any single-polarization beamformer $\bc_\mathrm{single}$ and its function output $\bV_\mathrm{dual} (\bc_\mathrm{single} )$. Hence, we can apply the function $\bV_{\mathrm{dual}}(\cdot)$ on the single-polarization beamformer that is already designed to have a proper beam pattern shape. This means that although the SE and MIP criteria are considered in this paper as particular examples, the function $\bV_{\mathrm{dual}}(\cdot)$ can be applied for any effective single-polarization beamformers with other criteria, e.g., the criteria defined in \cite{J.Zhang:2017,Z.Xiao:2017}.

%%%%%%%%%%%%%%%%%%%%%%%%%%%%%%%%%%%%%%%%%%%%%%%%%%%%%%%%%%%%%

\subsection{Receive beamforming designs}\label{Rx beamforming}

	Until now, we considered the Tx side channel assuming single-polarization UPA at the Rx BS. In this subsection, the original system with dual-polarization UPAs at both the Tx and Rx BSs is considered for the Rx beamformer design. We now recover the subscript $\mathrm{tx}$ to indicate the Tx side beamforming and channel. 
	
	For a given Tx beamformer, i.e., SE or MIP codeword or any other beamformer by $\bV_\mathrm{dual}(\cdot)$ in \eqref{dualize_tx}, the beamforming gain with the channel in \eqref{channel model} can be decomposed similarly as in \eqref{beamforming gain decomposition}
	\begin{align}\label{Rx beamforming gain decomposition}
		&
		\Bigg| \bc_{\mathrm{rx}}^\mathrm{H} h_0 \Bigg\{ \begin{bmatrix} \sqrt{\frac{1}{1+\chi}}e^{j\angle\zeta_0^{vv}} & \sqrt{\frac{\chi}{1+\chi}}e^{j\angle\zeta_0^{vh}} \\\sqrt{\frac{\chi}{1+\chi}}e^{j\angle\zeta_0^{hv}} & \sqrt{\frac{1}{1+\chi}}e^{j\angle\zeta_0^{hh}} \end{bmatrix} 
		\notag\\
		&\quad~
		\otimes \left( 
		\bd_{\mathrm{rx}}(\psi_{\mathrm{rx},0}^\mathrm{az},\psi_{\mathrm{rx},0}^\mathrm{el}) \bd_{\mathrm{tx}}(\psi_{\mathrm{tx},0}^\mathrm{az},\psi_{\mathrm{tx},0}^\mathrm{el})^\mathrm{H}
%		\ba_{\mathrm{rx}}(\theta_{\mathrm{rx},0}^\mathrm{az},\theta_{\mathrm{rx},0}^\mathrm{el}) \ba_{\mathrm{tx}}(\theta_{\mathrm{tx},0}^\mathrm{az},\theta_{\mathrm{tx},0}^\mathrm{el})^\mathrm{H}
		\right) \Bigg\} \bR(\phi) \bc_{\mathrm{tx}}
		\Bigg|^2
		\notag\\
		&\stackrel{(a)}{=}
		\Bigg| h_0\bc_{\mathrm{rx}}^\mathrm{H} \Bigg\{ \begin{bmatrix} \rho_{vv} & \rho_{vh} \\ \rho_{hv} & \rho_{hh} \end{bmatrix} 
		\otimes \big( 
		\bd_{\mathrm{rx}}(\psi_{\mathrm{rx},0}^\mathrm{az},\psi_{\mathrm{rx},0}^\mathrm{el})
%		\ba_{\mathrm{rx}}(\theta_{\mathrm{rx},0}^\mathrm{az},\theta_{\mathrm{rx},0}^\mathrm{el})
		\notag\\
		&\quad~
		\cdot 
		\bd_{\mathrm{tx}}(\psi_{\mathrm{tx},0}^\mathrm{az},\psi_{\mathrm{tx},0}^\mathrm{el})^\mathrm{H}
%		\ba_{\mathrm{tx}}(\theta_{\mathrm{tx},0}^\mathrm{az},\theta_{\mathrm{tx},0}^\mathrm{el})^\mathrm{H}
		 \big) \Bigg\}
		\bR(\phi) b\bR(\phi)^\mathrm{H}\left( \begin{bmatrix} \rho_{vv}^*\\ \rho_{vh}^* \end{bmatrix} \otimes\bc_\mathrm{tx,single} \right) \Bigg|^2
		\notag\\
		&\stackrel{(b)}{=}
		|bh_0|^2|
		\bd_{\mathrm{tx}}(\psi_{\mathrm{tx},0}^\mathrm{az},\psi_{\mathrm{tx},0}^\mathrm{el})^\mathrm{H}
%		\ba_{\mathrm{tx}}(\theta_{\mathrm{tx},0}^\mathrm{az},\theta_{\mathrm{tx},0}^\mathrm{el})^\mathrm{H}
		\bc_{\mathrm{tx,single}}|^2
		\notag\\
		&\quad~
		\cdot\left| \bc_{\mathrm{rx}}^\mathrm{H} \left( \begin{bmatrix} |\rho_{vv}|^2+|\rho_{vh}|^2 \\ \rho_{vv}^*\rho_{hv}+\rho_{vh}^*\rho_{hh} \end{bmatrix} 
		\otimes 
		\bd_{\mathrm{rx}}(\psi_{\mathrm{rx},0}^\mathrm{az},\psi_{\mathrm{rx},0}^\mathrm{el})
%		\ba_{\mathrm{rx}}(\theta_{\mathrm{rx},0}^\mathrm{az},\theta_{\mathrm{rx},0}^\mathrm{el})
		 \right) \right|^2
		\notag\\
		&=
		|bh_0|^2G_\mathrm{tx}
		\underbrace{ \underbrace{ \Big| \left( \begin{bmatrix} \xi_v & \xi_h \end{bmatrix} 
		\otimes  
		\bd_{\mathrm{rx}}(\psi_{\mathrm{rx},0}^\mathrm{az},\psi_{\mathrm{rx},0}^\mathrm{el})^\mathrm{H} 
%		\ba_{\mathrm{rx}}(\theta_{\mathrm{rx},0}^\mathrm{az},\theta_{\mathrm{rx},0}^\mathrm{el})^\mathrm{H} 
		\right) }_{\text{Rx side channel}} \bc_{\mathrm{rx}} \Big|^2
		}_{\text{Rx side beamforming gain}}
	\end{align}
	where $\bc_{\mathrm{tx}}=\bV_\mathrm{dual}(\bc_{\mathrm{tx,single}})$, $\rho_{hv}=\sqrt{\frac{\chi}{1+\chi}}e^{j\angle\zeta_{0}^{hv}}$, $\rho_{hh}=\sqrt{\frac{1}{1+\chi}}e^{j\angle\zeta_{0}^{hh}}$,  $G_\mathrm{tx}=|\bd_{\mathrm{tx}}(\psi_{\mathrm{tx},0}^\mathrm{az},\psi_{\mathrm{tx},0}^\mathrm{el})^\mathrm{H} \bc_{\mathrm{tx,single}}|^2$, $\xi_v = |\rho_{vv}|^2+|\rho_{vh}|^2$, and $\xi_h = \rho_{vv}\rho_{hv}^*+\rho_{vh}\rho_{hh}^*$. The equality $(a)$ is by the structure of the Tx beamformer $\bV_{\mathrm{dual}}(\bc_{\mathrm{tx,single}})$, and $(b)$ is by the property of the Kronecker product $(\bW\otimes \bX)(\bY\otimes \bZ)=(\bW\bY)\otimes(\bX\bZ)$ and the unitary matrix $\bR(\phi)\bR(\phi)^\mathrm{H}=\bI_{M_\mathrm{tx}}$.  For the Rx beamforming, we can simply consider the Rx side beamforming gain and Rx side channel with proper normalization.
	
	The Rx side channel in \eqref{Rx beamforming gain decomposition} has the same structure with the Tx side channel in \eqref{beamforming gain decomposition}. The orientation difference is already considered at the Tx side, and the Rx side beamforming gain in \eqref{Rx beamforming gain decomposition} is independent from $\bR(\phi)$. Therefore, the Tx beamforming methods explained in Section \ref{SE} and \ref{MIP} can be used for the Rx beamforming by replacing $\begin{bmatrix} \rho_{vv} & \rho_{vh} \end{bmatrix}$, $\bd_{\mathrm{tx}}(\psi_{\mathrm{tx},0}^\mathrm{az},\psi_{\mathrm{tx},0}^\mathrm{el})$, and $\bR(\phi)$ with $\begin{bmatrix} \xi_v & \xi_h \end{bmatrix}$, $\bd_{\mathrm{rx}}(\psi_{\mathrm{rx},0}^\mathrm{az},\psi_{\mathrm{rx},0}^\mathrm{el})$, and $\bI_{M_\mathrm{rx}}$. The following SE and MIP Rx codewords are
	\begin{align}
		\bc_{\mathrm{rx},SE,\mathrm{dual}}
		&
		=b_{\mathrm{rx}}\begin{bmatrix} \xi_v^* \\ \xi_h^* \end{bmatrix} \otimes \frac{\bD_{\mathrm{rx},h,1}\bq_{L_{\mathrm{rx},h}}\otimes \bD_{\mathrm{rx},v,1}\bq_{L_{\mathrm{rx},v}}}{\lVert \bD_{\mathrm{rx},h,1}\bq_{L_{\mathrm{rx},h}}\otimes \bD_{\mathrm{rx},v,1}\bq_{L_{\mathrm{rx},v}}\rVert_2}		
		\notag\\
		&
		=b_{\mathrm{rx}} \begin{bmatrix} \xi_v^* \\ \xi_h^* \end{bmatrix} \otimes \bc_{\mathrm{rx},SE,\mathrm{single}}		,
		\\
%	\end{align}
%	\begin{align}
		\bc_{\mathrm{rx},MIP,\mathrm{dual}}
		&=b_{\mathrm{rx}} \begin{bmatrix} \xi_v^*\\ \xi_h^* \end{bmatrix} \otimes \boldsymbol{\mathfrak{v}}_\mathrm{max}\left( \bD_{\mathrm{rx},1}\bD_{\mathrm{rx},1}^\mathrm{H}\right) 
		\notag\\
		&=b_{\mathrm{rx}} \begin{bmatrix} \xi_v^*\\ \xi_h^* \end{bmatrix} \otimes \bc_{\mathrm{rx},MIP,\mathrm{single}},
	\end{align}
	where $b_\mathrm{rx} = \left(|\xi_v|^2+|\xi_h|^2\right)^{-\frac{1}{2}}$, $\bc_{\mathrm{rx},SE,\mathrm{single}} = \frac{\bD_{\mathrm{rx},h,1}\bq_{L_{\mathrm{rx},h}}\otimes \bD_{\mathrm{rx},v,1}\bq_{L_{\mathrm{rx},v}}}{\lVert \bD_{\mathrm{rx},h,1}\bq_{L_{\mathrm{rx},h}}\otimes \bD_{\mathrm{rx},v,1}\bq_{L_{\mathrm{rx},v}}\rVert_2}$, and $\bc_{\mathrm{rx},MIP,\mathrm{single}}=\boldsymbol{\mathfrak{v}}_\mathrm{max}\left( \bD_{\mathrm{rx},1}\bD_{\mathrm{rx},1}^\mathrm{H}\right) $. By the structure of the Rx beamformers, it is also possible to generate the dual-polarization Rx beamformer from the single-polarization Rx beamformer with the same optimality as  
	\begin{align}\label{dualize_rx}
		\bc_{\mathrm{rx,dual}}=\bV_{\mathrm{rx,dual}}(\bc_{\mathrm{rx,single}} )=b_\mathrm{rx} \begin{bmatrix} \xi_v^*\\ \xi_h^*\end{bmatrix} \otimes \bc_{\mathrm{rx,single}}. 
	\end{align}
	The final hybrid Rx beamformers, then, can be obtained by applying OMP-based algorithms as Algorithm \ref{SE algorithm} for the hybrid Rx beamformers.
%	The details of codewords derivation are similar to the Tx beamformer case with difference in dimension of several matrices. 
%	The dual-polarization beamforming design from single-polarization beamformer, hence, is possible for both Tx and Rx beamforming with dimension adjustment. 

%%%%%%%%%%%%%%%%%%%%%%%%%%%%%%%%%%%%%%%%%%%%%%%%%%%%%%%%%%%%%%%%%%%%%%
%%%%%%%%%%%%%%%%%%%%%%%%%%%%%%%%%%%%%%%%%%%%%%%%%%%%%%%%%%%%%%%%%%%%%%

	\subsection{Beam alignment and channel information acquisition}\label{sec3-d}
	
	The proposed Tx and Rx beamforming methods with the SE or MIP criteria need partial knowledge of the dual-polarization channel. 
%	One is the XPD of channel. Because the XPD value does not change much over time, we assume the value is known both at the Tx and Rx BSs. Another is the orientation difference, which is assumed to be fix in Section \ref{sec2}, and we assume it is also known at the Tx BS.\footnote{The Tx BS may not have accurate knowledge of these parameters in practice. 
	The XPD and orientation difference, which are assumed to be fixed in Section \ref{sec2}, are assumed to be known at the Tx BS.\footnote{The Tx BS may not have accurate knowledge of these parameters in practice. 
	We take the randomness of these parameters into account in Section \ref{sec4} for numerical studies.}
	The rest of channel variables are the complex gains $e^{j\angle\zeta_{0}^{vv}}$ and $e^{j\angle\zeta_{0}^{vh}}$ in $\rho_{vv}=\sqrt{\frac{1}{1+\chi}}e^{j\angle\zeta_{0}^{vv}}$ and $\rho_{vh}=\sqrt{\frac{1}{1+\chi}}e^{j\angle\zeta_{0}^{vh}}$ for the Tx beamforming in \eqref{dualize_tx} and $\xi_h$ and $\xi_v$ for the Rx beamforming in \eqref{dualize_rx}. Although the four values are determined by small-scale channel fading, it is well known that the channel coherence time in mmWave communications can become quite large after proper beam alignment \cite{V.Va:2017}. Therefore, the knowledge of the four values can be updated infrequently for the backhaul communications with fixed Tx and Rx BS.

	To obtain these values, the Tx BS first needs to perform the beam alignment \cite{S.Hur:2013,S.Noh:2017,J.Song:2015}. 
	Without the information of the gains, the codeword for the beam alignment is defined as
	\begin{align}\label{codeword for alignment}
		\bc^{(p,q)}_\mathrm{tx,align} &= b\bR(\phi)^\mathrm{H}\left( \begin{bmatrix} \sqrt{\frac{1}{1+\chi}}\alpha\\ \sqrt{\frac{\chi}{1+\chi}}\beta \end{bmatrix} \otimes \bc_\mathrm{tx,single}^{(p,q)} \right),
		\notag\\
		\bc^{(p,q)}_\mathrm{rx,align} &= b_\mathrm{rx}\begin{bmatrix} \omega \\  \upsilon \end{bmatrix} \otimes \bc_\mathrm{rx,single}^{(p,q)} ,
	\end{align}
	where $\alpha, \beta, \omega, \upsilon \in\mathbb{C}$ are set to be constant numbers. $\bc_\mathrm{ax,single}^{(p,q)}$ is $\bc_{\mathrm{ax},SE,\mathrm{single}}^{(p,q)}$ for the SE codeword or $\bc_{\mathrm{ax},MIP,\mathrm{single}}^{(p,q)}$ for the MIP codeword for alignment with $\mathrm{ax}\in\{\mathrm{tx},\mathrm{rx}\}$. Using this temporary codewords, the Tx and the Rx BSs conduct the beam alignment as in \eqref{align}. Note that the beam alignment does not have to be done frequently since the Tx and Rx BSs are fixed in backhaul communications.
	
	After the beam alignment, it is possible to obtain the four values $\alpha$, $\beta$, $\omega$, $\upsilon$ for beamforming codewords optimization. What we actually need is the ratios $\frac{e^{j\angle\zeta_{0}^{vv}}}{e^{j\angle\zeta_{0}^{vh}}}$ and $\frac{\xi_v}{\xi_h}$, as codewords are unit-norm constrained. A possible way to figure out the two ratios is using pilots. We fix the transmit symbol $s=1$ and use pilot sequences on $(\alpha,\beta)$ and $(\omega,\upsilon)$ in \eqref{codeword for alignment} instead. The length $4J$ pilot sequences $(\alpha_j,\beta_j)$ for $(\alpha,\beta)$ can be any sequences satisfying
	\begin{align}\label{pilot requirement}
		\sum_{j=1}^J\alpha_j&=\sum_{j=2J+1}^{3J}\alpha_j=\varsigma,\ \ \sum_{j=J+1}^{2J}\alpha_j=\sum_{j=3J+1}^{4J}\alpha_j=0,
		\notag\\
		 \sum_{j=1}^J\beta_j&=\sum_{j=2J+1}^{3J}\beta_j=0,~~ \sum_{j=J+1}^{2J}\beta_j=\sum_{j=3J+1}^{4J}\beta_j=\varsigma 
		,
	\end{align} 
	%where \eqref{sequence power constraint} is for unit-norm constraint of beamformers. 
%	where $\frac{|\alpha_j|^2+\chi|\beta_j|^2}{1+\chi}=1,~ \forall j$. 
	where $\varsigma\in\mathbb{C}$ is a constant.
	The length $4J$ pilot sequences $(\alpha_j,\beta_j)$ are used simultaneously with pilot sequences $(\omega_j,\upsilon_j)$ for $(\omega,\upsilon)$, where $(\omega_j,\upsilon_j)=(\kappa_1,0)$ for $j\in\{1,\cdots,2J\}$ and $(\omega_j,\upsilon_j)=(0,\kappa_2)$ for $j\in\{2J+1,\cdots,4J\}$ with constant $\kappa_1, \kappa_2 \in\mathbb{C}$.

	With these sequences, it is possible to figure out the ratio $\frac{e^{j\angle\zeta_{0}^{vv}}}{e^{j\angle\zeta_{0}^{vh}}}$ by using the properties of the sequences
	\begin{align}
		y_j &= \sqrt{P} \bc_{\mathrm{rx},j}^\mathrm{H} \bH \bc_{\mathrm{tx},j} s + \bc_{\mathrm{rx},j}^\mathrm{H}\bn_j,
		\\
		Y_1&=\sum_{j=1}^J y_j
		\notag\\ 
		&
		\approx b_\mathrm{rx}b\sqrt{P}\frac{e^{j\angle\zeta_{0}^{vv}}}{1+\chi} \bc_{\mathrm{rx,single}}^\mathrm{H}
		\bd_{\mathrm{rx}}(\psi_{\mathrm{rx}}^\mathrm{az},\psi_{\mathrm{rx}}^\mathrm{el})
%		\ba_{\mathrm{rx}}(\theta_{\mathrm{rx}}^\mathrm{az},\theta_{\mathrm{rx}}^\mathrm{el})		
		\notag\\
		&\quad~
		\cdot
		\bd_{\mathrm{tx}}(\psi_{\mathrm{tx}}^\mathrm{az},\psi_{\mathrm{tx}}^\mathrm{el})^\mathrm{H}
%		\ba_{\mathrm{tx}}(\theta_{\mathrm{tx}}^\mathrm{az},\theta_{\mathrm{tx}}^\mathrm{el})^\mathrm{H}
		\bc_\mathrm{tx,single}\kappa_1\sum_{j=1}^J\alpha_j\notag\\
		&= b_\mathrm{rx}b\sqrt{P}\frac{e^{j\angle\zeta_{0}^{vv}}}{1+\chi}G_\mathrm{rx,tx}\kappa_1\varsigma ,\label{pilot seq1}
		\\
%	\end{align}
%	\begin{align}
		Y_2&=\sum_{j=J+1}^{2J} y_j
		\notag\\
		&
		\approx b_\mathrm{rx}b\sqrt{P}\frac{\chi e^{j\angle\zeta_{0}^{vh}}}{1+\chi} \bc_{\mathrm{rx,single}}^\mathrm{H}
		\bd_{\mathrm{rx}}(\psi_{\mathrm{rx}}^\mathrm{az},\psi_{\mathrm{rx}}^\mathrm{el})
%		\ba_{\mathrm{rx}}(\theta_{\mathrm{rx}}^\mathrm{az},\theta_{\mathrm{rx}}^\mathrm{el})
		\notag\\
		&\quad~
		\cdot 
		\bd_{\mathrm{tx}}(\psi_{\mathrm{tx}}^\mathrm{az},\psi_{\mathrm{tx}}^\mathrm{el})^\mathrm{H}
%		\ba_{\mathrm{tx}}(\theta_{\mathrm{tx}}^\mathrm{az},\theta_{\mathrm{tx}}^\mathrm{el})^\mathrm{H}
		\bc_\mathrm{tx,single} \kappa_1\sum_{j=J+1}^{2J}\beta_j\notag\\
		&= b_\mathrm{rx}b\sqrt{P}\frac{\chi e^{j\angle\zeta_{0}^{vh}}}{1+\chi}G_\mathrm{rx,tx}\kappa_1\varsigma, \label{pilot seq2}
	\end{align}
	where $G_\mathrm{rx,tx}= \bc_{\mathrm{rx,single}}^\mathrm{H} \bd_{\mathrm{rx}}(\psi_{\mathrm{rx}}^\mathrm{az},\psi_{\mathrm{rx}}^\mathrm{el})  \bd_{\mathrm{tx}}(\psi_{\mathrm{tx}}^\mathrm{az},\psi_{\mathrm{tx}}^\mathrm{el})^\mathrm{H} \bc_\mathrm{tx,single}$. \red{The approximations \eqref{pilot seq1} and \eqref{pilot seq2} come from the high beamforming gain after the beam alignment, which gives large effective SNR.} Using \eqref{pilot seq1} and \eqref{pilot seq2}, we have 
	\begin{align}
%		\frac{\chi\sum_{j=1}^{J} y_j}{\sum_{j=J+1}^{2J} y_j}&\simeq\frac{e^{j\zeta^{vv}}}{e^{j\zeta^{vh}}}.
		\frac{\chi Y_1}{Y_2}&\approx\frac{e^{j\zeta_{0}^{vv}}}{e^{j\zeta_{0}^{vh}}}.
	\end{align}
	The final Tx codeword can be attained as
	\begin{align}
		\bc_\mathrm{tx,dual}^{(p,q)}=b\bR(\phi)^\mathrm{H}\left( \begin{bmatrix} \sqrt{\frac{1}{1+\chi}}\cdot \left(\frac{e^{j\zeta_{0}^{vv}}}{e^{j\zeta_{0}^{vh}}}\right)^*\\ \sqrt{\frac{\chi}{1+\chi}}\cdot 1  \end{bmatrix} \otimes \bc_\mathrm{tx,single}^{(p,q)} \right),
	\end{align}
	where $b\in\mathbb{C}$ is to satisfy $\lVert\bc_{\mathrm{tx,dual}}\rVert_2^2=1$.
	
	To optimize the Rx codeword, we need to find the ratio $\frac{\xi_v}{\xi_h}$. With the rest of pilot sequences, we have
	\begin{align}
		Y_3&=\sum_{j=2J+1}^{3J} y_j
		\notag\\
		&
		\approx b_\mathrm{rx}b\sqrt{P}\frac{\sqrt{\chi}e^{j\angle\zeta_{0}^{hv}}}{1+\chi} \bc_{\mathrm{rx,single}}^\mathrm{H} 
		\bd_{\mathrm{rx}}(\psi_{\mathrm{rx}}^\mathrm{az},\psi_{\mathrm{rx}}^\mathrm{el})
%		\ba_{\mathrm{rx}}(\theta_{\mathrm{rx}}^\mathrm{az},\theta_{\mathrm{rx}}^\mathrm{el})
		\notag\\
		&\quad~
		\cdot 
		\bd_{\mathrm{tx}}(\psi_{\mathrm{tx}}^\mathrm{az},\psi_{\mathrm{tx}}^\mathrm{el})^\mathrm{H}
%		\ba_{\mathrm{tx}}(\theta_{\mathrm{tx}}^\mathrm{az},\theta_{\mathrm{tx}}^\mathrm{el})^\mathrm{H}
		\bc_\mathrm{tx,single}\kappa_2\sum_{j=2J+1}^{3J}\alpha_j
		\notag\\
		&
		= b_\mathrm{rx}b\sqrt{P}\frac{\sqrt{\chi}e^{j\angle\zeta_{0}^{hv}}}{1+\chi}G_\mathrm{rx,tx}\kappa_2\varsigma ,\label{pilot seq3}
		\\
%	\end{align}
%	\begin{align} 
		Y_4&=\sum_{j=3J+1}^{4J} y_j
		\notag\\
		&
		\approx b_\mathrm{rx}b\sqrt{P}\frac{\sqrt{\chi}e^{j\angle\zeta_{0}^{hh}}}{1+\chi} \bc_{\mathrm{rx,single}}^\mathrm{H} 
		\bd_{\mathrm{rx}}(\psi_{\mathrm{rx}}^\mathrm{az},\psi_{\mathrm{rx}}^\mathrm{el})
%		\ba_{\mathrm{rx}}(\theta_{\mathrm{rx}}^\mathrm{az},\theta_{\mathrm{rx}}^\mathrm{el})
		\notag\\
		&\quad~
		\cdot 
		\bd_{\mathrm{tx}}(\psi_{\mathrm{tx}}^\mathrm{az},\psi_{\mathrm{tx}}^\mathrm{el})^\mathrm{H}
%		\ba_{\mathrm{tx}}(\theta_{\mathrm{tx}}^\mathrm{az},\theta_{\mathrm{tx}}^\mathrm{el})^\mathrm{H}
		\bc_\mathrm{tx,single} \kappa_2\sum_{j=3J+1}^{4J}\beta_j
		\notag\\
		&
		= b_\mathrm{rx}b\sqrt{P}\frac{\sqrt{\chi}e^{j\angle\zeta_{0}^{hh}}}{1+\chi}G_\mathrm{rx,tx}\kappa_2\varsigma, \label{pilot seq4}
	\end{align}
	where the approximations \eqref{pilot seq3} and \eqref{pilot seq4} are by the high beamforming gain after the beam alignment \red{that gives large effective SNR}. The ratio $\frac{\xi_v}{\xi_h}$ is attained by using the whole pilot sequences
	\begin{align}
		\frac{\xi_v}{\xi_h}
		&
		\approx\frac{|\rho_{vv}|^2+|\rho_{vh}|^2}{\rho_{vv}\rho_{hv}^*+\rho_{vh}\rho_{hh}^*}
		\notag\\
		&
%		=\Bigg\{\left| \frac{1}{\kappa_1}\sum_{j=1}^{J} y_j\right|^2+\left| \frac{1}{\kappa_1}\sum_{j=J+1}^{2J} y_j\right|^2\Bigg\}
%		\notag\\
%		&\quad~
%		\div\Bigg\{\left(\frac{1}{\kappa_1}\sum_{j=1}^{J} y_j\right)\left(\frac{1}{\kappa_2}\sum_{j=2J+1}^{3J} y_j\right)^*
%		\notag\\
%		&\quad~
%		+\frac{1}{\sqrt{\chi}}\left(\frac{1}{\kappa_1}\sum_{j=J+1}^{2J} y_j\right)\left(\frac{1}{\kappa_2}\sum_{j=3J+1}^{4J} y_j\right)^*\Bigg\} .		
		=\Bigg\{\left| \frac{1}{\kappa_1} Y_1 \right|^2+\frac{1}{\chi}\left| \frac{1}{\kappa_1} Y_2\right|^2\Bigg\}
		\notag\\
		&\quad~
		\div\Bigg\{\left(\frac{1}{\kappa_1} Y_1\right)\left(\frac{1}{\kappa_2} Y_3 \right)^*
%		\notag\\
%		&\quad~
		+\frac{1}{\chi} \left(\frac{1}{\kappa_1} Y_2\right)\left(\frac{1}{\kappa_2} Y_4\right)^*\Bigg\} .
	\end{align}
	The final Rx codeword is 
	\begin{align}
		\bc^{(p,q)}_\mathrm{rx,dual} &= b_\mathrm{rx}\begin{bmatrix} \left(\frac{\xi_v}{\xi_h}\right)^* \\  1 \end{bmatrix} \otimes \bc_\mathrm{rx,single}^{(p,q)}, 
	\end{align}
	where $b_\mathrm{rx}\in\mathbb{C}$ is to satisfy $\lVert\bc_{\mathrm{rx,dual}}\rVert_2^2=1$. It is important to note that the effective use of pilot sequences, i.e., embedding the pilots in the beamformer, is possible by the structure of the dual-polarization beamformers \eqref{dualize_tx}, \eqref{dualize_rx}.
	
	In Section \ref{sec4}, the numerical results in Fig. \ref{data rate over J} show that the length of pilot sequences can be small, e.g., the data rates with $J=1$ and $64$ are quite similar at very low SNR and the same at moderate SNR values. Therefore, this additional pilot overhead can be marginal.

	\begin{figure*}[!t]
		\centering
		\subfloat[SE codeword]{
			\includegraphics[width=.32\textwidth]{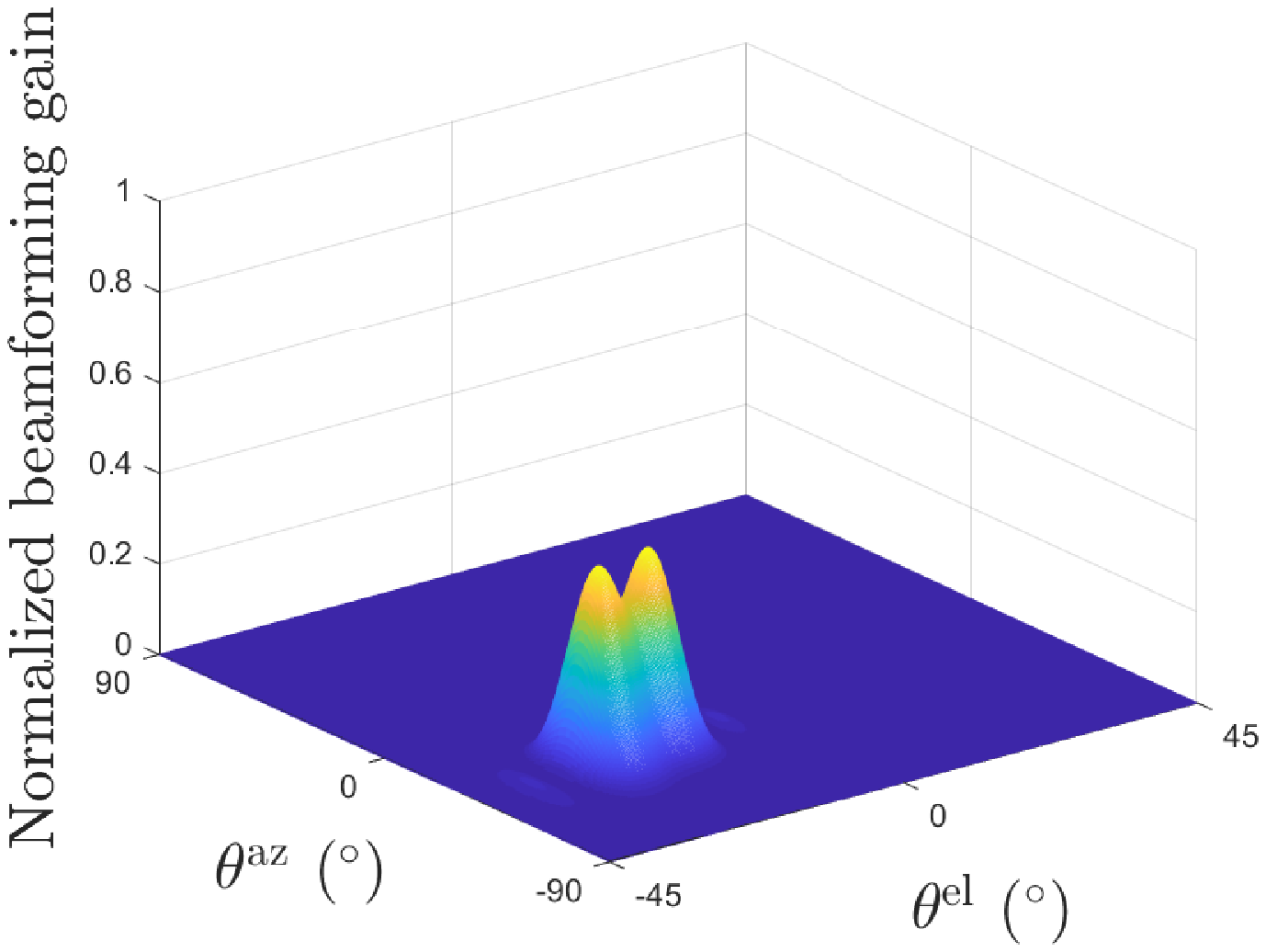}
			\label{SE codeword}
		}
		\subfloat[MIP codeword]{
			\includegraphics[width=.32\textwidth]{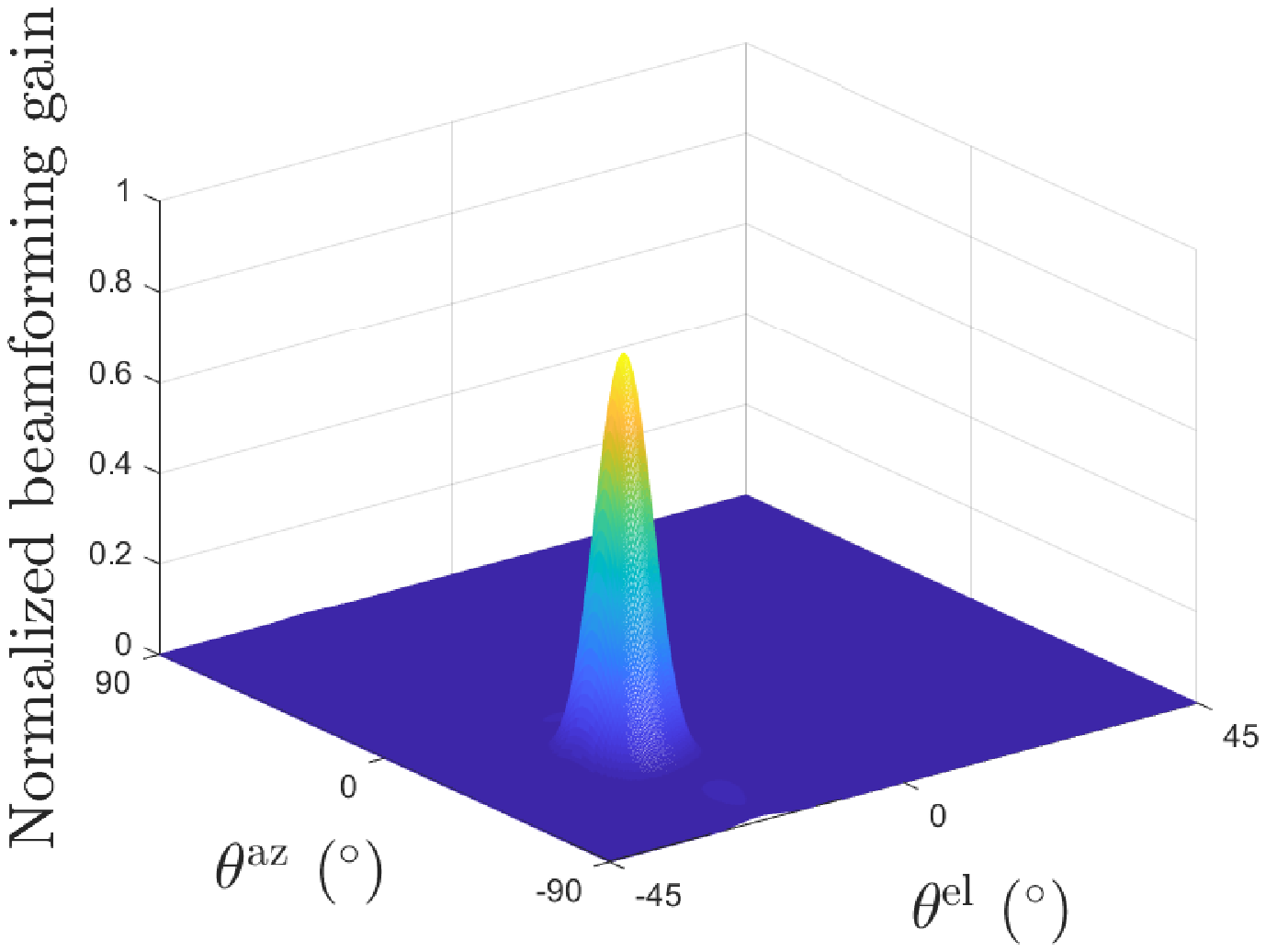}
			\label{MIP codeword}
		}
		\subfloat[Codeword in \cite{B.Clerckx:2008}]{
			\includegraphics[width=.32\textwidth]{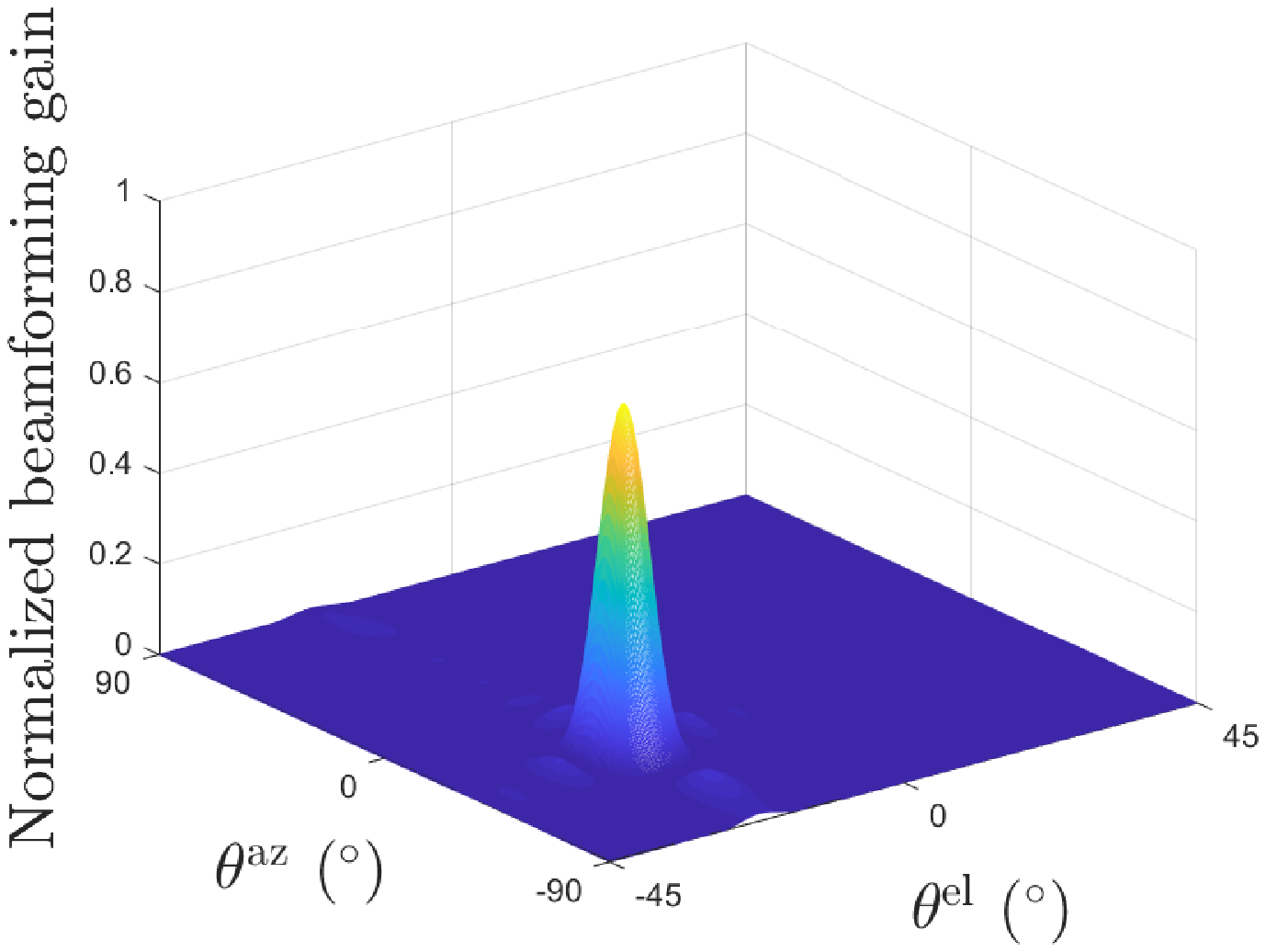}
			\label{RBD codeword}
		}
		\caption{Normalized beamforming gains at the region $B^{(2,2)}$ with $(M_h,M_v)=(8,16),\ (Q_h,Q_v)=(6,6)$.}
		\label{codeword}
	\end{figure*}
	% 아무튼 간에 이 페이지에 넣고 싶어
	\begin{figure}[t]
		\centering
		\includegraphics[width=1\linewidth]{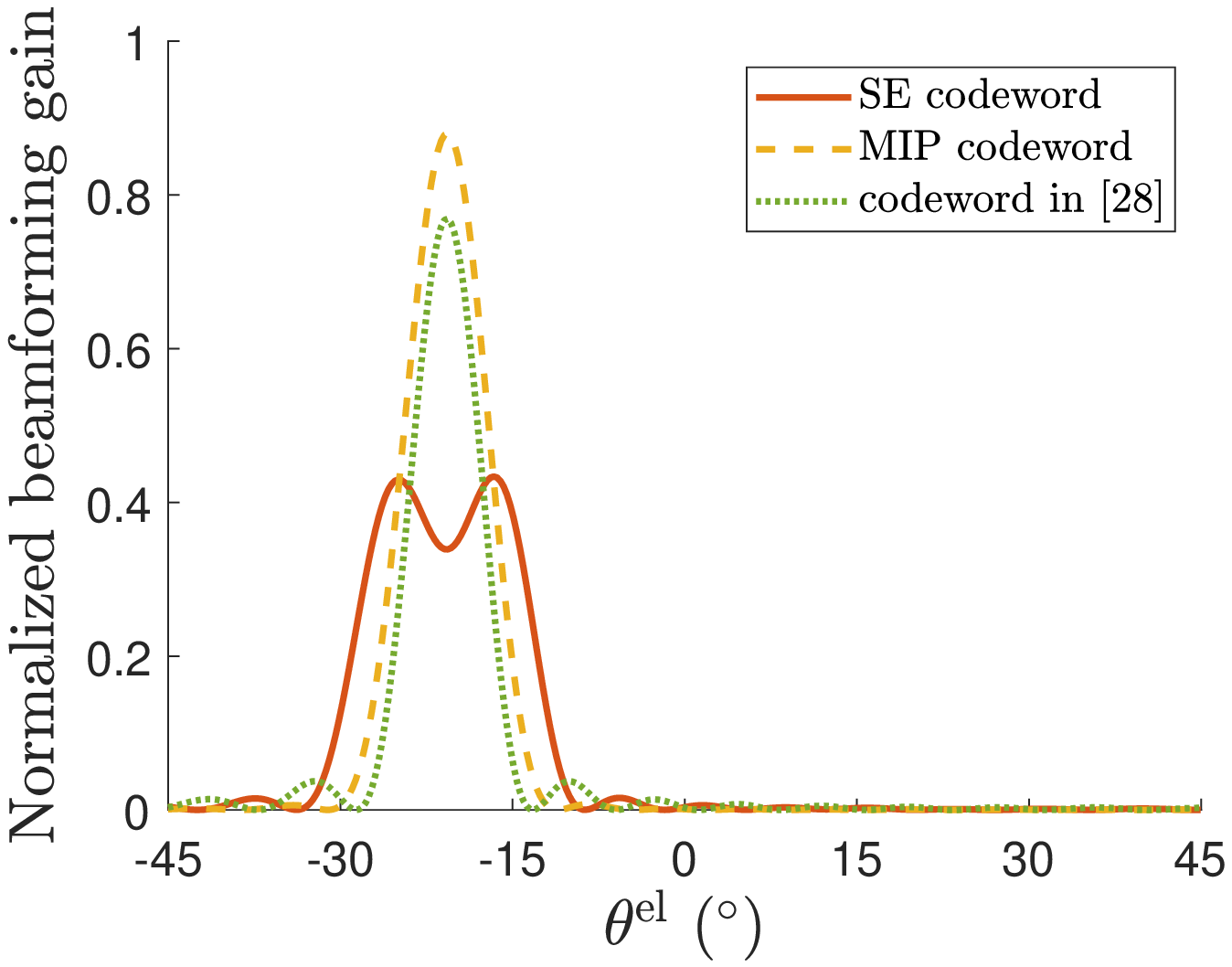}
		\caption{Normalized beamforming gains of different beamformers at the region $B^{(2,2)}$ with $(M_h,M_v)=(8,16),\ (Q_h,Q_v)=(6,6)$.}
		\label{superposition}
	\end{figure}

	\section{Simulation results}\label{sec4}
	
	\begin{figure*}[t]
		\centering
		\subfloat[SE codebook]{
			\includegraphics[width=.32\textwidth]{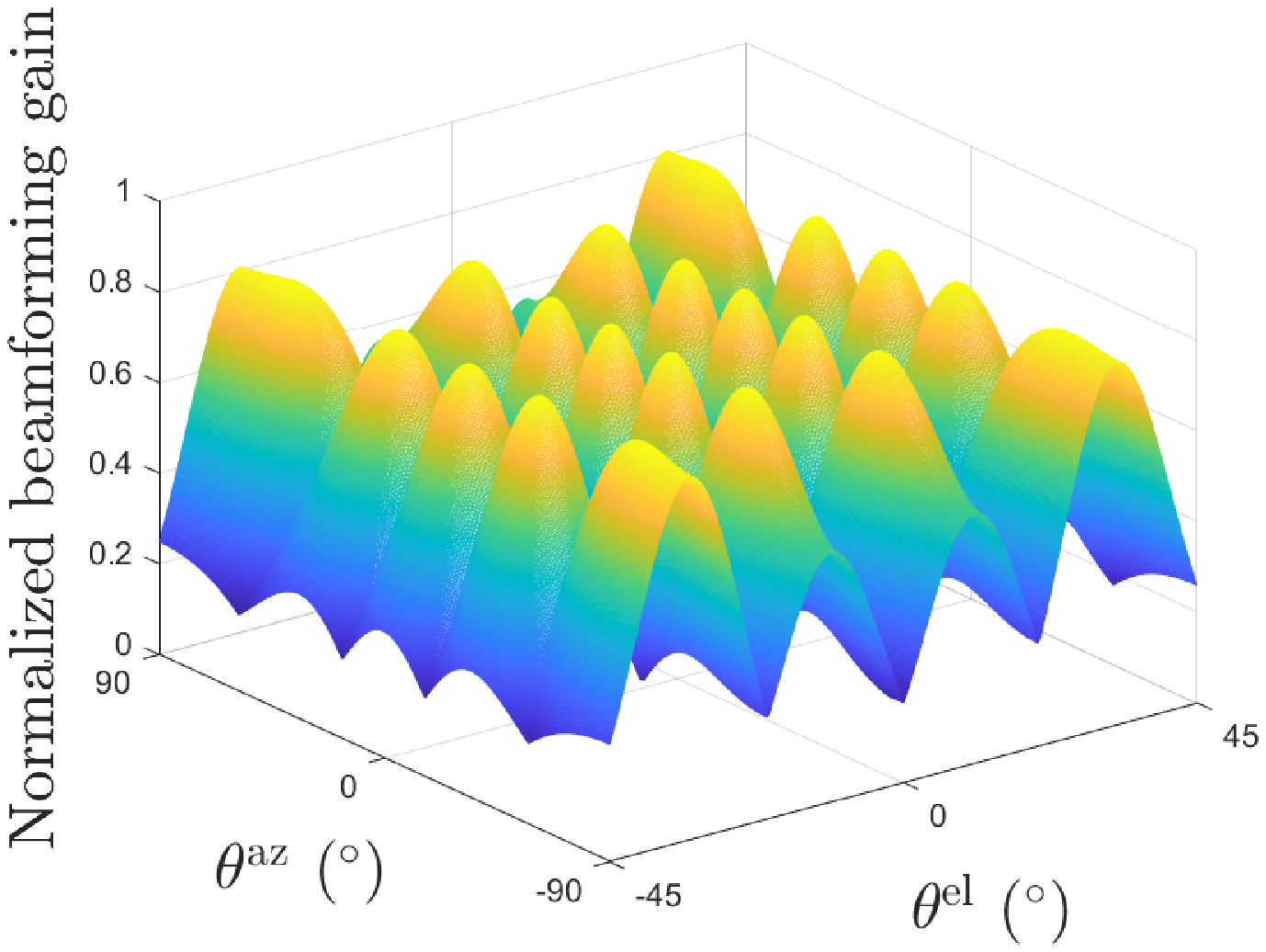}
			\label{SE codebook}
		}
		\subfloat[MIP codebook]{
			\includegraphics[width=.32\textwidth]{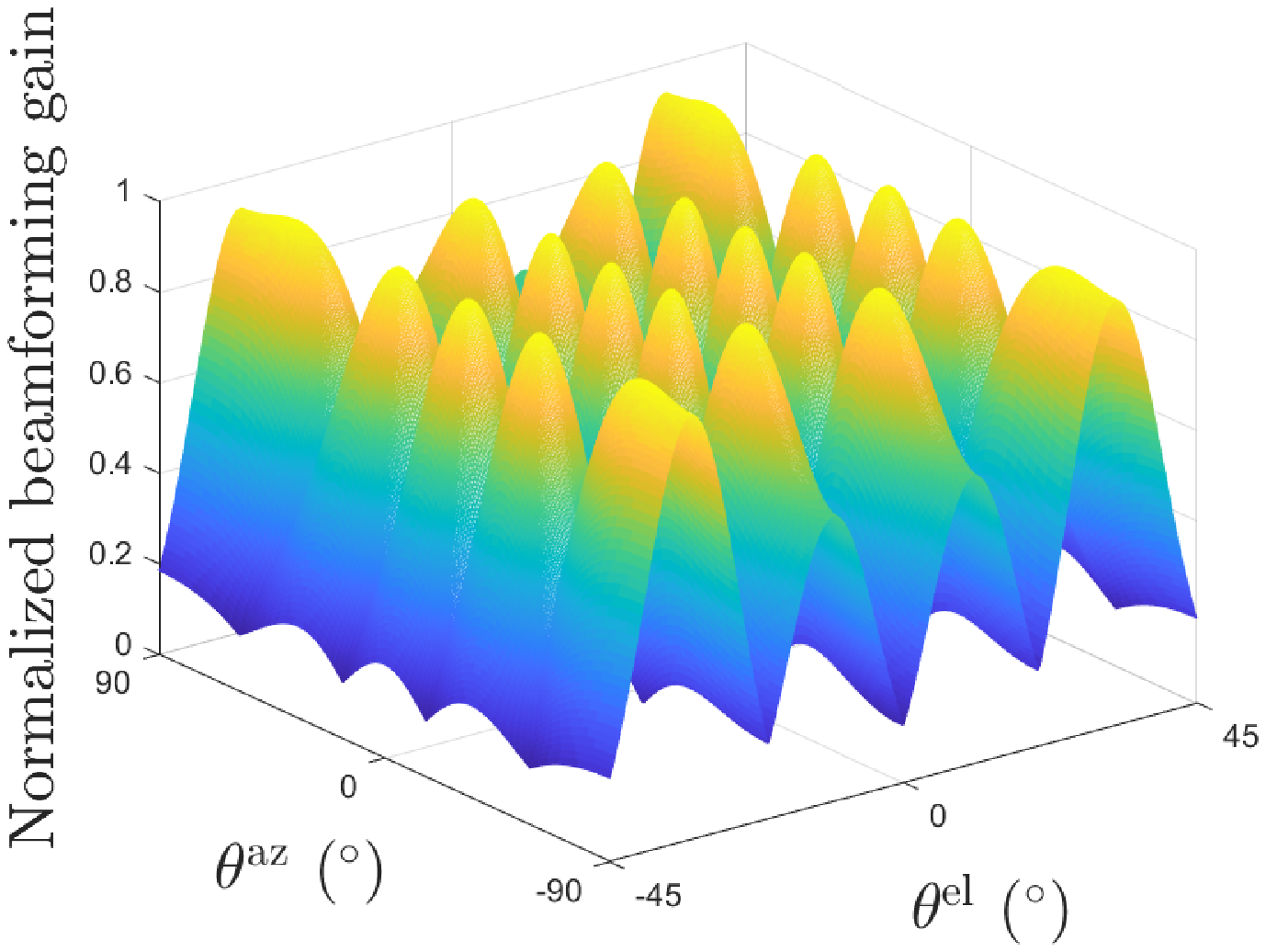}
			\label{MIP codebook}
		}
		\subfloat[codebook in \cite{B.Clerckx:2008}]{
			\includegraphics[width=.32\textwidth]{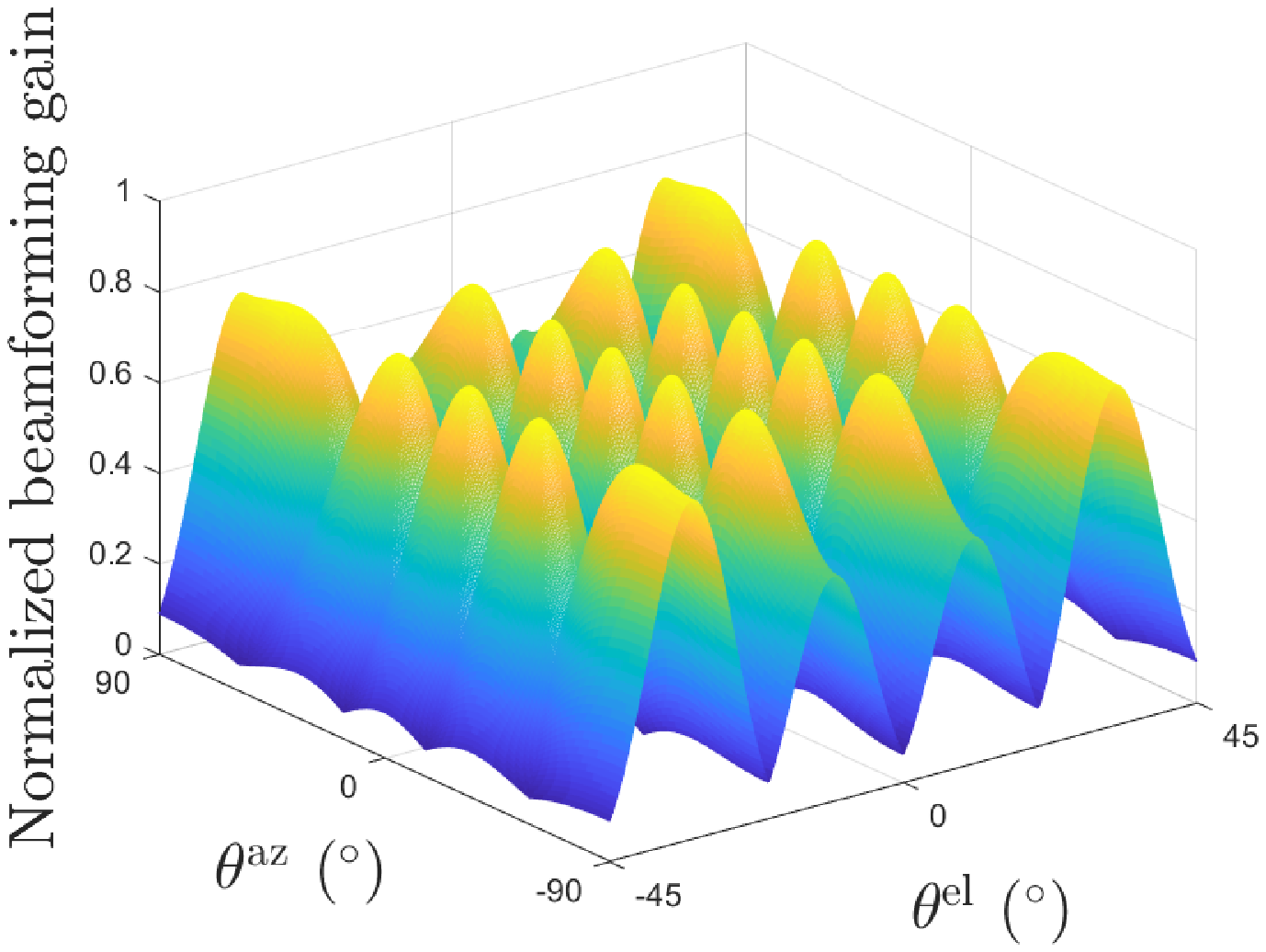}
			\label{RBD codebook}
		}
		\caption{Normalized beamforming gains with $(M_h,M_v)=(4,8),\ (Q_h,Q_v)=(4,4)$.}
		\label{codebook}
	\end{figure*}

	This section presents the numerical results of beam patterns and data rates of proposed beamformers and simple, yet effective dual-polarization beamformers proposed in \cite{B.Clerckx:2008}.
%	In this subsection, Tx beam patterns of SE and MIP codebooks are compared with that of the dual-polarization codebook in \cite{B.Clerckx:2008}. 
%	The codebook in \cite{B.Clerckx:2008} is the extension of the DFT codebook into a dual-polarization version, where a modified discrete Fourier transform (DFT) matrix is used for each block diagonal matrices.
	The codebook in \cite{B.Clerckx:2008} is constructed by the rotated block diagonal matrix, where each block diagonal matrix corresponds to a modified discrete Fourier transform (DFT) matrix. This means that the codebook in \cite{B.Clerckx:2008} is the extension of the DFT codebook into a dual-polarization version.
%	This section presents the numerical results of beam patterns and data rates. We compare the proposed beamformers with the beamformer in \cite{B.Clerckx:2008}, which is simple yet effective dual-polarization beamformers and the single-polarization beamformer in \cite{J.Zhang:2017} after applying the functions $V_\mathrm{dual}(\cdot)$ and $V_\mathrm{rx,dual}(\cdot)$. 
	We consider $4$-RF chains $(N=4)$ and UPA with $M=2M_h M_v$ dual-polarization antennas for both the Tx and Rx BSs, where the phases of analog beamformer are quantized with $4$-bit resolution per phase. For the realization of the dual-polarization channel, we set $\chi=0.3$ for the XPD value and $\phi=0$ for the orientation difference for the beam pattern comparison in Section \ref{sec4-a}, while we take the randomness of these parameters into consideration for the data rate comparison in Section \ref{sec4-b}. The spatial frequency range is quantized into $Q_h\times Q_v$ regions, and each region is divided into $L_h\times L_v=7\times7$ sections for both the Tx and Rx BSs. In the SE codebook design, we used the candidate set of $(\bq_{L_h},\bq_{L_v} )$ as 
	\begin{align}
	\mathcal{G}_h\times \mathcal{G}_v=\left\{ (\by,\bz):y_i=e^{-\pi+\frac{2\pi}{B}\ell},\ z_j=e^{-\pi+\frac{2\pi}{B} m}\right\},
	\end{align} 
	where $\ell\in\{1,\cdots,B\}$, $m\in\{1,\cdots,B\}$, $i\in\{1,\cdots,L_h \}$, and $j\in\{1,\cdots,L_v \}$ with $B=3$.

	\subsection{Beam pattern comparison}\label{sec4-a}
	
	We only consider the Tx beam patterns since the Rx beam patterns have the same results. As in \eqref{reference gain}, a beam pattern shape of a Tx codeword $\bc$ is described by using the reference gain
	\begin{align} 
	&g_\mathrm{ref} 
	(\theta_{\mathrm{tx}}^\mathrm{az},\theta_{\mathrm{tx}}^\mathrm{el}, \bc) 			
	\notag\\
	&\quad
	=\Bigg| b_\mathrm{tx} \Bigg\{ \begin{bmatrix} \rho_{vv} & \rho_{vh} \end{bmatrix} 	
%	\notag\\ 
%	&\quad~~
	\otimes   \ba_{\mathrm{tx}}(\theta_{\mathrm{tx}}^\mathrm{az},\theta_{\mathrm{tx}}^\mathrm{el} )^\mathrm{H} \Bigg\}\bR(\phi) \bc\Bigg|^2,
	\end{align}
	while the reference gain now takes into account the original paired array response vector $\ba_{\mathrm{tx}}(\theta_{\mathrm{tx}}^\mathrm{az},\theta_{\mathrm{tx}}^\mathrm{el})$ with paired spatial frequencies, i.e., $\psi_{\mathrm{tx}}^\mathrm{az}=\pi  \sin \theta_{\mathrm{tx}}^\mathrm{az} \cos\theta_{\mathrm{tx}}^\mathrm{el}$ and $\psi_{\mathrm{tx}}^\mathrm{el}=\pi\sin\theta_{\mathrm{tx}}^\mathrm{el}$.
%	where $(b, \eta_{v}, \eta_{h}, \ba_h(\theta^\mathrm{az},\theta^\mathrm{el}), \ba_v(\theta^\mathrm{el}), \bR)$ are $(b_\mathrm{tx}, \rho_{vv}, \rho_{vh}, $$\ba_{\mathrm{tx},h}(\theta_{\mathrm{tx}}^{\mathrm{az}}, \theta_{\mathrm{tx}}^\mathrm{el}), \ba_{\mathrm{tx},v}(\theta_{\mathrm{tx}}^\mathrm{el}),\bR(\phi))$ for the Tx side reference gain or $(b_\mathrm{rx}, \xi_{v}, \xi_{h},$$ \ba_{\mathrm{rx},h}(\theta_{\mathrm{rx}}^{\mathrm{az}}, \theta_{\mathrm{rx}}^\mathrm{el}), \ba_{\mathrm{rx},v}(\theta_{\mathrm{rx}}^\mathrm{el}),\bI_{M_\mathrm{rx}})$ for the Rx side reference gain. 
	%only covers the region of interest. 
	%
	%\begin{figure}[t]
	%	\centering
	%	\subfloat[SE codeword]{
	%		\includegraphics[width=.45\linewidth]{SE_single_B22.eps}
	%		\label{SE codeword}
	%	}
	%	\subfloat[MIP codeword]{
	%		\includegraphics[width=.45\linewidth]{MIP_single_B22.eps}
	%		\label{MIP codeword}
	%	}\\
	%	\subfloat[RVQ codeword]{
	%	\includegraphics[width=.45\linewidth]{RVQ_single_B22.eps}
	%	\label{RVQ codeword}
	%	}
	%	\subfloat[codeword in \cite{B.Clerckx:2008}]{
	%	\includegraphics[width=.45\linewidth]{RBD_single_B22.eps}
	%	\label{RBD codeword}
	%	}
	%	\caption{Normalized beamforming gains at the region $B^{(2,2)}$ with $(M_h,M_v)=(8,16),\ (Q_h,Q_v)=(6,6)$.}
	%	\label{codeword}
	%\end{figure}

	In Fig. \ref{codeword}, the Tx beam patterns of codewords, which are for the region $B^{(2,2)}$, are depicted. The beam pattern of SE codeword is more uniform than any other beam patterns, since the SE criterion forces the beam pattern to distribute its power uniformly inside the covering region. The beam pattern of the MIP codeword has the highest peak gain due to its designing criterion, i.e., maximizing the inner product. 
	%The RVQ codeword has its maximum gain inside the covering region, but the energy is scattered over the entire region. 
	The beam pattern of the codeword in \cite{B.Clerckx:2008} is similar to the beam pattern of the MIP codeword but has more narrow shape and lower peak gain compared to that of the MIP codeword. These features can be observed remarkably in Fig. \ref{superposition}. The elevation angle range of the region $B^{(2,2)}$ is $[-\frac{2\pi}{12},-\frac{\pi}{12}]$. The MIP codeword has the maximum peak gain with wider covering range than the codeword in \cite{B.Clerckx:2008}, but its gain is lower near the edge of region $\theta^\mathrm{el}\simeq-\frac{2\pi}{12}$. The SE codeword, by comparison, retains higher gain near the edge of the region resulting the most uniform beam pattern. 
	%The small main lobe of the codeword in \cite{B.Clerckx:2008} means that the remainder of the energy is directing outside of the covering region, and the energy leakage can cause the interference. 
	In summary, the MIP codeword can be used for high peak gain, and the SE codeword can be used for the uniform gain.
	%
	%\begin{figure}[t]
	%	\centering
	%	\subfloat[SE codebook]{
	%		\includegraphics[width=.45\linewidth]{SE.eps}
	%		\label{SE codebook}
	%	}
	%	\subfloat[MIP codebook]{
	%		\includegraphics[width=.45\linewidth]{MIP.eps}
	%		\label{MIP codebook}
	%	}\\
	%	\subfloat[RVQ codebook]{
	%		\includegraphics[width=.45\linewidth]{RVQ.eps}
	%		\label{RVQ codebook}
	%	}
	%	\subfloat[codebook in \cite{B.Clerckx:2008}]{
	%		\includegraphics[width=.45\linewidth]{RBD.eps}
	%		\label{RBD codebook}
	%	}
	%	\caption{Normalized beamforming gains with $(M_h,M_v)=(4,8),\ (Q_h,Q_v)=(4,4)$.}
	%	\label{codebook}
	%\end{figure}

	In Fig. \ref{codebook}, the Tx side reference gain of the entire region is plotted.
	Even for different antenna dimension and division of entire region compared to Fig. \ref{codeword}, each codebook shows the same trend, i.e., the SE codebook has the most uniform beam pattern and the highest minimum gain while the MIP codebook has the highest average and peak gain at each region.
	% by selecting the codeword at each region as
	%\begin{align} 
	%	\bc(\theta_h,\theta_v )&=\argmax_{\bc\in C} \Bigg| \Bigg\{ b\bR(\phi)\Bigg( \begin{bmatrix} \sqrt{\frac{1}{1+\chi}} \zeta^{vv} \\ \sqrt{\frac{\chi}{1+\chi}} \zeta^{hv} \end{bmatrix}	\notag\\
	%	&\qquad  \otimes \bd_h (\pi \sin\theta_h \cos\theta_v)\otimes\bd_v (\pi \sin\theta_v) \Bigg)\Bigg\}^\mathrm{H} \bc\Bigg|^2, 
	%\end{align}
	%where $C=\left\{\bc:\bc=\bc^{(p,q)},p\in\{1,\cdots,Q_h \},q\in\{1,\cdots,Q_v \} \right\}$ is the codebook. 
	%With different antenna dimension and division of entire region from Fig. \ref{codeword}, consistent characteristics of each codebook can be seen. The SE codebook has the most uniform beam pattern and the highest minimum gain. The MIP codebook has the highest average gain and the highest peak gain at each region. Both the MIP codebook and the codebook in \cite{B.Clerckx:2008} have high peak gain, but the peak gain of the MIP codebook is higher than that of the codebook in \cite{B.Clerckx:2008}. In terms of the minimum gain, that of the MIP codebook is also higher than the codebook in \cite{B.Clerckx:2008}. 
	%It is obvious that the RVQ codebook is not suitable to the dual-polarization UPA scenarios. 

	\subsection{Data rate comparison}\label{sec4-b}
	
	We consider channels with a dominant LOS component and three NLOS components to compare data rates of different codebooks. The Rician $K$-factor is $K=13.2$ dB, and the codeword selection is based on the hard beam alignment using the receive signal power \eqref{align} as in \cite{J.Song:2015,S.Hur:2013}. We compute data rates based on the selected codewords
	\begin{align} 
	R_\mathrm{rate}={\E}\left[\log_2 \left(1+\frac{P}{\sigma^2} |\bc_{\mathrm{rx}}^\mathrm{H}\bH\bc_{\mathrm{tx}}|^2 \right) \right].	\end{align} 
	Although the orientation difference between the Tx and Rx antennas is assumed to be fixed in Section \ref{sec2}, we reflect the random movement of the actual antenna arrays, e.g., caused by wind turbulence, by setting $\phi$ as a uniform random variable in $[- \frac{\pi}{36}, \frac{\pi}{36}]$. The XPD $\chi$, similarly, considered as a random variable in $[0.25,0.35]$. The Tx and Rx BSs are assumed to know the mean values of the orientation difference and the XPD.

\begin{figure}[!t]
		\centering
		\subfloat[SNR $-15$ dB]{
			\includegraphics[width=1\columnwidth]{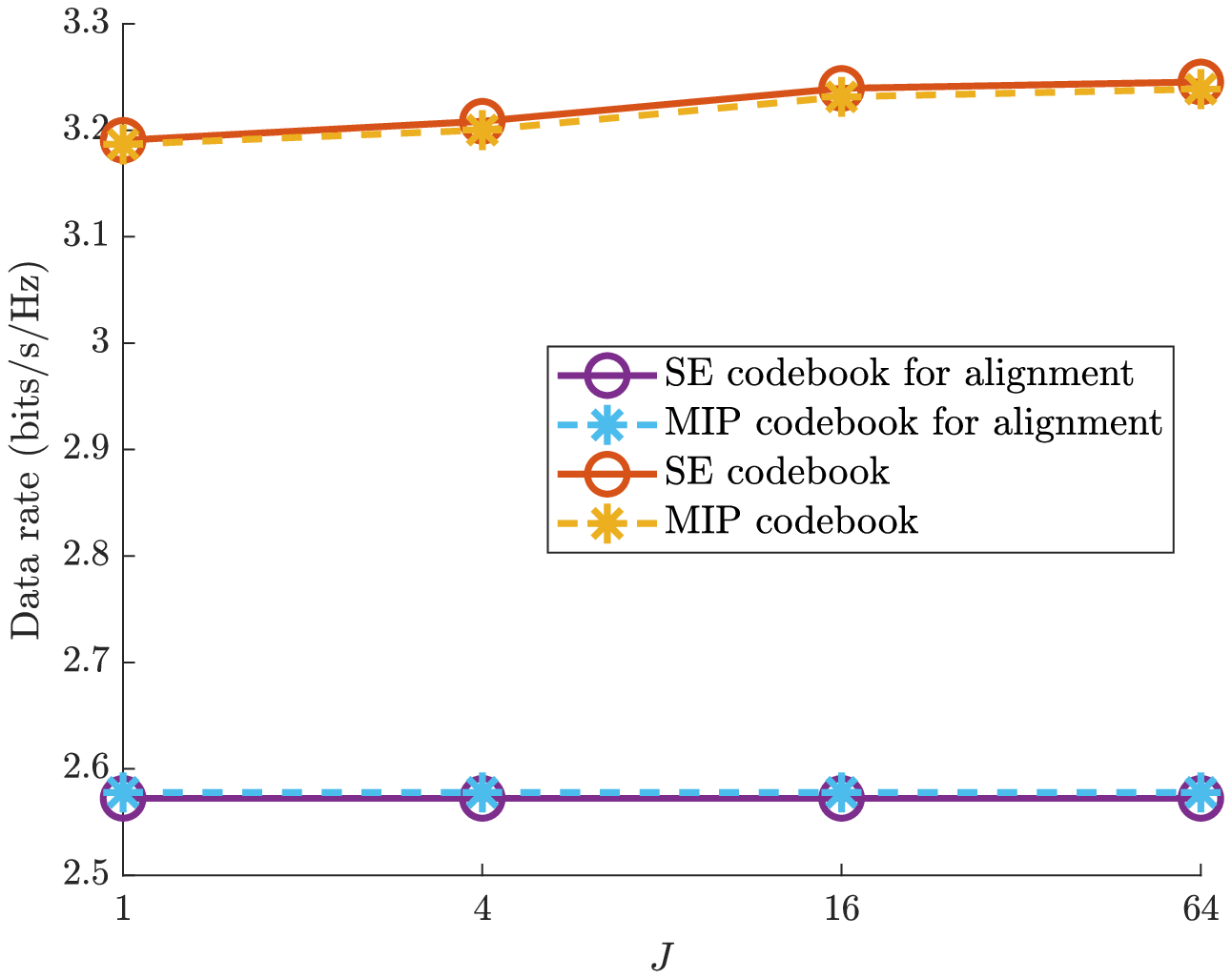}
			\label{data rate over J a}
		}\\
		\subfloat[SNR $-5$ dB]{
			\includegraphics[width=1\columnwidth]{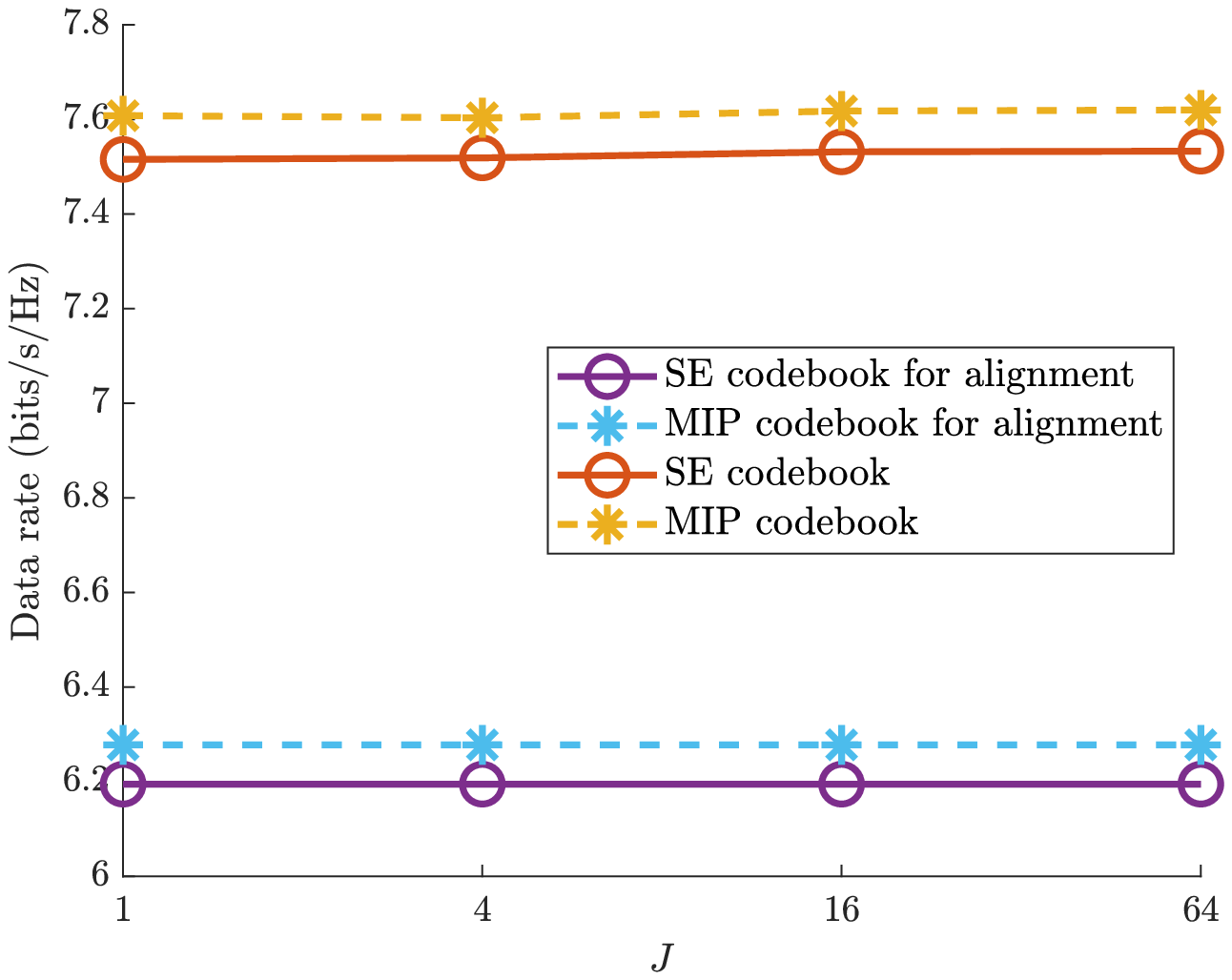}
			\label{data rate over J b}
		}
		\caption{Data rates of proposed codebooks with $(M_h,M_v)=(8,8),\ (Q_h,Q_v)=(6,6)$.}
		\label{data rate over J}
	\end{figure}

	The SE and MIP codebooks for beam alignment are constructed by setting $\alpha=1$, $\beta=1$, $\omega=1$, and $\upsilon=\sqrt{\chi}$. For each of the pilot sequences, we set $\alpha_j$'s and $\beta_j$'s as repetition of first two columns of $J\times J$ DFT matrix in order and inverse order, and $\kappa_1$ and $\kappa_2$ are set to be $1$. In Fig. \ref{data rate over J}, the data rates of proposed codebooks over the pilot length $J$ are depicted at SNR $-15$ dB and $-5$ dB. The data rates of SE and MIP codebooks for alignment, which use constant $\alpha$, $\beta$, $\omega$, and $\upsilon$, are illustrated for references. The data rates of proposed codebooks increase over $J$ with more accurate estimation of the ratios $\frac{e^{j\angle\zeta_{0}^{vv}}}{e^{j\angle\zeta_{0}^{vh}}}$ and $\frac{\xi_v}{\xi_h}$ at very low SNR of $-15$ dB as in Fig. \ref{data rate over J a}, but the increment is marginal. With $-5$ dB SNR in Fig. \ref{data rate over J b}, the data rate is almost the same over $J\ge1$. \red{This is because the ratio estimation with the pilot is conducted after the beam alignment, which gives the large effective training SNR even with a small pilot length.} This means that, with very small pilot overhead, the data rates of proposed codebooks outperform those of the codebooks for the beam alignment.

\begin{figure}[!t]
		\centering
		%	\subfloat[Data rate with 1 LOS and 3 NLOS]{
%		\includegraphics[width=1\linewidth]{D_data_rate_LOS_48_54_XPD_P.eps}
		\includegraphics[width=1\linewidth]{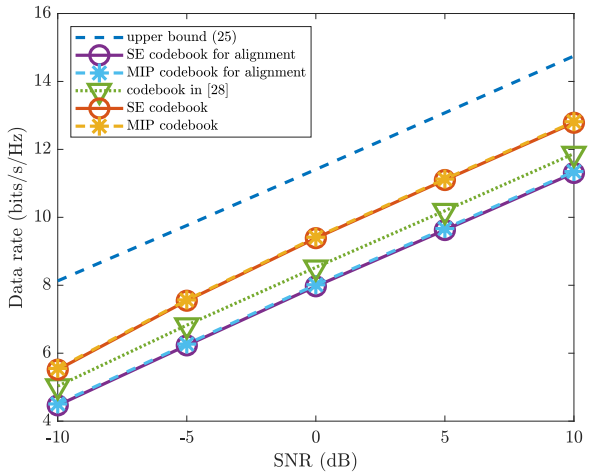}
		%		\label{data rate LOS}
		%	}\\
		%	\subfloat[Data rate with 3 NLOS]{
		%		\includegraphics[width=.95\linewidth]{D_data_rate_NLOS_48_54_new.eps}
		%		\label{data rate NLOS}
		%	}
		\caption{Data rate of codebooks with $(M_h,M_v)=\red{(8,8)},\ (Q_h,Q_v)=\red{(6,6)},\ J=1$.}
		\label{data rate of double}
	\end{figure}

	In Fig. \ref{data rate of double}, the data rates of proposed SE and MIP codebooks are compared with that of the codebook in \cite{B.Clerckx:2008}, and the upper bound \eqref{upper bound} is plotted for the reference. Before using pilot sequences to figure out the ratios of complex channel gains, the data rates of proposed codebooks for the beam alignment are lower than that of codebook in \cite{B.Clerckx:2008}. After the estimation of the ratios, the data rates of the proposed SE and MIP codebooks both outperform the codebook in \cite{B.Clerckx:2008}. This superiority of data rates is from fully exploiting the dual-polarization structure with partial channel information. 
	%In Fig. \ref{data rate NLOS}, meanwhile, the data rate of originally intended codebooks are slightly lower than that of codebooks for align near $0$ dB. This is by the inaccuracy of the ratio estimation at low SNR, and this can be overcome by pilot length adjustment depending on the SNR. 
	
	%Fig. vxvxvxvxvxvx shows data rates of codebooks, which are not doubled, in two scenarios, and the upper bound \eqref{upper bound} is depicted for comparison. The MIP codebook has the highest data rate over all the SNR range since it maximizes the beamforming gain at each covering region. Over all the SNR range, the data rate of SE codebook is very close to the MIP codebook and higher than other competing codebooks. In Fig. vxvxvxvxvxvxvx, the data rates of codebooks in the second scenario are depicted. The data rate slightly decreased overall, but the superiority of the SE and MIP codebooks is maintained. 
	
\begin{figure}[!t]
		\centering
		%	\subfloat[CDF of data rate with 1 LOS and 3 NLOS]{
		\includegraphics[width=1\linewidth]{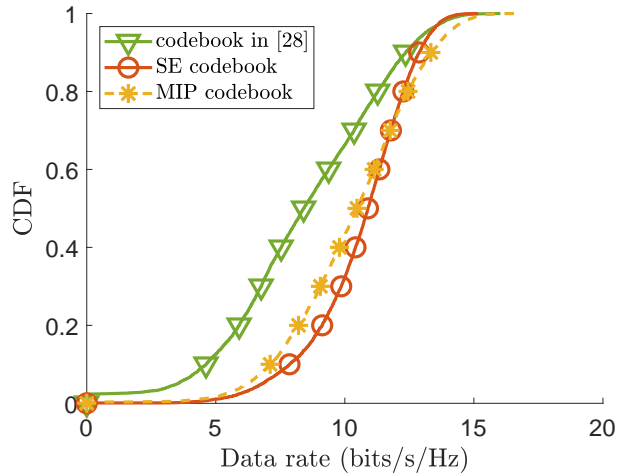}
		%		\label{CDF LOS}
		%	}\\
		%	\subfloat[CDF of data rate with 3 NLOS]{
		%		\includegraphics[width=1\linewidth]{CDF_NLOS_SNR10_168_88_new.eps}
		%		\label{CDF NLOS}
		%	}
		\caption{Data rate CDF of codebooks with $(M_h,M_v)=(16,8),\ (Q_h,Q_v)=(8,8),\ J=1$, SNR $0$ dB.}
		\label{CDF}
	\end{figure}

	In Fig. \ref{CDF}, the cumulative distribution function (CDF) of data rates of proposed codebooks, which are after the ratio estimation using the pilot sequences, are depicted at $0$ dB SNR. With its uniform beam pattern, the SE codebook has the most gentle slope at low data rates. The high average and peak gain of MIP codebook gives the high data rates. With very little exception at CDF around one, the proposed two codebooks outperform the one from \cite{B.Clerckx:2008}. 
	%Although the low minimum gain of the MIP codebook increases the slope at low data rate, the CDF of MIP codebook is always on the right of that of codebook in \cite{B.Clerckx:2008}.
	
	\section{Conclusion}\label{sec5}
	
	In this paper, we proposed the beamformer designs for mmWave MIMO backhaul systems with dual-polarization UPAs. The SE and MIP criteria are considered as specific examples for beam design optimizations. For each criterion, the reformulated optimization problem is solved with constraints regarding the dual-polarization UPA structure. The resulting beamformers depend on the partial channel information, and we also proposed the use of pilot sequences to figure out the required channel information. While solving the optimization problems, we have shown that, for both the Tx and Rx beamformings, the optimal dual-polarization UPA beamformers can be directly built from the optimal beamformers of single-polarization UPA sharing the same optimality. This finding shows that any previously designed beamformers for the single-polarization can be used to efficiently build dual-polarization beamformers satisfying the same performance metric by simple matrix operations.
	%The construction of dual-polarization beamformers from single-polarization beamformers means that any previously developed beamformers for the single-polarization can be used to build dual-polarization beamformers by simple matrix operations. 
	
	%\section*{Acknowledgment}
	%
	%acknowledgement

	\bibliographystyle{IEEEtran}
%	\bibliography{dual_pol_references}
	\bibliography{dual_pol_references_abbreviation}
	%\bibliography{dual_pol_references_original}

\end{document}